%% file: karoubi_3.tex
\documentclass[11pt,a4paper]{article}
\pdfoutput=1

\usepackage{./jheppub}
\usepackage[T1]{fontenc} 
\usepackage{graphicx}
\usepackage{amsfonts,amsmath,amssymb}
\usepackage{verbatim}
\usepackage{array}
\usepackage{mathrsfs}
\usepackage{mathtools}
\usepackage{yfonts}
\usepackage{dsfont}
\usepackage{bbm}
\usepackage{colonequals}
\usepackage{amscd}
\usepackage{slashed}
\usepackage{extarrows}
\usepackage{float}
\usepackage{afterpage}
\usepackage{subcaption}
\usepackage{wrapfig}

\usepackage{suffix,mathtools,cancel,bbm}
\usepackage{caption}
\usepackage{multirow}
\usepackage{enumerate}

	\usepackage{tikz-cd}
    \usetikzlibrary{decorations.pathmorphing}
\usepackage{scalerel}
\usepackage{stackengine}

\input{./auxx}

\usepackage{colortbl}
\definecolor{mygray}{gray}{0.93}

\usepackage{sectsty}
\allsectionsfont{\boldmath}

\usepackage{comment}

\usepackage{sectsty}
\allsectionsfont{\boldmath}

\usepackage{adjustbox}

\allowdisplaybreaks

\title{\boldmath On the Physics of Higher Condensation Defects}

\author[a]{Ibrahima Bah,}
\author[b]{Enoch Leung,}
\author[a]{and Thomas Waddleton} 
\affiliation[a]{William H.~Miller III Department of Physics and Astronomy, Johns Hopkins University, 3400 North Charles Street, Baltimore, MD 21218, U.S.A.}
\affiliation[b]{Max Planck Institute for Mathematics in the Sciences, Inselstraße 22, 04103 Leipzig, Germany}

\emailAdd{iboubah@jhu.edu, enoch.leung@mis.mpg.de, twaddle1@jhu.edu}

\abstract{We study the structure of topological defects for finite Abelian symmetries in quantum field theories, and argue on physical grounds that they satisfy the definition of a higher fusion category proposed by Johnson-Freyd. Our primary focus is on the requirement of Karoubi completeness, i.e.~the factorization conditions on higher condensation defects. We demonstrate this on a tree of such defects, constructed by successive higher gauging, explicitly using Lagrangian techniques in a concrete four-dimensional example, before turning to more general field theories. Along the way we also comment on the phenomenon where decoupled topological field theories appear as fusion coefficients. We further discuss the categorical role of anomalies, and how they may affect the properties of (higher) condensation defects.}


\newcommand{\cond}{\mathrel{\,\hspace{.75ex}\joinrel\rhook\joinrel\hspace{-.75ex}\joinrel\rightarrow}}
\newcommand{\acts}{\,\rotatebox[origin=c]{-90}{$\circlearrowright$}\,}

\begin{document}


\maketitle
\flushbottom

\addtocontents{toc}{\protect\setcounter{tocdepth}{2}}


\section{Introduction and Summary}\label{sec:introduction}
 
Symmetry has a rich history in physics where it underpins the best and most practically useful descriptions of physical systems. In the last decade and a half, there has been an explosion of research into the {\it structure} of global symmetries in a quantum field theory (QFT). Sparked from \cite{Gaiotto:2014kfa}, it is now understood that global symmetries can be studied through topological defects acting on a QFT. This has allowed for significant generalizations of the concept of symmetry, including higher-form symmetries \cite{Aharony:2013hda,Gaiotto:2017yup,Hsin:2018vcg,Hsieh:2020jpj}, higher-group symmetries \cite{Tachikawa:2017gyf,Cordova:2018cvg,Benini:2018reh}, and ultimately categorical symmetries \cite{Bhardwaj:2017xup,Thorngren:2019iar,Thorngren:2021yso,Bhardwaj:2022yxj,Delmastro:2022pfo,Brennan:2022tyl}.

Of these various generalizations, the latter is the most mathematically sophisticated and the broadest in scope, thus proving the least straightforward to realize in physical systems. While hints of categorical structures have been known in different corners of physics for some time \cite{Moore:1988qv,Dijkgraaf:1989pz,Kitaev:2005hzj}, a complete picture in physical settings has remained elusive, especially in (spacetime) dimensions higher than two. In spite of this, there has been considerable progress made in two main directions: studying such structures in physical examples, and understanding them from a deeper mathematical point of view. The former has found significant success in many sub-fields of physics including particle physics \cite{Choi:2022jqy,Cordova:2022fhg,Cordova:2022qtz,Choi:2023pdp,Cordova:2024ypu}, string theory \cite{Apruzzi:2021nmk,Apruzzi:2022rei,Heckman:2022muc,Heckman:2022xgu,Bah:2023ymy,Apruzzi:2023uma,Waddleton:2024iiv,Cvetic:2025kdn}, condensed matter \cite{Kong:2013aya,Barkeshli:2014cna,Kong:2017hcw,Kong:2020cie,Seiberg:2023cdc,Pace:2024oys,zhang2025hierarchyconstructionnonabelianfractional}, and quantum computing \cite{Tantivasadakarn:2022hgp,Gorantla:2024ocs}.

The latter direction has similarly found success in the rapidly developing language of higher fusion categories in mathematics \cite{2018arXiv181211933D,Decoppet:2022dnz,Decoppet:2024htz,Ferrer:2024vpn}. Along these lines, in \cite{Johnson-Freyd:2020usu} the author proposed a definition for a separable weak fusion $n$-category via a list of conditions to be satisfied. There, the definition was introduced to classify $(n+1)$-dimensional topological orders. Here, we adapt it to discuss the symmetry structure of an $(n+1)$-dimensional QFT. If the symmetry structure of a physical QFT naturally recovers the properties in this proposed definition, it can then immediately be analyzed with all the developed tools of higher category theory. It thus proves crucial to confirm that this is the case. 

Most of the necessary conditions for defining higher fusion categories, described in Section \ref{sec:4d_example}, are readily justified in a physical QFT. The requirement of Karoubi completeness, however, is less apparent from the outset. A primary aim in this paper is to elucidate what this requirement means, and how it manifests in the operator content of a physical QFT.

Our overall goal is an extension of this point: verify that the symmetry defects of a physical QFT naturally, from purely {\it physical} arguments, satisfy the definition of a fusion $n$-category as put forth in \cite{Johnson-Freyd:2020usu}. Importantly, we do not aim to {\it prove} that any QFT must satisfy this definition, and indeed we do not claim any such thing. We instead illustrate this structure through explicit operator manipulations, and explain the physical origin and interpretation of each aspect.

\subsection*{Summary of Main Results}

Starting with a concrete example of a 4d Euclidean QFT $\cT$ in the continuum, equipped with an anomaly-free 1-form global symmetry $\bZ_N^{(1)}$, we investigate some algebraic properties of the symmetry generators realized as codimension-2 topological defects $\cU_n(\Sigma^2)$ with $n \in \bZ_N$. They act on Wilson lines through linking in the 4d ambient space. The behavior of these defects under addition and fusion are analyzed at the level of correlators.

Beyond the standard symmetry generators, these exists a special family of topological defects known as {\it condensation defects} \cite{Roumpedakis:2022aik}. For any subgroup $\bZ_K^{(1)} \subseteq \bZ_N^{(1)}$, such a defect is geometrically constructed by summing over insertions of $\cU_n(\Sigma^2)$ over a submanifold $M^3$ of spacetime, i.e.
\begin{equation}
    \cC_K^{[1]}(M^3) \coloneqq \bigoplus_{\Sigma^2\in H_2(M^{3};\bZ_K)}\cU_1(\Sigma^2) \, ,
\end{equation}
where the sum is implicitly understood to correspond to those generating the subgroup $\bZ_K^{(1)}$. Equivalently, a continuum path integral expression is given by
\begin{equation}
    \cC_K^{[1]}(M^3) = \int\cD a_1\cD c_1\,\exp\left(2\pi i\int_{M^3} K a_1\wedge dc_1 + a_1\wedge B_2\right) \, ,
\end{equation}
where $a_1$ is a dynamical gauge field localized to $M^3$, and $B_2$ is the some background gauge field living in the bulk. The physical interpretation of the defect $\cC_K^{[1]}(M^3)$ is a gauging of the $\bZ_K^{(1)}$ symmetry with respect to the codimension-1 submanifold $M^3$, often referred to as a {\it 1-gauging}.

Recall that at the quantum level, an (ordinary) gauging is not simply a quotienting of the degrees of freedom in the 4d QFT. It also gives rise to new twisted sectors labeled by a (1-form) quantum symmetry corresponding to the Pontryagin dual, i.e.~the group of continuous maps from $\bZ_K$ to $U(1)$. We denote it as $\widehat{\bZ}_K$, which is also a $\bZ_K$ group.  
Before gauging, the Pontryagin dual describes the group that labels the charges of the original global symmetry $\bZ_K^{(1)}$. Physically, this is a generalized version of Fourier transform, which does not remove any information from the system, but rather rearrange it in some specific manner. After gauging, the resultant theory, which we denote as $\cT/\bZ_K^{(1)}$, has a $\widehat{\bZ}_K^{(1)} \times \bZ_{N/K}^{(1)}$ symmetry. It follows that one may gauge any anomaly-free subgroup of this pair of symmetries in $\cT/\bZ_K^{(1)}$, and particularly, gauging the entire $\widehat{\bZ}_K^{(1)}$ will return us to the original theory $\cT$.

The situation is analogous for 1-gauging, with minor modifications. When viewed as a 3d worldvolume theory in its own right, the condensation defect $\cC_K^{[1]}(M^3)$ admits a pair of symmetries, $\widehat{\bZ}_K^{(1)} \times \overline{\bZ}_{N/K}^{(0)}$, respectively generated by codimension-2 and codimension-1 defects relative to $M^3$, i.e.~lines and surfaces. The former factor corresponds to the Pontryagin dual symmetry, while the latter factor corresponds to the bulk 1-form symmetry restricted to the worldvolume of $\cC_K^{[1]}(M^3)$ (hence the shift in degree). This leads to a vast generalization of condensation defects. One can freely 1-gauge any anomaly-free subgroup of $\widehat{\bZ}_K^{(1)} \times \bZ_{N/K}^{(0)}$ on a codimension-1 submanifold $M^2 \subset M^3$ to obtain what we call a 2-condensation defect. For example, suppose we denote the symmetry generators of $\widehat{\bZ}_K^{(1)}$ as $\widehat{\cU}_{\hat{k}}(\gamma^1)$, then for any subgroup $\bZ_{\dot{K}}^{(1)} \subset \widehat{\bZ}_K^{(1)}$ of the Pontryagin dual symmetry, we define
\begin{equation}
    \cC_{\dot{K}}^{[2]}(M^2) \coloneqq \bigoplus_{\gamma^1\in H_1(M^2;\bZ_{\dot{K}})}\widehat{\cU}_{1}(\gamma^1) \, .
\end{equation}
Likewise, we may construct 2-condensation defects $\cC_{\dot{N}}^{[2]}(M^2)$ for subgroups $\bZ_{\dot{N}}^{(0)} \subseteq \overline{\bZ}_{N/K}^{(0)}$ of the residual symmetry. The superscript signifies the codimension of the condensation defect with respect to the total spacetime. Applying the same argument as before, these 2-condensations defects as individual worldvolume theories have their own symmetries.

\begin{figure}
    \centering
\tikzset{every picture/.style={line width=0.75pt}} 

\begin{tikzpicture}[x=0.75pt,y=0.75pt,yscale=-1,xscale=1]

\draw  [color={rgb, 255:red, 208; green, 2; blue, 27 }  ,draw opacity=1 ][fill={rgb, 255:red, 208; green, 2; blue, 27 }  ,fill opacity=0.2 ] (413.83,6.41) -- (413.83,188.33) -- (288.34,227.36) -- (288.34,45.43) -- cycle ;
\draw [color={rgb, 255:red, 144; green, 19; blue, 254 }  ,draw opacity=1 ]   (288.34,136.4) -- (413.83,97.37) ;
\draw  [fill={rgb, 255:red, 0; green, 0; blue, 0 }  ,fill opacity=1 ] (346.46,116.88) .. controls (346.46,114.33) and (348.53,112.26) .. (351.08,112.26) .. controls (353.64,112.26) and (355.71,114.33) .. (355.71,116.88) .. controls (355.71,119.44) and (353.64,121.51) .. (351.08,121.51) .. controls (348.53,121.51) and (346.46,119.44) .. (346.46,116.88) -- cycle ;

\draw (384.1,24) node [anchor=north west][inner sep=0.75pt]  [color={rgb, 255:red, 208; green, 2; blue, 27 }  ,opacity=1 ]  {$\mathcal{C}_{K}^{[ 1]}$};
\draw (295.34,136.17) node [anchor=north west][inner sep=0.75pt]  [color={rgb, 255:red, 144; green, 19; blue, 254 }  ,opacity=1 ]  {$\mathcal{C}_{\dot{K}}^{[ 2]}$};
\draw (338.17,83.14) node [anchor=north west][inner sep=0.75pt]  [color={rgb, 255:red, 0; green, 0; blue, 0 }  ,opacity=1 ]  {$\mathcal{C}_{\ddot{K}}^{[ 3]}$};
\end{tikzpicture}
    \caption{Condensation defects within condensation defects. Here the codimension-1 condensation defect $\cC_K^{[1]}$ is an operator in the ambient QFT, and is dressed with a codimension-2 condensation defect $\cC_{\dot{K}}^{[2]}$. This is itself dressed with {\it another} codimension-3 condensation defect $\cC_{\ddot{K}}^{[3]}$.}
    \label{fig:summary_condensation_of_condensation}
\end{figure}
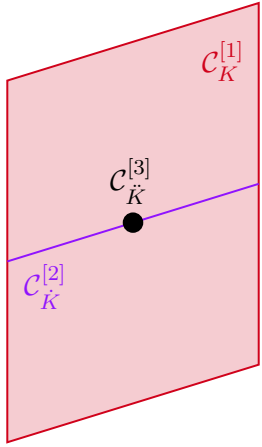

In fact, the construction above can be carried out in a nested fashion by successively 1-gauging the new set of symmetries arising in each step. Such a process typically terminates only when we reach the lowest possible dimension, i.e.~points. As a result, one obtains a rich hierarchy of higher condensation defects that are localized to the worldvolumes of their respective parent defects (e.g.~see Figure \ref{fig:summary_condensation_of_condensation} for an illustration). Even without the presence of more exotic non-invertible symmetries, the nested structure of this hierarchy alone already reveals that symmetries in QFTs exhibit a higher-categorical nature.

Moreover, we also want the higher category to encode the fusion data of our symmetry defects, and hence a natural candidate is a higher fusion category. Fusion 1-categories are known to describe 2d rational conformal field theories \cite{Moore:1988qv}. For higher-dimensional theories, the analogous language of higher fusion categories is still undergoing development. In this paper, we work with the construction by \cite{Gaiotto:2019xmp,Johnson-Freyd:2020usu}, where they employed a recursive definition of weak fusion $n$-categories for arbitrary $n$. This is to be contrasted with the definition of strict fusion $2$-categories in \cite{2018arXiv181211933D}.\footnote{See also \cite{2021arXiv210315150D} for a weakened notion of fusion 2-categories.}

In order to properly account for the symmetry data of a generic $d$-dimensional QFT $\cT$, we should assemble a $d$-category $\fC$. The collections $\fC_m$ of its $m$-morphisms, with $0 \leq m \leq d$, are schematically given by
\begin{itemize}
    \item $\fC_0 = \{\text{theories related to $\cT$ by (ordinary) gauging}\}$,
    \item $\fC_1 = \{\text{codimension-1 defects (in all theories) and interfaces between theories}\}$,
    \item $\fC_2 = \{\text{codimension-2 defects and interfaces between codimension-1 defects}\}$,
    \item $\fC_m = \{\text{codimension-$m$ defects and interfaces between codimension-$(m-1)$ defects}\}$,
    \item $\fC_d = \{\text{point defects and junctions between line defects}\}$.
\end{itemize}
Particularly, for $\fC$ to be a fusion $d$-category,\footnote{More precisely, we only demand $\text{End}(\cT)$ to be a fusion $(d-1)$-category for the purposes of this paper.} it has to satisfy a property known as {\it Karoubi completeness}. Other names used in the literature for the same notion, or closely related ones, include condensation/idempotent/orbifold/Cauchy completeness \cite{Frohlich:2009gb,Carqueville:2012dk,Carqueville:2017aoe,2018arXiv181211933D,Carqueville:2023aak,Kong:2024ykr,Carqueville:2025kxt,Muller:2025ext}. They serve as complementary perspectives on the issue in question.

Karoubi completeness of $\fC$ requires each of its idempotent $m$-morphisms for $1 \leq m \leq d$, which in our case correspond to (higher) condensation defects of various codimensions, to admit some factorization into a pair of $m$-morphisms satisfying certain relations. We physically identify such a pair as gauging interfaces between a given theory $\cT$ and its gauged counterpart, and more generally between a pair of parent condensation defects. On top of that, concrete Lagrangian descriptions of the defects and interfaces involved are presented to motivate our identifications.

Importantly, we elucidate in this paper the precise relation between the physical notion of condensation defects \cite{Roumpedakis:2022aik} and the mathematical notion of (categorical) condensation \cite{Gaiotto:2019xmp}. Simply put, the 1-condensation defect $\cC_K^{[1]}$ (constructed by 1-gauging a $p$-form $K^{(p)}$ symmetry in the bulk theory $\cT$)\footnote{For simplicity, here we assume $K$ to be a finite Abelian group.} is exactly the idempotent 1-morphism participating in the categorical condensation from the theory $\cT$ to the theory $\cT/K^{(p)}$. Analogous statements hold for other higher condensation defects localized to their respective parent condensation defects. Not only are condensation defects resulted from 1-gauging compatible with the notion of categorical condensations, but those obtained by $q$-gauging \cite{Roumpedakis:2022aik}, for $1 \leq q  \leq p+1$, also naturally fall under the same framework. We therefore see that Karoubi completeness acts as an organizing principle for us to establish the ladder of relations across different levels in our symmetry $d$-category $\fC$, i.e.~theories and topological defects of various codimensions. If the symmetry involved turns out to be anomalous, it provides more refined constraints on these components, and we analyze the subsequent outcomes.

Along the way, we also give, from a higher-categorical viewpoint, a sharper characterization of decoupled topological quantum field theories (TQFTs) as coefficients in the fusion product of two given topological defects. The initial observation in higher dimensional theories, that fusion coefficients for defects are decoupled TQFTs but not integers \cite{Roumpedakis:2022aik}, has been intriguing. We spell out some of the subtleties that arise when working with higher categories, and explain why decoupled TQFTs, rather than numbers, are good candidates for fusion coefficients in this context. In fact, such a realization constitutes an integral part of our argument for the Karoubi completeness of $\fC$.

The rest of this paper is organized as follows. In Section \ref{sec:4d_example}, we study the algebraic properties of topological defects in an example of a continuum 4d QFT, and explicitly construct the hierarchy of higher condensation defects present in the theory. We then continue to investigate the topological interfaces between these condensation defects in Section \ref{sec:karoubi_completeness}, and see how they play a role in the splitting property of idempotent condensation defects. In Section \ref{sec:categorical_structures}, we describe some higher-categorical ingredients paving the way for our subsequent analysis. The generalization of the results in our previous 4d example, to arbitrary spacetime dimensions and arbitrary anomaly-free finite Abelian symmetries, is presented in Section \ref{sec:gen_anomaly_free}, and phrased in the language of categorical condensations and Karoubi completeness. In Section \ref{sec:anomalous_symmetries}, we follow up by an interpretation of (higher) anomalies in terms of categorical data, and explore how they potentially modify the idempotence and splittability of condensation defects. Last but not least, we conclude in Section \ref{sec:discussion} with some discussion and outlook.


\section{Continuum 4d Example}\label{sec:4d_example}


As discussed in the previous section, we wish to ascertain the proper mathematical structure of symmetry operators in a QFT. In $d$-dimensions, a general QFT can be equipped with $p$-form global symmetries for any $0\leq p\leq d-1$, some of which may interact non-trivially to form the structure of a higher group \cite{Tachikawa:2017gyf,Benini:2018reh,Cordova:2018cvg,Kang:2023uvm}.\footnote{While we do discuss the presence and structure of $(d-1)$-form symmetries in the present work, we do not extend this to dual $(-1)$-form symmetries.} In terms of the operator content, this is the statement that there are codimension-$(p+1)$ topological operators that generate the symmetries of a QFT. For every codimension, there are at least two such operators: an identity operator $1^{[p+1]}$ and a zero operator $0^{[p+1]}$. As we will see, a QFT also generally has other non-trivial topological operators corresponding to global symmetries not necessarily described by standard groups.

We argue in this paper that these operators fit into a mathematical object known as a {\it higher fusion category}. Following the definition laid out by \cite{Johnson-Freyd:2020usu}, such an object is a higher category which is
\begin{itemize}
    \item $\bC$-linear,
    \item additive,
    \item monoidal,
    \item Karoubi-complete,
    \item fully dualizable.
\end{itemize}
The present section will focus on understanding from a {\it physical} perspective how the structures above emerge naturally. In order to set the notation and terminology in a familiar setting, we are going to first study a specific example to get our bearings straight before jumping into more general discussions. 

To this end, let us specialize to a 4d QFT $\cT$, defined on a spacetime $W^4$, with an anomaly-free $\bZ_N^{(1)}$ symmetry, where the superscript denotes the form degree of the symmetry.\footnote{For simplicity, we assume all manifolds to be spin, unless otherwise specified.} The theory $\cT$ admits the following operators:
\begin{itemize}
    \item identity $1^{[p+1]}(M^{d-p-1})$ and zero $0^{[p+1]}(M^{d-p-1})$ operators, for $0 \leq p \leq d-1$, supported on codimension-$p+1$ submanifolds $M^{d-p-1} \subset W^4$,
    \item symmetry generators $\cU_n(\Sigma^2)$ for the 1-form symmetry supported on codimension-2 submanifolds $\Sigma^2 \subset W^4$,
    \item condensations of the 1-form symmetry generators, and higher condensations atop these, which we will discuss in detail later.
\end{itemize}
The symmetry generators $\cU_n(\Sigma^2)$ can be expressed as 
\begin{equation}\label{eq:example_one-form_defect}
    \cU_{n}(\Sigma^2) = \exp\left(\frac{2\pi i n}{N}\int_{\Sigma^2}\cB_2\right)\,,
\end{equation}
where $\cB_2 \in H^2(W^4;\bZ_N)$ is a dynamical 2-form gauge field in $\cT$, $\Sigma^2\in H_2(W^4;\bZ_N)$ is a 2-cycle, and $n \in \bZ_N$ labels how the generator acts on charged operators. The charged operators are lines $\cW_q(\gamma^1)$ supported on a 1-cycle $\gamma^1\in H_1(W^4;\bZ_N)$ such that $\gamma^1$ and $\Sigma^2$ link in $W^4$. The label $q \in \bZ_N$ denotes the charge of the associated line operator that determines the correlation function,
\begin{equation}\label{eq:example_linking_correlator}
    \langle\cdots\cU_n(\Sigma^2)\,\cW_q(\gamma^1)\cdots\rangle = e^{\frac{2\pi i q n}{N}\,{\rm Link}(\gamma^1,\Sigma^2)}\langle\cdots \cW_q(\gamma^1)\,\cU_n(\Sigma^2)\cdots\rangle\,,
\end{equation}
where ${\rm Link}(-,-)$ denotes the linking number of its two arguments, and the dots denote other operator insertions that do not link with $\Sigma^2$. On the RHS, the operator $\cU_n(\Sigma^2)$ may be commuted to the right past all other insertions, where it acts trivially on the vacuum.\footnote{We do not consider spontaneous symmetry breaking in the present paper.}

By construction, $\cU_n(\Sigma^2)$ is a topological operator in the sense that the correlation functions are only sensitive to the topology of the support $\Sigma^2$. As a continuum description, we can model $\cB_2$ with a flat connection with $\bZ_N$-valued holonomies, rather than a discrete cocycle as above, as 
\begin{equation}
    \cU_n(\Sigma^2) = \exp\left(2\pi i n\int_{\Sigma^2} B_2\right)\,,
\end{equation}
where $dB_2 = 0$ \cite{Witten:1998wy}. Such data can be encoded through a $BF$-type action that constrains the holonomy values through a Lagrange multiplier field \cite{Maldacena:2001ss,Banks:2010zn}. If $B_2$ is flat, then $\cU_n(\Sigma^2)$ is invariant under smooth deformations of $\Sigma^2$ by virtue of Stokes' Theorem.

It is worth noting that \eqref{eq:example_linking_correlator} may be modified in the presence of an anomaly for the $\bZ_N^{(1)}$ symmetry, such as if the $\cW_q(\gamma^1)$ operators belong to a projective representation of the symmetry group. However, as mentioned previously, we will for now assume that the $\bZ_N^{(1)}$ symmetry is non-anomalous, and will return to discuss anomaly-related considerations in Section \ref{sec:anomalous_symmetries}.

\subsection{Linearity, Additivity, and Monoidality}

Before turning our attention to the construction of (higher) condensation operators, we can already deduce some non-trivial structure in the symmetry operators of $\cT$, relating to fundamental properties of a higher fusion category.

\paragraph{Monoidality:} Given two symmetry generators $\cU_n(\Sigma^2)$ and $\cU_{n'}(\Sigma^2)$, a natural question is whether there exists a notion of their product, corresponding to a composition of the two symmetry actions, i.e.~we wish to construct a new operator
\begin{equation}
    \left(\cU_n\otimes_2\cU_{n'}\right)(\Sigma^2) = \cU_n(\Sigma^2) \circ \cU_{n'}(\Sigma^2) \, .
\end{equation}
To be precise, let $\Sigma^2_x$ and $\Sigma^2_y$ denote two copies of $\Sigma^2$ given by the boundary components of the cylinder $\Sigma^2\times[x,y]\subset W^4$. We may then define the product operator through an operator product expansion (OPE) as follows,
\begin{equation}\label{eq:example_monoidal}
    \left(\cU_{n}\otimes_2\cU_{n'}\right)(\Sigma^2_x) := \lim_{y\to x}\left(\cU_n(\Sigma^2_x) \circ \cU_{n'}(\Sigma^2_y)\right)
\end{equation}
Particularly, this OPE denotes the {\it parallel} fusion of the two operators, such that the resulting product is defined on the same support $\Sigma^2_x$.

In a general QFT, the OPE \eqref{eq:example_monoidal} is not necessarily well-defined, and is generically sensitive to short-distance physics. Here, the symmetry generators $\cU_n(\Sigma^2)$, which commute with the Hamiltonian of $\cT$, are topological, so their OPE is insensitive to the short-distance physics. Since these operators generate a $\bZ_N^{(1)}$ symmetry, the parallel fusion respects the multiplication rule of $\bZ_N$, i.e.
\begin{equation}
    \left(\cU_n\otimes_2\cU_{n'}\right)(\Sigma^2) = \cU_{n+n'}(\Sigma^2) \, .
\end{equation}
Note that the fusion rule depends on the support only through its topology, independent of its position. We will discuss more examples and properties of fusion rules below in Section \ref{sec:decoupled_TQFTs}.\footnote{One may also be interested in the ``partial fusion'' of topological operators. To perform this, we can consider breaking a given defect up into two components separated by an identity interface, and performing a parallel fusion on only one of the components. This naturally gives a physical interpretation to graphical calculus diagrams, such as those in \cite{2018arXiv181211933D}.}

We can analogously define a parallel fusion $\otimes_{p+1}$ for any codimension-$(p+1)$ operators in $\cT$. The definition of each {\it monoidal product} $\otimes_{p+1}$ is similar, but the encoded data differs depending on the codimension of the extended operators. For $p \geq 1$, the product $\otimes_{p+1}$ will be Abelian, as operators of higher codimension have more possible transverse dimensions into which they can move.

However, this also allows for more possible modifications to the data of the product. Such data is related to the commutativity of the fusion as the operators move around one another in $W^4$. The possible modifications are related to (higher) anomalies of $\cT$, and we will discuss them later in Section \ref{sec:anomalous_symmetries}. In what follows, we will typically omit the subscript $p$ when the context is clear.

Although the OPE above has only been defined between operators of the same dimension, in many cases one may wish to consider the ``product'' of two operators defined with supports of {\it different} dimensions. For example, we may wish to compute the fusion of a surface and a line, $\cO_1(\Sigma^2)\otimes\cO_2(\gamma^1)$, with $\gamma^1 \subset \Sigma^2$, which will play a crucial role later as we discuss the properties of condensation defects.

To make sense of such a product in the context of the present discussion, we note that $\cO_2(\gamma^1)$ is a non-trivial line operator that can be used to ``dress'' the identity surface operator $1^{[2]}(\Sigma^2)$ along the codimension-1 locus $\gamma^1\subset \Sigma^2$. More formally, since $\cO_2(\gamma^1)$ acts as a map from $1^{[2]}(\Sigma^2)$ to itself, we can define a {\it new} surface operator denoted as
\begin{equation}\label{eq:dressed_operator}
    (\cO_2\acts 1^{[2]})(\Sigma^2)\,.
\end{equation}
The notation is chosen to emphasize that the operator in question should be considered as a 2-dimensional operator, but dressed with additional data on its worldvolume. Once we have constructed the dressed operator as in \eqref{eq:dressed_operator}, we can compute the desired OPE to be
\begin{align}
    \cO_1(\Sigma^2)\otimes \cO_2(\gamma^1) & \coloneqq \cO_1(\Sigma^2) \otimes_2 (\cO_2\acts 1^{[2]})(\Sigma^2)\nonumber\\
    & = (\cO_2\acts(\cO_1 \otimes_2 1^{[2]}))(\Sigma^2)\nonumber\\
    & = (\cO_2\acts\cO_1)(\Sigma^2)\,,
\end{align}
and find the final result as another dressed operator supported on $\Sigma^2$, again by regarding $\cO_2(\gamma^1)$ formally as some endomorphism of $\cO_1(\Sigma^2)$.

It should be noted that the dressed operator construction seems to provide an alternative fusion for operators of codimension greater than 1. For example, we may consider the fusion of two surface operators, $\cU_n(\Sigma^2)$ and $\cU_{n'}(\Sigma^2)$, not as surfaces but as dressed 3-dimensional operators,
\begin{equation}\label{eq:Eckmann-Hilton_product}
    \cU_n(\Sigma^2) \otimes' \cU_{n'}(\Sigma^2) \coloneqq (\cU_n\acts 1^{[1]})(M^3) \otimes_1 (\cU_{n'}\acts 1^{[1]})(M^3) = ((\cU_n\otimes_2\cU_{n'})\acts 1^{[1]})(M^3) \, ,
\end{equation}
with $\Sigma^2 \subset M^3$. Naïvely, this produces a new monoidal operation $\otimes'$ distinct from $\otimes_2$ and the OPE \cite{Bhardwaj:2022yxj}, but since the two operations distribute over one another as in \eqref{eq:Eckmann-Hilton_product}, one can show that what the above produces is equivalent to the one coming from the OPE \eqref{eq:example_monoidal}.\footnote{In mathematics, this equivalence is known as the Eckmann-Hilton argument.}

\paragraph{$\bC$-linearity:} Let us now focus on the topological operators supported on points. In our theory $\cT$, there are only the two topological local operators $0^{[4]}$ and $1^{[4]}$, but we may consider more general $\cO_i(x)$ here. As these operators are topological, it should be noted that the exact value $x$ takes is irrelevant for any physical measurements. Given any two local operators $\cO_1$ and $\cO_2$, trivial or not, we can take arbitrary linear combinations of them to create new local operators $\alpha\cO_1 + \beta\cO_2$ for $\alpha,\beta\in \bC$, such that correlation functions decompose linearly as
\begin{equation}\label{eq:linearity_linearcombo}
    \langle\cdots(\alpha\cO_1+\beta\cO_2)(x)\cdots\rangle = \alpha\langle\cdots\cO_1(x)\cdots\rangle + \beta\langle\cdots\cO_2(x)\cdots\rangle \, .
\end{equation}
Physically, this amounts to considering superpositions of (local) operators. Mathematically, this is the statement that the set of topological local operators has the structure of a complex vector space.



\paragraph{Additivity:} We can consider the superposition of operators beyond that of the topological local operators. For example, given two generators $\cU_n(\Sigma^2)$ and $\cU_{n'}(\Sigma^2)$ of the $\bZ_N^{(1)}$ symmetry in $\cT$, we can construct a new operator $(\cU_n\oplus\cU_{n'})(\Sigma^2)$ whose correlation functions satisfy
\begin{equation}
    \langle\cdots\left(\cU_{n}\oplus\cU_{n'}\right)(\Sigma^2)\cdots\rangle \coloneqq \langle\cdots\cU_{n}(\Sigma^2)\cdots\rangle + \langle\cdots\cU_{n'}(\Sigma^2)\cdots\rangle \, .
\end{equation}
Similarly, we can consider this notion of addition for operators of all codimensions. For two operators $\cO_1(M^{4-p-1})$ and $\cO_2(M^{4-p-1})$ supported on a codimension-$(p+1)$ manifold $M^{4-p-1} \subset W^4$, we define
\begin{equation}\label{eq:example_additivity}
    \langle\cdots\left(\cO_1\oplus\cO_2\right)(M^{4-p-1})\cdots\rangle \coloneqq \langle\cdots\cO_1(M^{4-p-1})\cdots\rangle + \langle\cdots\cO_2(M^{4-p-1})\cdots\rangle\,.
\end{equation}
Strictly speaking, there is a distinct $\oplus_{p+1}$ operation for each codimension, but to prevent notational clutter we will denote all such operations simply as $\oplus$.

The expressions above defining the operations $\oplus$ are very similar to the addition of topological local operators as in \eqref{eq:linearity_linearcombo}, but there is an important subtlety when considering the ``addition'' of higher-dimensional operators in a Euclidean QFT. If an extended operator is truly topological, then it can be placed along any submanifold of the spacetime $W^4$ including the timelike direction. Consequently, the interpretation of such an extended object is no longer a symmetry {\it operator} acting on states in a given Hilbert space, but a symmetry {\it defect} determining the boundary conditions of states in a {\it twisted} Hilbert space. In other words, if $\cO_1(M^{4-p-1})$ is a topological defect with one dimension along the time direction, a correlation function with an insertion of $\cO_1$ is computed as
\begin{equation}
    \langle\cdots\cO_1(M^{4-p-1})\cdots\rangle := \langle\cdots\rangle_{\cH_1}\,,
\end{equation}
where $\cH_1$ is the twisted Hilbert space associated with the operator $\cO_1(M^{4-p-1})$. One may also say that the presence of a defect specializes to a twisted sector of the overall Hilbert space for the QFT $\cT$.

Now let us consider two extended topological defects $\cO_{1,2}(M^{4-p-1})$ corresponding respectively to the twisted Hilbert spaces $\cH_{1,2}$. The direct sum $\cO_1\oplus\cO_2$ is characterized by a decomposition into different twisted sectors as
\begin{equation}\label{eq:example_non-neg_coeffs}
    \langle\cdots\left(\cO_1\oplus\cO_2\right)(M^{4-p-1})\cdots\rangle = \langle\cdots\rangle_{\cH_1} + \langle\cdots\rangle_{\cH_2} \, .
\end{equation}
In general, one can consider multiple copies of each twisted sector, which is equivalent to taking linear combinations of defects $\cO_1$ and $\cO_2$ with non-negative integer coefficients,
\begin{equation}
    \langle\cdots\left(n\cO_1\oplus m\cO_2\right)\cdots\rangle := \underbrace{\langle\cdots\rangle_{\cH_1} + \dots + \langle\cdots\rangle_{\cH_1}}_{n\text{ copies}}+ \underbrace{\langle\cdots\rangle_{\cH_2} + \dots + \langle\cdots\rangle_{\cH_2}}_{m\text{ copies}}\,,
\end{equation}
or simply as
\begin{equation}
    n\cO_1\oplus m\cO_2 := \underbrace{\cO_1\oplus\cdots\oplus\cO_1}_{n\text{ copies}}\oplus\underbrace{\cO_2\oplus\cdots\oplus\cO_2}_{m\text{ copies}}\,.
\end{equation}
Here we see that the coefficients $n$ and $m$ count the number of copies of $\cH_1$ and $\cH_2$, respectively, that appear in the decomposition. More precisely, they label the number of linearly independent projection and inclusion maps $\rho_i: \cH \leftrightarrows \cH_i: \iota_i$ between the total Hilbert space $\cH$ and the twisted sectors $\cH_i$. This justifies the requirement that $n$ and $m$ are non-negative integers, otherwise we cannot make sense of projecting into, say, half of a twisted sector.

The fact that the topological extended operators $\cO_i$ can always be supported on arbitrary timelike submanifolds (of the appropriate dimensions) means they can be regarded as defects which define twisted Hilbert spaces. This implies that it makes sense to construct direct sums of symmetry operators only with non-negative integer coefficients.\footnote{As we will see in Section \ref{sec:decoupled_TQFTs}, the precise interpretation of these coefficients needs to be refined when considering the higher-categorical structure of the symmetry defects.}

\begin{figure}
    \centering
\tikzset{every picture/.style={line width=0.75pt}} 
\begin{tikzpicture}[x=0.75pt,y=0.75pt,yscale=-1,xscale=1]
\draw  [fill={rgb, 255:red, 144; green, 19; blue, 254 }  ,fill opacity=0.5 ] (211.66,117.78) .. controls (211.66,124.63) and (238.6,130.18) .. (271.83,130.18) .. controls (305.06,130.18) and (332,124.63) .. (332,117.78) .. controls (332,110.93) and (358.94,105.38) .. (392.17,105.38) .. controls (425.4,105.38) and (452.34,110.93) .. (452.34,117.78) -- (452.34,216.97) .. controls (452.34,210.12) and (425.4,204.57) .. (392.17,204.57) .. controls (358.94,204.57) and (332,210.12) .. (332,216.97) .. controls (332,223.82) and (305.06,229.37) .. (271.83,229.37) .. controls (238.6,229.37) and (211.66,223.82) .. (211.66,216.97) -- cycle ;
\draw  [fill={rgb, 255:red, 208; green, 2; blue, 27 }  ,fill opacity=0.5 ] (90,117.78) .. controls (90,124.64) and (103.62,130.21) .. (120.41,130.21) .. controls (137.21,130.21) and (150.83,124.64) .. (150.83,117.78) .. controls (150.83,110.92) and (164.45,105.35) .. (181.24,105.35) .. controls (198.04,105.35) and (211.66,110.92) .. (211.66,117.78) -- (211.66,217.21) .. controls (211.66,210.34) and (198.04,204.78) .. (181.24,204.78) .. controls (164.45,204.78) and (150.83,210.34) .. (150.83,217.21) .. controls (150.83,224.07) and (137.21,229.63) .. (120.41,229.63) .. controls (103.62,229.63) and (90,224.07) .. (90,217.21) -- cycle ;
\draw  [fill={rgb, 255:red, 74; green, 144; blue, 226 }  ,fill opacity=0.5 ] (452.34,117.55) .. controls (452.34,124.41) and (465.96,129.97) .. (482.76,129.97) .. controls (499.55,129.97) and (513.17,124.41) .. (513.17,117.55) .. controls (513.17,110.68) and (526.79,105.12) .. (543.59,105.12) .. controls (560.38,105.12) and (574,110.68) .. (574,117.55) -- (574,216.97) .. controls (574,210.11) and (560.38,204.54) .. (543.59,204.54) .. controls (526.79,204.54) and (513.17,210.11) .. (513.17,216.97) .. controls (513.17,223.84) and (499.55,229.4) .. (482.76,229.4) .. controls (465.96,229.4) and (452.34,223.84) .. (452.34,216.97) -- cycle ;
\draw [color={rgb, 255:red, 0; green, 0; blue, 0 }  ,draw opacity=1 ][line width=2.25]    (211.66,117.78) -- (211.66,216.97) ;
\draw [color={rgb, 255:red, 0; green, 0; blue, 0 }  ,draw opacity=1 ][line width=2.25]    (452.34,117.78) -- (452.34,216.97) ;
\draw    (222.45,105.67) .. controls (236.6,93.79) and (223.3,128.02) .. (217.7,147.65) ;
\draw [shift={(217.2,149.43)}, rotate = 285.26] [color={rgb, 255:red, 0; green, 0; blue, 0 }  ][line width=0.75]    (10.93,-3.29) .. controls (6.95,-1.4) and (3.31,-0.3) .. (0,0) .. controls (3.31,0.3) and (6.95,1.4) .. (10.93,3.29)   ;
\draw    (437.17,102.17) .. controls (419.06,96.51) and (432.79,132.03) .. (444.28,150.14) ;
\draw [shift={(445.34,151.77)}, rotate = 236.31] [color={rgb, 255:red, 0; green, 0; blue, 0 }  ][line width=0.75]    (10.93,-3.29) .. controls (6.95,-1.4) and (3.31,-0.3) .. (0,0) .. controls (3.31,0.3) and (6.95,1.4) .. (10.93,3.29)   ;

\draw (299.79,160.46) node [anchor=north west][inner sep=0.75pt]    {$n\mathcal{O}_{1} \oplus m\mathcal{O}_{2}$};
\draw (502.15,161.96) node [anchor=north west][inner sep=0.75pt]    {$\mathcal{O}_{j}$};
\draw (140.22,159.96) node [anchor=north west][inner sep=0.75pt]    {$\mathcal{O}_{i}$};
\draw (206.7,92.4) node [anchor=north west][inner sep=0.75pt]    {$\iota _{i}$};
\draw (441.07,85.69) node [anchor=north west][inner sep=0.75pt]    {$\rho _{j}$};
\end{tikzpicture}
    \caption{The direct sum operator $n\cO_1\oplus m\cO_2$ and interfaces to constituent operators. The interface $\iota_i$ mediates from $\cO_i$ to the sum $n\cO_1\oplus m\cO_2$, while the interface $\rho_j$ mediates from the sum $n\cO_1\oplus m\cO_2$ to $\cO_j$. Both $i$ and $j$ can take values $1$ or $2$.}
    \label{fig:direct_sum}
\end{figure}
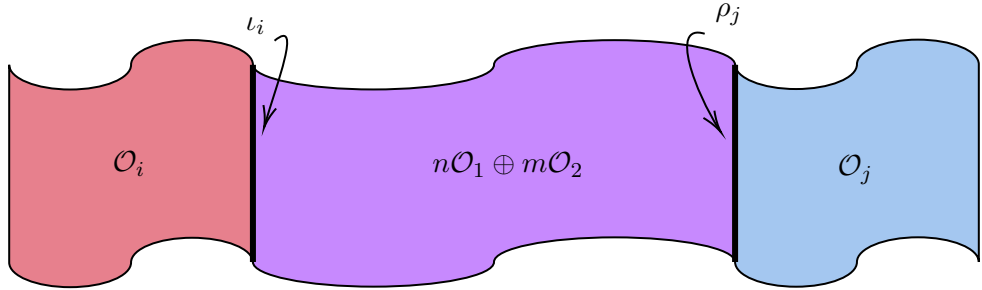

Geometrically, the maps $\rho_i$ and $\iota_i$ realize interfaces between the sum $n\cO_1\oplus m\cO_2$ and the constituent defects $\cO_1$ and $\cO_2$ as in Figure \ref{fig:direct_sum}. Specifically, there are $n$ possible pairs of interfaces $(\iota_1,\rho_1)$ and $m$ possible pairs of interfaces $(\iota_2,\rho_2)$. In the same way we cannot make sense of a non-integer number of projection maps into a twisted sector, we cannot make sense of a non-integer number of interfaces between two symmetry defects.


\paragraph{A tale of two additions:} The special case of topological local operators/defects deserves special attention. In general, the interpretations as operators or defects are distinguished by whether they are supported on spacelike or timelike submanifolds, but there is no well-defined notion for a point to be spacelike or timelike. Nonetheless, we can effectively differentiate between these two notions (i.e.~operators vs.~defects) by studying how they behave under the two inequivalent ``additions,'' $+$ and $\oplus$, respectively describing superposition of operators and decomposition of Hilbert spaces.

If we regard two local operators as linear transformations, say, some $n \times n$ matrices $A,B$, acting on states $|\varphi\rangle$ in a vector space $V$, then we define the local operator
\begin{equation}
    A + B: V \to V
\end{equation}
as the $n \times n$ matrix given by the entry-wise addition of $A$ and $B$. At the level of correlators, we obtain
\begin{equation}
    \langle A+B \rangle: V^\ast \otimes V \to \mathbb{C}
\end{equation}
as a single complex number. Here $V^\ast$ denotes the vector space dual to $V$, in which the dual vectors $\langle\varphi|$ live. Therefore, $A+B$ should be viewed as a linear transformation acting on a {\it single} Hilbert space.

On the contrary, given a state $|\varphi\rangle$ as a column vector (with $n$ entries) in $V$, there is a diagonal map sending $|\varphi\rangle$ to a column vector in $V \oplus V$, by populating the first and second $n$ entries respectively with $|\varphi\rangle$. We then define the direct sum of $A$ and $B$ to be
\begin{equation}
    A \oplus B: V \oplus V \to V \oplus V \, ,\label{eq:direct_sum_vector_spaces}
\end{equation}
where $A \oplus B = \text{diag}(A,B)$ is a block-diagonal $2n \times 2n$ matrix. The corresponding correlator
\begin{equation}
    \langle A \oplus B \rangle: (V^\ast \oplus V^\ast) \otimes (V \oplus V) \to \mathbb{C}
\end{equation}
gives a complex number again since $V \oplus V$ is itself a vector space. In other words, $A \oplus B$ should be viewed as a linear transformation acting on a ``total Hilbert space'' consisting of two {\it identical} copies of the original Hilbert space.

The construction above clarifies the statement that non-trivial topological local operators in a $d$-dimensional QFT imply a $(d-1)$-form symmetry, in which case the theory decomposes into its constituent {\it universes}. The total Hilbert space of the theory can be written as a disjoint union of Hilbert spaces of the individual universes \cite{Hellerman:2006zs, Tanizaki:2019rbk, Cherman:2020cvw, Komargodski:2020mxz, Sharpe:2022ene}. This is exactly the analogue of \eqref{eq:example_non-neg_coeffs} but for topological local operators, so that the RHS decomposes into universes rather than sectors of a single given Hilbert space, which is sharply different from the standard addition in \eqref{eq:linearity_linearcombo}.

Similarly to before, the argument above shows that it only makes sense to consider direct sums of ``universes'' with non-negative integer coefficients, as reflected in the number of copies of $V$ in \eqref{eq:direct_sum_vector_spaces}. It is worth noting that the operators $A+B$ and $A \oplus B$ are technically distinct in nature, but from a practical perspective, the correlators (which are simply {\it numbers}) computed using the two notions of additions agree with each other, at least when the coefficients are non-negative integers.

Systematically, for a given (co)dimension, it is convenient for one to fix a basis of symmetry defects, in which arbitrary {\it defects} can be expressed as non-negative integer linear combinations of.\footnote{There is a natural notion of basis based on whether its elements are {\it simple} in a suitably categorical sense, as described in Appendix \ref{app:categorical_notions}.} However, one may also define {\it operators} as more general linear combinations with complex-number coefficients, provided that they are strictly supported on spacelike submanifolds. The relations between correlators involving defects and operators can be deduced using (analogues of) the aforementioned maps. For simplicity, we will henceforth restrict our attention only to topological defects, i.e.~those constructed as non-negative integer linear combinations, but use the terms ``operators'' and ``defects'' synonymously unless otherwise specified.

\subsection{Higher Condensation Defects}\label{subsec:condensation_defects}

Using the defects $\mathcal{U}_n(\Sigma^2)$ generating the $\mathbb{Z}_K^{(1)}$ symmetry in $\mathcal{T}$, one can build a special family of topological defects, i.e.~{\it condensation defects}, via a procedure known as higher gauging \cite{Roumpedakis:2022aik}. Specifically, a $q$-gauging of a (finite) global symmetry $G^{(p)}$, for $0 \leq q \leq p+1$, is defined as a sum over insertions of codimension-$(p+1)$ $G^{(p)}$-defects on a codimension-$q$ submanifold $M^{d-q} \subseteq W^d$. In this terminology, an ordinary gauging is simply a 0-gauging.

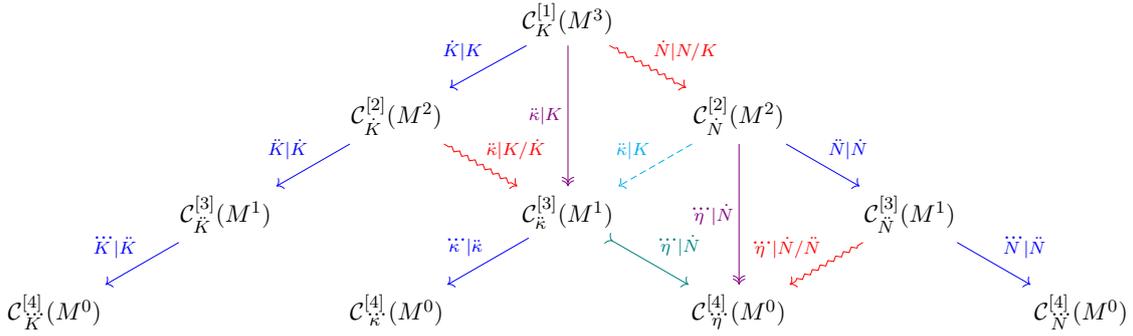
\begin{figure}[t!]
	\centering
        \adjustbox{scale=0.83,center}{%
	\begin{tikzcd}
		& & & \mathcal{C}_K^{[1]}(M^3) \arrow[dl, blue, "\dot{K}|K"'] \arrow[dr, red, squiggly, "\dot{N}|N/K"] \arrow[dd, violet, two heads, "\ddot{\kappa}|K"'] & & &\\
		& & \mathcal{C}_{\dot{K}}^{[2]}(M^2) \arrow[dl, blue, "\ddot{K}|\dot{K}"'] \arrow[dr, red, squiggly, "\ddot{\kappa}|K/\dot{K}"] & & \mathcal{C}_{\dot{N}}^{[2]}(M^2) \arrow[dl, cyan, dashed, "\ddot{\kappa}|K"'] \arrow[dr, blue, "\ddot{N}|\dot{N}"] \arrow[dd, violet, two heads, "\dddot{\eta}|\dot{N}"'] & &\\
		& \mathcal{C}_{\ddot{K}}^{[3]}(M^1) \arrow[dl, blue, "\dddot{K}|\ddot{K}"'] & & \mathcal{C}_{\ddot{\kappa}}^{[3]}(M^1) \arrow[dl, blue, "\dddot{\kappa}|\ddot{\kappa}"'] \arrow[dr, teal, tail, "\dddot{\eta}|\dot{N}"] & & \mathcal{C}_{\ddot{N}}^{[3]}(M^1) \arrow[dl, red, squiggly, "\dddot{\eta}|\dot{N}/\ddot{N}"'] \arrow[dr, blue, "\dddot{N}|\ddot{N}"]&\\
		\mathcal{C}_{\dddot{K}}^{[4]}(M^0) & & \mathcal{C}_{\dddot{\kappa}}^{[4]}(M^0) & & \mathcal{C}_{\dddot{\eta}}^{[4]}(M^0) & & \cC^{[4]}_{\dddot{N}}(M^0)
	\end{tikzcd}
        }
	\caption{\label{fig:condensation_defects_diagram} The nested relations between various condensation defects of different dimensions. \textcolor{blue}{Solid} and \textcolor{violet}{double-headed} arrows respectively denote 1- and 2-gauging of the Pontryagin dual symmetry in the relevant worldvolume theory. \textcolor{red}{Squiggly} arrows denote the 1-gauging of the residual symmetry that is present on the condensation defect. The \textcolor{cyan}{dashed} arrow denotes a 1-gauging of a global symmetry descending from a dual symmetry in the original 3-condensation defect $\mathcal{C}_K^{[1]}(M^3)$. The arrow \textcolor{teal}{with a tail} is a similar special case, denoting a 1-gauging of the residual symmetry arising from the dashed arrow. Along each arrow, we have indicated the divisibility conditions for the gauging. For example, on the top right we must have that $\dot{N}$ divides $N/K$ in order to 1-gauge the subgroup $\bZ_{\dot{N}}\subseteq\overline{\bZ}_{N/K}^{(0)}$.}
\end{figure}

The key idea of this subsection is that the process of higher gauging can be carried out iteratively all the way down to codimension-$d$ submanifolds, i.e.~points. In each step, the condensation defect obtained from the previous step is treated as a (T)QFT in its own right, admitting its own global symmetries. With our 4-dimensional QFT $\mathcal{T}$, we are going to explicitly construct below a rich family of condensation defects spanning all codimensions, as depicted graphically in Figure \ref{fig:condensation_defects_diagram}. Importantly, all these defects will stem only from the input data of $\cT$ without any additional assumption.

For notational convenience, we will hereafter refer to condensation defects supported on codimension-$n$ submanifolds (with respect to $d$-dimensional spacetime) as {\it n-condensation defects}. The relation between $n$-condensation defects and $q$-gauging is as follows. A 1-condensation defect is necessarily obtained by 1-gauging some symmetry of $\mathcal{T}$. On the other hand, a generic $n$-condensation defect typically does not come from a direct $n$-gauging in $\cT$. For example, it may be obtained by going through $n$ recursive stages of 1-gauging, as we will see below. Another motivation for the choice of labeling is that $n$-condensation defects will always be $n$-morphisms in the higher category that we introduce later in Section \ref{sec:categorical_structures}, regardless of the sequences of higher gauging used to construct them.\footnote{Note that the notion of (categorical) $n$-condensation as defined in \cite{Gaiotto:2019xmp}, which we will also review in Section \ref{sec:categorical_structures}, is determined by the dimension of the defect, rather than its codimension. In other words, an $n$-condensation defect corresponds (primarily) to the data of a $(d+1-n)$-condensation but not an $n$-condensation. To avoid confusion, we will refrain from using the latter terminology.}

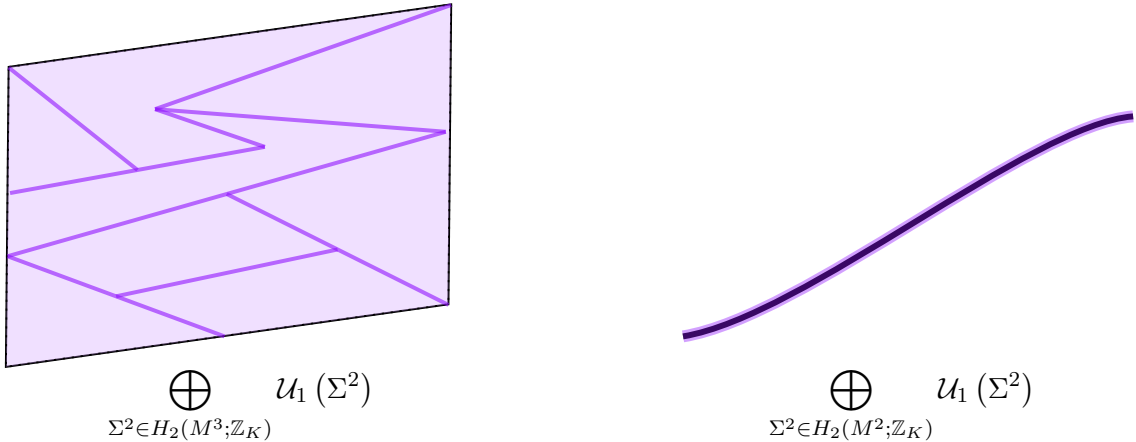
\begin{figure}
    \centering
\tikzset{every picture/.style={line width=0.75pt}} 

\begin{tikzpicture}[x=0.75pt,y=0.75pt,yscale=-1,xscale=1]

\draw [color={rgb, 255:red, 0; green, 0; blue, 0 }  ,draw opacity=1 ][line width=2.25]    (388.67,196) .. controls (449.33,186) and (562,89.33) .. (613.33,85.33) ;
\draw  [color={rgb, 255:red, 0; green, 0; blue, 0 }  ,draw opacity=1 ][fill={rgb, 255:red, 255; green, 255; blue, 255 }  ,fill opacity=1 ] (52.05,60.46) -- (272.93,28.98) -- (271.55,179.86) -- (50.66,211.33) -- cycle ;
\draw  [color={rgb, 255:red, 0; green, 0; blue, 0 }  ,draw opacity=0.6 ][fill={rgb, 255:red, 208; green, 152; blue, 255 }  ,fill opacity=0.3 ][dash pattern={on 0.84pt off 2.51pt}] (52.05,60.46) -- (272.93,28.98) -- (271.55,179.86) -- (50.66,211.33) -- cycle ;
\draw [color={rgb, 255:red, 144; green, 19; blue, 254 }  ,draw opacity=0.6 ][line width=1.5]    (272.94,29.31) -- (125.11,81.7) ;
\draw [color={rgb, 255:red, 144; green, 19; blue, 254 }  ,draw opacity=0.6 ][line width=1.5]    (125.11,81.7) -- (179.94,100.67) ;
\draw [color={rgb, 255:red, 144; green, 19; blue, 254 }  ,draw opacity=0.6 ][line width=1.5]    (179.94,100.67) -- (52.74,123.91) ;
\draw [color={rgb, 255:red, 144; green, 19; blue, 254 }  ,draw opacity=0.6 ][line width=1.5]    (270.25,93.08) -- (125.11,81.7) ;
\draw [color={rgb, 255:red, 144; green, 19; blue, 254 }  ,draw opacity=0.6 ][line width=1.5]    (51.5,155.69) -- (270.25,93.08) ;
\draw [color={rgb, 255:red, 144; green, 19; blue, 254 }  ,draw opacity=0.6 ][line width=1.5]    (271.56,180.17) -- (160.88,124.38) ;
\draw [color={rgb, 255:red, 144; green, 19; blue, 254 }  ,draw opacity=0.6 ][line width=1.5]    (159.81,196) -- (51.5,155.69) ;
\draw [color={rgb, 255:red, 144; green, 19; blue, 254 }  ,draw opacity=0.6 ][line width=1.5]    (105.66,175.84) -- (216.22,152.28) ;
\draw [color={rgb, 255:red, 144; green, 19; blue, 254 }  ,draw opacity=0.6 ][line width=1.5]    (116.34,112.29) -- (52.05,60.78) ;
\draw [color={rgb, 255:red, 144; green, 19; blue, 254 }  ,draw opacity=0.4 ][line width=4.5]    (388.67,196) .. controls (449.33,186) and (562,89.33) .. (613.33,85.33) ;

\draw (100,210) node [anchor=north west][inner sep=0.75pt]    {$\displaystyle\bigoplus _{\Sigma ^{2} \in H_{2}\left( M^{3};\bZ_K\right)}\mathcal{U}_{1}\left( \Sigma ^{2}\right)$};
\draw (430,210) node [anchor=north west][inner sep=0.75pt]    {$\displaystyle\bigoplus _{\Sigma ^{2} \in H_{2}\left( M^{2};\bZ_K\right)}\mathcal{U}_{1}\left( \Sigma ^{2}\right)$};
\end{tikzpicture}
\caption{\label{fig:n-condensation_picture}Depictions of higher gauging for a subgroup $\bZ_K^{(1)}\subseteq\bZ_N^{(1)}$ of the 1-form symmetry in $\cT$. On the left, a 1-condensation defect made from 1-gauging the symmetry on a 3-dimensional submanifold $M^3\subset W^4$. On the right, a 2-condensation defect made from a 2-gauging of the same symmetry on a 2-dimensional submanifold $M^2\subset W^4$. In both cases a transverse direction has been suppressed.}
\end{figure}

\paragraph{1-condensation defects:} Starting with a subgroup $\bZ_K^{(1)}\subseteq\bZ_N^{(1)}$, we may either 1- or 2-gauge the symmetry to construct a 1- or 2-condensation defect respectively. The two analyses are analogous, so to be concrete let us focus on the former. To construct a 1-condensation defect from an arbitrary subgroup of the 1-form symmetry, we sum over insertions of the $\bZ_K^{(1)}$ generators on a codimension-$1$ submanifold $M^{3}\subset W^4$,
\begin{equation}\label{eq:example_condensation_defect}
    \cC^{[1]}_K(M^{3}) \coloneqq \bigoplus_{\Sigma^2\in H_2(M^{3};\bZ_K)}\cU_1(\Sigma^2)\,,
\end{equation}
where we have omitted an overall normalization factor.\footnote{There can be an obstruction to such a 1-gauging, or more general $q$-gauging, coming from a higher anomaly in $\cT$. As before, we postpone such subtleties to Section \ref{sec:anomalous_symmetries}.} Here we expressed the condensation defect $\cC^{[1]}_K(M^{3})$ as a formal sum of the operators $\cU_1(\Sigma^2)$ with {\it different} supports, namely, elements (representatives) of the homology group $H_2(M^{3};\bZ_K)$. Geometrically, one may visualize the former as a mesh of the latter localized on $M^3$, as in Figure \ref{fig:n-condensation_picture}.

It is convenient for us to realize the formal expression above by means of a path integral,
\begin{equation}\label{eq:example_continuum}
    \cC_K^{[1]}(M^3) = \int\cD a_1\cD c_1\,\exp\left(2\pi i\int_{M^3} K a_1\wedge dc_1 + a_1\wedge B_2\right)\,,
\end{equation}
where $a_1$ and $c_1$ are dynamical gauge fields localized on $M^3$, and $B_2$ is the continuum analogue of $\cB_2$ in \eqref{eq:example_one-form_defect}. The classical equation of motion for $c_1$ is $K da_1 = 0$, so we obtain a cocycle $K a_1 \sim \mathcal{A}_1 \in Z^1(M^3;\mathbb{Z}_K)$ on shell. Up to an overall normalization, the path integral then localizes to a finite sum over representatives of the cohomology group $H^1(M^3;\mathbb{Z}_k)$, i.e.
\begin{equation}
    \mathcal{C}_K^{[1]}(M^3) = \sum_{\mathcal{A}_1 \in H^1(M^3;\mathbb{Z}_K)} \exp\bigg(\frac{2\pi i}{K} \int \mathcal{A}_1 \cup \mathcal{B}_2\bigg) \, .
\end{equation}
Via Poincaré duality (with respect to $M^3$), $\mathcal{A}_1$ is dual to a 2-cycle which we identify as $\Sigma^2$ in \eqref{eq:example_condensation_defect}.

The action \eqref{eq:example_continuum} describes an (untwisted) 3d Dijkgraaf-Witten theory coupled to $\cT$ through the ``background'' gauge field $B_2$ \cite{Dijkgraaf:1989pz,Kapustin:2014gua}. As a 3-dimensional QFT, $\cC_K^{[1]}(M^3)$ has the following two global symmetries living on its worldvolume.
\begin{itemize}
    \item A 1-form symmetry $\widehat{\bZ}_K^{(1)}$ arising as the Pontryagin dual, or quantum symmetry, of the $\bZ_K^{(1)}$ symmetry that was 1-gauged to produce the condensation defect \cite{Gaiotto:2014kfa, Roumpedakis:2022aik}. This symmetry can be understood as the electromagnetic-dual of the $\bZ_K$ symmetry, but suitably adjusted to higher gauging. The quantum symmetry $\widehat{\bZ}_K^{(1)}$ is generated by the line defects
    \begin{equation}\label{eq:quantum_symmetry_step1}
        \widehat{\cU}_{\hat{k}}(\gamma^1) \coloneqq \exp\bigg(\frac{2\pi i \hat{k}}{K} \, \langle\gamma^1,\Sigma^2\rangle\bigg) = \exp\bigg(\frac{2\pi i \hat{k}}{K}\int_{\gamma^1}\cA_1\bigg)\,,
    \end{equation}
    where $\langle-,-\rangle: H_1(M^3;\mathbb{Z}_K) \times H_2(M^3;\mathbb{Z}_K) \to \mathbb{Z}_K$ is the intersection product on $M^3$, and $\mathcal{A}_1 \coloneqq \text{PD}(\Sigma^2) \in H^1(M^3;\mathbb{Z}_K)$ is the Poincaré dual of $\Sigma^2$ with respect to $M^3$. In fact, the $\mathbb{Z}_K$-valued 1-cocycle $\mathcal{A}_1$ is the discrete analogue of the dynamical gauge field $a_1$ in \eqref{eq:example_continuum}, whereas $B_2$ is a ``background'' gauge field for the Pontryagin dual symmetry $\widehat{\bZ}_K^{(1)}$. Note that the latter is not summed over in the path integral for $\mathcal{C}_K^{[1]}(M^3)$.
    
    \item A residual 0-form symmetry $\overline{\bZ}_{N/K}^{(0)}$ coming from the remaining global symmetry in $\cT$ that is not gauged on $\cC^{[1]}_K(M^3)$. These are generated by $\overline{\cU}_{\overline{n}}(\Sigma^2)$ operators, with $\overline{n}$ defined modulo $N/K$, which are a subset of the original operators $\cU_k(\Sigma^2)$ restricted to the $\cC^{[1]}_K(M^3)$ theory. The shift in form-degree, namely, the residual global symmetry is a 0-form symmetry, is due to the worldvolume theory being defined in one less dimension than $\cT$.
\end{itemize}

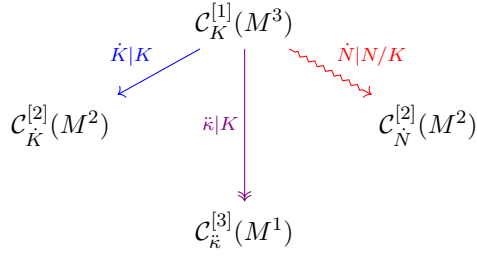
\begin{figure}[t!]
	\centering
        \adjustbox{scale=0.9,center}{%
	\begin{tikzcd}
		& & & \mathcal{C}_K^{[1]}(M^3) \arrow[dl, blue, "\dot{K}|K"'] \arrow[dr, red, squiggly, "\dot{N}|N/K"] \arrow[dd, violet, two heads, "\ddot{\kappa}|K"'] & & &\\
		& & \mathcal{C}_{\dot{K}}^{[2]}(M^2) & & \mathcal{C}_{\dot{N}}^{[2]}(M^2) & &\\
		& & & \mathcal{C}_{\ddot{\kappa}}^{[3]}(M^1) & & &
	\end{tikzcd}
        }
	\caption{\label{fig:condensation_defects_first_step} Condensation defects directly obtainable in the $\cC^{[1]}_K(M^3)$ worldvolume theory. The \textcolor{blue}{solid} and \textcolor{violet}{double-headed} arrows respectively denote a 1- and 2-gauging of (subgroups of) the Pontryagin dual symmetry $\widehat{\bZ}_K^{(1)}$, while the \textcolor{red}{squiggly} arrow denotes a 1-gauging of (subgroups of) the residual $\overline{\bZ}_{N/K}^{(0)}$ symmetry.}
\end{figure}

\paragraph{2-condensation defects:} We can further consider the higher-gauging of the pair of global symmetries in the $\cC^{[1]}_K(M^3)$. To do this, let us first note that we may use Poincar\'e duality to express $\widehat{\cU}_{\hat{k}}(\gamma^1)$ as
\begin{equation}
    \widehat{\mathcal{U}}_{\hat{k}}(\gamma^1) = \exp\bigg(\frac{2\pi i \hat{k}}{K} \int_{M^3} \mathcal{A}_1 \cup \mathcal{B}_2'\bigg) \, ,
\end{equation}
where $\mathcal{B}_2' \coloneqq \text{PD}(\gamma^1) \in H^2(M^3;\mathbb{Z}_K)$ is a {\it fixed} 2-cocycle. The precise meaning of inserting such a dual line defect on the condensation defect $\mathcal{C}_K^{[1]}$ is that its effective action gets modified to the following,
\begin{align}\label{eq:condensation_with_defect}
    \mathcal{C}_{K,\text{defect}}^{[1]}(M^3,\gamma^1) & \coloneqq \bigoplus_{\Sigma^2 \in H_2(M^3;\mathbb{Z}_K)} \mathcal{U}_{1}(\Sigma^2) \, \widehat{\mathcal{U}}_{1}(\gamma^1)\nonumber\\
    & = \int \mathcal{D}a_1 \mathcal{D}c_1 \exp\bigg(2\pi i \int_{M^3} K a_1 \wedge dc_1 + a_1 \wedge B_2 + a_1 \wedge B_2'\bigg) \, .
\end{align}

With this in mind, we can construct three different higher condensation defects.
\begin{itemize}
    \item Firstly, we may 1-gauge a subgroup $\bZ_{\dot{K}}^{(1)} \subseteq \widehat{\bZ}_K^{(1)}$ of the dual symmetry by summing over insertions of the subset of $\widehat{\cU}_{\hat{k}}(\gamma^1)$ defects generating the subgroup, on a 2-dimensional submanifold $M^2\subset M^3$. This gives us a 2-condensation defect $\cC^{[2]}_{\dot{K}}(M^2)$.
    
    \item Secondly, we may 2-gauge a subgroup $\bZ_{\ddot{\kappa}}^{(1)} \subseteq \widehat{\bZ}_K^{(1)}$ of the dual symmetry in a similar manner, on a 1-dimensional submanifold $M^1\subset M^3$. This gives us a 3-condensation defect $\cC^{[3]}_{\ddot{\kappa}}(M^1)$.
    
    \item Finally, we may 1-gauge instead a subgroup $\bZ_{\dot{N}}^{(0)} \subseteq \overline{\bZ}_{N/K}^{(0)}$ of the residual symmetry, which gives us a 2-condensation defect $\cC^{[2]}_{\dot{N}}(M^2)$. In the limiting case where $K=1$, the defect obtained in this way is equivalent to a direct 2-gauging of the subgroup $\bZ_{\dot{N}}^{(1)} \subseteq \bZ_N^{(1)}$ in the original 4d theory $\cT$.
\end{itemize}
In Table \ref{tab:child_condensations_1}, we collect the explicit expressions for these higher condensation defects, similar to those for $\cC^{[1]}_K(M^3)$ written in \eqref{eq:example_condensation_defect} and \eqref{eq:example_continuum}. The relation of each higher condensation defect to the parent theory on $\cC^{[1]}_K(M^3)$ is depicted graphically in Figure \ref{fig:condensation_defects_first_step}.

The defect $\cC^{[2]}_{\dot{N}}(M^2)$ should be seen as the ``electromagnetic'' dual of the defect $\cC^{[2]}_{\dot{K}}(M^2)$ in the following way. By construction, these two symmetry groups take part in a non-trivial group extension,
\begin{equation}
    0\to \widehat{\bZ}_K\to\bZ_N\to \overline{\bZ}_{N/K}\to 0\,,
\end{equation}
where the $\bZ_N$ symmetry is the one present in the 4d theory $\cT$. This structure introduces a mixed anomaly between the 0- and 1-form global symmetries on $\cC^{[1]}_K(M^3)$, much as in the case of the electric and magnetic 1-form symmetries in 4d Yang-Mills theory with $SU(N)/\bZ_K$ gauge group \cite{Aharony:2013hda}. One may also wish to construct condensation defects by simultaneously higher gauging subgroups of the $\widehat{\bZ}_K^{(1)}$ and $\overline{\bZ}_{N/K}^{(0)}$ symmetry groups. However, due to the mixed anomaly just mentioned, such a construction may be obstructed by higher anomalies. We will return to this point in Section \ref{sec:anomalous_symmetries} when discussing the effects of such obstructions.

On each of the three higher condensation defects constructed on $\cC^{[1]}_K(M^3)$, there are again new global symmetries arising from the higher gauging procedure. For instance, on the defect $\cC^{[2]}_{\dot{K}}(M^2)$ there are the following two global symmetries.
\begin{itemize}
    \item A 0-form symmetry $\widehat{\bZ}_{\dot{K}}^{(0)}$ coming from the Pontryagin dual of the $\bZ_{\dot{K}}^{(1)} \subset \widehat{\bZ}_K^{(1)}$ symmetry that was 1-gauged to produce $\cC^{[2]}_{\dot{K}}(M^2)$ in the first place. This symmetry is generated by the following topological line operators,
    \begin{equation}\label{eq:quantum_symmetry_step2}
        \widehat{\cU}_{\dot{k}}(\Gamma^1) \coloneqq \exp\bigg(\frac{2\pi i \dot{k}}{\dot{K}} \, \langle\Gamma^1,\gamma^1\rangle\bigg) = \exp\bigg(\frac{2\pi i \dot{k}}{\dot{K}}\int_{\Gamma^1}\tilde{\cA}_1\bigg) \, .
    \end{equation}
    Here, $\Gamma^1\in H_1(M^2;\bZ_{\dot{K}})$ and $\tilde{\cA}_1 := {\rm PD}(\gamma^1)\in H^1(M^2;\bZ_{\dot{K}})$ is the Poincar\'e dual of $\gamma^1$ with respect to $M^2$, much like $\cA_1$ in \eqref{eq:quantum_symmetry_step1}.
    
    \item A residual 0-form symmetry $\overline{\bZ}_{K/\dot{K}}^{(0)}$ inherited from the subgroup of $\widehat{\bZ}_K^{(1)}$ in $\cC^{[1]}_K(M^3)$ that was {\it not} gauged. This symmetry is generated by defects $\overline{\cU}_{\bar{k}}(\gamma^1)$, where $\bar{k}$ is defined modulo $K/\dot{K}$, regarded as (a subset of) the defects $\widehat{\cU}_{\hat{k}}(\gamma^1)$ restricted to $\cC^{[2]}_{\dot{K}}(M^2)$. There is again a shift in degree for dimensional reasons.
\end{itemize}

The two relevant global symmetries on the ``dual'' defect $\cC^{[2]}_{\dot{N}}(M^2)$ are similar but with minor modifications.
\begin{itemize}
    \item A 1-form symmetry $\widehat{\bZ}_{\dot{N}}^{(1)}$ arising as the Pontryagin dual of the $\bZ_{\dot{N}}^{(0)}$ symmetry used to construct $\cC^{[2]}_{\dot{N}}(M^2)$ in the first place. The defects generating this symmetry are defined similarly to \eqref{eq:quantum_symmetry_step1} and \eqref{eq:quantum_symmetry_step2}.
    \begin{equation}
        \widehat{\cU}_{\dot{n}}(x) \coloneqq \exp\bigg(\frac{2\pi i\dot{n}}{\dot{N}} \, \langle x,\Sigma^2\rangle\bigg) = \exp\bigg(\frac{2\pi i \dot{n}}{\dot{N}}\int_x \cA_0\bigg) \equiv \exp\bigg(\frac{2\pi i \dot{n}}{\dot{N}} \, \cA_0(x)\bigg) \, .
    \end{equation}
    Here, $x\in H_0(M^2;\bZ_{\dot{N}})$ and $\cA_0 \coloneqq {\rm PD}(\Sigma^2) \in H^0(M^2;\bZ_{\dot{N}})$.
    
    \item A 0-form symmetry $\overline{\bZ}_K^{(0)}$ originating from the residual symmetry of the {\it Pontryagin dual symmetry} $\widehat{\bZ}_K^{(1)}$ in the parent $\cC^{[1]}_K(M^3)$ defect. The defects generating this are exactly all the $\widehat{\cU}_{\hat{k}}(\gamma^1)$ defects restricted to $\cC^{[2]}_{\dot{N}}(M^2)$. There are no additional restrictions as the $\widehat{\bZ}_K^{(1)}$ symmetry was unaffected in constructing this 2-condensation defect.
\end{itemize}

\begin{figure}[t!]
	\centering
        \adjustbox{scale=0.9,center}{%
	\begin{tikzcd}
		& & \mathcal{C}_{\dot{K}}^{[2]}(M^2) \arrow[dl, blue, "\ddot{K}|\dot{K}"'] \arrow[dr, red, squiggly,"\ddot{\kappa}|K/\dot{K}"] & & \mathcal{C}_{\dot{N}}^{[2]}(M^2) \arrow[dl, cyan, dashed, "\ddot{\kappa}|K"'] \arrow[dr, blue, "\ddot{N}|\dot{N}"] \arrow[dd, violet, two heads, "\dddot{\eta}|\dot{N}"']& &\\
		& \mathcal{C}_{\ddot{K}}^{[3]}(M^1) & & \mathcal{C}_{\ddot{\kappa}}^{[3]}(M^1) & & \mathcal{C}_{\ddot{N}}^{[3]}(M^1) &\\
		& & & & \mathcal{C}_{\dddot{\eta}}^{[4]}(M^0) & &
	\end{tikzcd}
        }
	\caption{\label{fig:condensation_defects_second_step} Higher condensations defects directly obtainable in the $\cC^{[2]}_{\dot{K}}$ and $\cC^{[2]}_{\dot{N}}$ worldvolume theories. The \textcolor{blue}{solid}, \textcolor{violet}{double-headed}, and \textcolor{red}{squiggly} arrows denote the same as above: 1-, 2-gauging of Pontryagin dual symmetry subgroups and 1-gauging of residual symmetry subgroups respectively. The new \textcolor{cyan}{dashed} arrow denotes a 1-gauging of a subgroup of the residual Pontryagin dual symmetry coming from the ambient $\cC^{[1]}_K(M^3)$ worldvolume. If $\cC^{[2]}_{\dot{N}}$ is directly constructed in the $\cT$ theory, this arrow is not present.}
\end{figure}
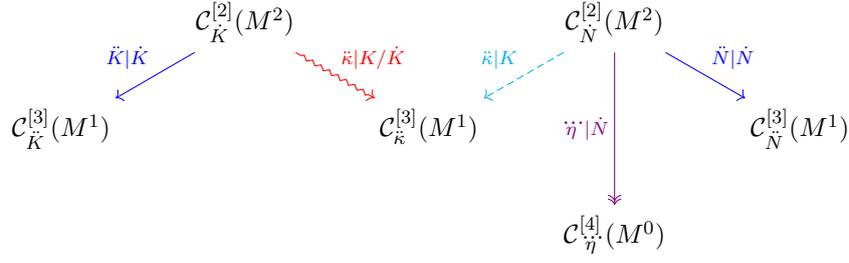

\paragraph{3-condensation defects:} We can now iterate the previous procedure on each of the two 2-condensation defects $\cC^{[2]}_{\dot{K}}(M^2)$ and $\cC^{[2]}_{\dot{N}}(M^2)$ constructed on $\cC^{[1]}_K(M^3)$, by higher-gauging either of the two relevant global symmetries.
\begin{itemize}
    \item On $\cC^{[2]}_{\dot{K}}(M^2)$, we can 1-gauge a subgroup $\bZ_{\ddot{K}}^{(1)} \subseteq \widehat{\bZ}_{\dot{K}}^{(1)}$ of the dual symmetry by summing over (suitable) insertions of $\widehat{\cU}_{\dot{k}}(\Gamma^1)$ on a 1-dimensional submanifold $M^1 \subset M^2$, producing a 3-condensation defect $\cC^{[3]}_{\ddot{K}}(M^1)$. Alternatively, we can 1-gauge a subgroup $\bZ_{\ddot{\kappa}}^{(0)} \subseteq \overline{\bZ}_{K/\dot{K}}^{(0)}$ of the residual symmetry on $\cC^{[2]}_{\dot{K}}(M^2)$, producing a 3-condensation defect $\cC^{[3]}_{\ddot{\kappa}}(M^1)$.
    
    \item On $\cC^{[2]}_{\dot{N}}(M^2)$, we can 1-gauge a subgroup $\bZ_{\ddot{N}}^{(1)} \subseteq  \widehat{\bZ}_{\dot{N}}^{(1)}$ of the dual symmetry, producing a 3-condensation defect $\cC^{[3]}_{\ddot{N}}(M^1)$. Alternatively, we can 1-gauge a subgroup $\bZ_{\ddot{\kappa}}^{(0)} \subseteq \overline{\bZ}_K^{(0)}$ of the residual symmetry on $\cC^{[2]}_{\dot{N}}(M^2)$, producing a 3-condensation defect $\cC^{[3]}_{\ddot{\kappa}}(M^1)$.
    
    \item Finally, we may also 2-gauge a subgroup $\bZ_{\dddot{\eta}}^{(0)} \subseteq \widehat{\bZ}_{\dot{N}}^{(0)}$ of the dual symmetry on $\cC^{[2]}_{\dot{N}}(M^2)$, producing a 4-condensation defect $\cC^{[4]}_{\dddot{\eta}}(M^0)$.
\end{itemize}
In Table \ref{tab:child_condensations_2}, we collect all the expressions for these various higher condensation defects, including both the formal sum and the path integral realizations, while in Figure \ref{fig:condensation_defects_second_step}, we illustrate the relations between all these higher condensation defects and their parent defects.


\begin{figure}[t!]
	\centering
        \adjustbox{scale=0.9,center}{%
	\begin{tikzcd}
		& & \mathcal{C}_K^{[1]}(M^3) \arrow[dl, blue, "\dot{K}|K"'] \arrow[dr, red, squiggly, "\dot{N}|N/K"] \arrow[dd, violet, two heads, "\ddot{\kappa}|K"'] & &\\
		& \mathcal{C}_{\dot{K}}^{[2]}(M^2) \arrow[dr, red, squiggly, "\ddot{\kappa}|K/\dot{K}"] & & \mathcal{C}_{\dot{N}}^{[2]}(M^2) \arrow[dl, cyan, dashed, "\ddot{\kappa}|K"']&\\
	    & & \mathcal{C}_{\ddot{\kappa}}^{[3]}(M^1)& &
	\end{tikzcd}
        }
	\caption{\label{fig:condensation_defects_diamond} The three different ways to construct the 3-condensation defect $\cC^{[3]}_{\ddot{\kappa}}(M^1)$, starting from the 1-condensation defect $\cC^{[1]}(M^3)$.}
\end{figure}
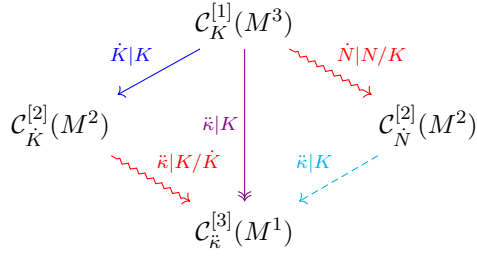

Let us take a moment to examine the interesting case of $\cC^{[3]}_{\ddot{\kappa}}(M^1)$. This defect has been constructed in three distinct ways thus far.
\begin{itemize}
    \item A 1-gauging of $\bZ_{\ddot{\kappa}}^{(0)} \subseteq \overline{\bZ}_{K/\dot{K}}^{(0)}$ on $\cC^{[2]}_{\dot{K}}(M^2)$.
    \item A 2-gauging of $\bZ_{\ddot{\kappa}}^{(1)} \subseteq \widehat{\mathbb{Z}}_K^{(1)}$ on $\cC^{[1]}_K(M^3)$.
    \item A 1-gauging of $\bZ_{\ddot{\kappa}}^{(0)} \subseteq \overline{\bZ}_K^{(0)}$ on $\cC^{[2]}_{\dot{N}}(M^2)$.
\end{itemize}
These constructions are generally inequivalent, since they descend from different parent defects, and hence have different symmetry structures. Schematically, from the three pathways above, the symmetries admitted by $\cC^{[3]}_{\ddot{\kappa}}(M^1)$ as 1d quantum mechanical theories are respectively
\begin{equation}
    \widehat{\bZ}_{\ddot{\kappa}}^{(0)} \times \overline{\bZ}_{\dot{K}}^{(-1)} \times \overline{\bZ}_{K/\ddot{\kappa}\dot{K}}^{(-1)} \, , \qquad \widehat{\bZ}_{\ddot{\kappa}}^{(0)} \times \overline{\bZ}_{K/\ddot{\kappa}}^{(-1)} \, , \qquad \widehat{\bZ}_{\ddot{\kappa}}^{(0)} \times \overline{\bZ}_{\dot{N}}^{(0)} \times \overline{\bZ}_{K/\ddot{\kappa}}^{(-1)} \, .
\end{equation}
We thus see that the former two are isomorphic only if $\gcd(\dot{K},K/\ddot{\kappa}\dot{K})=1$, in which case $\bZ_{K/\ddot{\kappa}} \cong \bZ_{\dot{K}} \times \bZ_{K/\ddot{\kappa}\dot{K}}$ by virtue of the Chinese remainder theorem.\footnote{While we generally do not discuss $(-1)$-form symmetries in this work, we keep them explicit here to illustrate the subtle difference between the various 1d theories. Such a symmetry does not have ``$(-1)$-dimensional'' charged operators in the conventional sense, but it can be realized by coupling to a background axion, i.e.~introducing a ``theta-term'' in the theory. Equivalently, the different $(-1)$-form symmetries simply signify the fact that these 1d theories are placed in different background 2d/3d theories as defect insertions.} Meanwhile, the latter two are isomorphic only if $\dot{N}=1$, in which case $\cC^{[2]}_{\dot{N}=1}(M^2)$ is actually an identity defect on $\cC^{[1]}_K(M^3)$, so the 3-condensation defect $\cC^{[3]}_{\ddot{\kappa}}(M^1)$ obtained by 1-gauging on this identity defect is equivalent to a direct 2-gauging on $\cC^{[1]}_K(M^3)$.

The subtlety discussed above will not affect the rest of this paper, unless otherwise specified. However, it serves to clarify the point that Figures \ref{fig:condensation_defects_diagram} and \ref{fig:condensation_defects_diamond} are not literally commutative diagrams.

\begin{figure}[t!]
	\centering
        \adjustbox{scale=0.83,center}{%
	\begin{tikzcd}
		& & & & & &\\
		& & & & & &\\
		& \mathcal{C}_{\ddot{K}}^{[3]}(M^1) \arrow[dl, blue, "\dddot{K}|\ddot{K}"'] & & \mathcal{C}_{\ddot{\kappa}}^{[3]}(M^1) \arrow[dl, blue, "\dddot{\kappa}|\ddot{\kappa}"'] \arrow[dr, teal, tail,"\dddot{\eta}|\dot{N}"] & & \mathcal{C}_{\ddot{N}}^{[3]}(M^1) \arrow[dl, red, squiggly, "\dddot{\eta}|\dot{N}/\ddot{N}"'] \arrow[dr, blue, "\dddot{N}|\ddot{N}"] & \\
		\mathcal{C}_{\dddot{K}}^{[4]}(M^0) & & \mathcal{C}_{\dddot{\kappa}}^{[4]}(M^0) & & \mathcal{C}_{\dddot{\eta}}^{[4]}(M^0) & & \cC^{[4]}_{\dddot{N}}(M^0)
	\end{tikzcd}
        }
	\caption{\label{fig:condensation_defects_third_step} Condensation defects directly obtainable in the 1-dimensional condensation defect worldvolume theories. As before, \textcolor{blue}{solid} and \textcolor{red}{squiggly} arrows respectively denote 1-gaugings of Pontryagin dual symmetries and residual symmetries. The arrow \textcolor{teal}{with a tail} is a special case that is relevant only if $\cC^{[3]}_{\ddot{\kappa}}(M^1)$ is constructed through the dashed arrow in Figure \ref{fig:condensation_defects_second_step}, when there is a residual $\bZ_{\dot{N}}$ symmetry to manipulate.}
\end{figure}
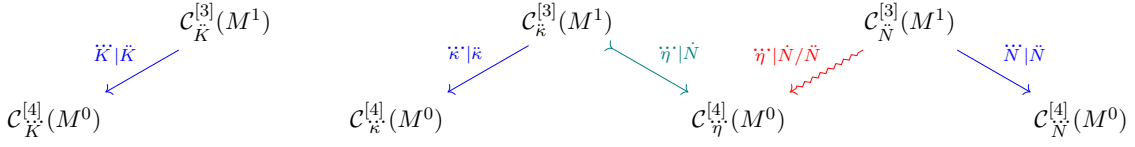

\paragraph{4-condensation defects:} We can repeat the drill to obtain 4-condensation defects. In the interest of brevity, we will not reproduce every worldvolume action and the associated symmetry defects, but the outline is as follows.
\begin{itemize}
    \item On $\cC^{[3]}_{\ddot{K}}(M^1)$, we can 1-gauge a subgroup $\bZ_{\dddot{K}}^{(0)} \subseteq \widehat{\bZ}_{\ddot{K}}^{(0)}$ to obtain a 4-condensation defect $\cC^{[4]}_{\dddot{K}}(M^0)$.
    
    \item On $\cC^{[3]}_{\ddot{\kappa}}(M^1)$, we can 1-gauge a subgroup $\bZ_{\dddot{\kappa}}^{(0)} \subseteq \widehat{\bZ}_{\ddot{\kappa}}^{(0)}$ to obtain a 4-condensation defect $\cC^{[4]}_{\dddot{\kappa}}(M^0)$. Additionally, if the residual $\overline{\bZ}_{\dot{N}}^{(0)}$ symmetry is present, meaning we had used the dashed arrow in Figure \ref{fig:condensation_defects_second_step}, we can consider 1-gauging a subgroup $\bZ_{\dddot{\eta}}^{(0)} \subseteq \overline{\bZ}_{\dot{N}}^{(0)}$ of this residual symmetry to construct a different 4-condensation defect $\cC^{[4]}_{\dddot{\eta}}(M^0)$.
    
    \item On $\cC^{[3]}_{\ddot{N}}(M^1)$, we can 1-gauge a subgroup $\bZ_{\dddot{N}}^{(0)} \subseteq \widehat{\bZ}_{\ddot{N}}^{(0)}$ to obtain a 4-condensation defect $\cC^{[4]}_{\dddot{N}}(M^0)$. We may also 1-gauge a subgroup $\bZ_{\dddot{\eta}}^{(0)} \subseteq \overline{\bZ}_{\dot{N}/\ddot{N}}^{(0)}$ to obtain a 4-condensation defect $\cC^{[4]}_{\dddot{\eta}}(M^0)$.
\end{itemize}
A graphical depiction of these processes can be found in Figure \ref{fig:condensation_defects_third_step}. Note that even though 4-condensation defects are point operators, none of them exists as a standalone bulk local operator, other than the trivial one. This is compatible with the fact that there is no $(d-1)$-form symmetry in $\mathcal{T}$, by construction, otherwise it would imply the existence of non-trivial topological point operators.

Combining Figures \ref{fig:condensation_defects_first_step}, \ref{fig:condensation_defects_second_step}, and \ref{fig:condensation_defects_third_step}, we can construct a ``tree'' of condensation defects obtainable within the theory $\cT$ as in Figure \ref{fig:condensation_defects_diagram}. More generally, one could also consider the higher gauging of mixed subgroups, rather than only one at a time as we have done here. Therefore, even for a ``simple'' theory with a single $\bZ_N^{(1)}$ symmetry, there is already a rich structure of condensation defects spanning all (co)dimensions.

\renewcommand{\arraystretch}{2.0}
\begin{table}
\begin{center}
    \begin{tabular}{|c|c|c|}
        \hline\vspace{-3pt}
        Defect & Formal Sum & Continuum Path Integral\\
        \hline
        $\cC^{[2]}_{\dot{K}}(M^2)$ & $\displaystyle\bigoplus_{\gamma^1\in H_1(M^2;\bZ_{\dot{K}})}\widehat{\cU}_{1}(\gamma^1)$ & $\displaystyle\int\cD a_1' \cD c_0'\,\exp\left(2\pi i\int_{M^2}\dot{K}a_1'\wedge dc_0' + a_1'\wedge a_1\right)$\\
        \hline
        $\cC^{[3]}_{\ddot{\kappa}}(M^1)$ & $\displaystyle\bigoplus_{\gamma^1 \in H_1(M^1;\bZ_{\ddot{\kappa}})}\widehat{\cU}_1(\gamma^1)$ & $\displaystyle\int\cD a_0'\cD c_0'\exp\left(2\pi i\int_{M^1}\ddot{\kappa}\, a_0' dc_0' + a_0' a_1\right)$\\
        \hline
        $\cC^{[2]}_{\dot{N}}(M^2)$ & $\displaystyle\bigoplus_{\Sigma^2\in H_2(M^2;\bZ_{\dot{N}})}\overline{\cU}_{1}(\Sigma^2)$ & $\displaystyle\int\cD a_0'\cD c_1'\exp\left(2\pi i\int_{M^2}\dot{N}a_0'\wedge dc_1' + a_0' B_2\right)$\\
        \hline
    \end{tabular}
    \caption{\label{tab:child_condensations_1}Higher condensation defects on $\cC^{[1]}_K(M^3)$, their expressions as formal sums, and continuum path integral realizations. In each path integral, the $a_i'$ and $c_i'$ fields are dynamical fields localized to the defect worldvolume.}
\end{center}
\end{table}

\renewcommand{\arraystretch}{2.0}
\begin{table}
\begin{center}
    \begin{tabular}{|c|c|c|}
        \hline\vspace{-3pt}
        Defect & Formal Sum & Continuum Path Integral\\
        \hline
        $\cC^{[3]}_{\ddot{K}}(M^1)$ & $\displaystyle\bigoplus_{\Gamma^1\in H_1(M^1;\bZ_{\ddot{K}})}\widehat{\cU}_{1}(\Gamma^1)$ & $\displaystyle\int\cD a'_0\cD c'_0\exp\left(2\pi i\int_{M^1}\ddot{K} a'_0dc'_0 + a'_0\tilde{a}_1\right)$\\
        \hline
        $\cC^{[3]}_{\ddot{\kappa}}(M^1)$ & $\displaystyle\bigoplus_{\gamma^1 \in H_1(M^1;\bZ_{\ddot{\kappa}})}\overline{\cU}_{1}(\gamma^1)$ & $\displaystyle\int\cD a_0'\cD c_0'\exp\left(2\pi i\int_{M^1}\ddot{\kappa}\, a_0' dc_0' + a_0' a_1\right)$\\
        \hline
        $\cC^{[3]}_{\ddot{\kappa}}(M^1)$ & $\displaystyle\bigoplus_{\gamma^1 \in H_1(M^1;\bZ_{\ddot{\kappa}})}\widehat{\cU}_{1}(\gamma^1)$ & $\displaystyle\int\cD a_0'\cD c_0'\exp\left(2\pi i\int_{M^1}\ddot{\kappa}\, a_0' dc_0' + a_0' a_1\right)$\\
        \hline
        $\cC^{[3]}_{\ddot{N}}(M^1)$ & $\displaystyle\bigoplus_{x \in H_0(M^1;\bZ_{\ddot{N}})}\widehat{\cU}_{1}(x)$ & $\displaystyle\int\cD a_1'\cD c_0'\exp\left(2\pi i\int_{M^1}\ddot{N}\, a_1' c_0' + a_1' a_0\right)$\\
        \hline
        $\cC^{[4]}_{\dddot{\eta}}(M^0)$ & $\displaystyle\bigoplus_{x \in H_0(M^0;\bZ_{\dddot{\eta}})}\widehat{\cU}_{1}(x)$ & $\displaystyle\int\cD a_0'\cD c_0'\exp\left(2\pi i\int_{M^0}\dddot{\eta}\, a_0' c_0' + a_0' a_0\right)$\\
        \hline
    \end{tabular}
    \caption{\label{tab:child_condensations_2}Higher condensation defects on $\cC^{[2]}_{\dot{K}}(M^2)$ and $\cC^{[2]}_{\dot{N}}(M^2)$, their expressions as formal sums, and continuum path integral realizations. As above, the $a_i'$ and $c_i'$ fields are dynamical fields localized to the defect worldvolume. The fields $\tilde{a}_1$, $a_1$, and $a_0$ are the continuum analogues of $\tilde{\cA}_1$, $\cA_1$, and $\cA_0$ respectively.}
\end{center}
\end{table}

\section{Interfaces and Condensation Completeness}\label{sec:karoubi_completeness}

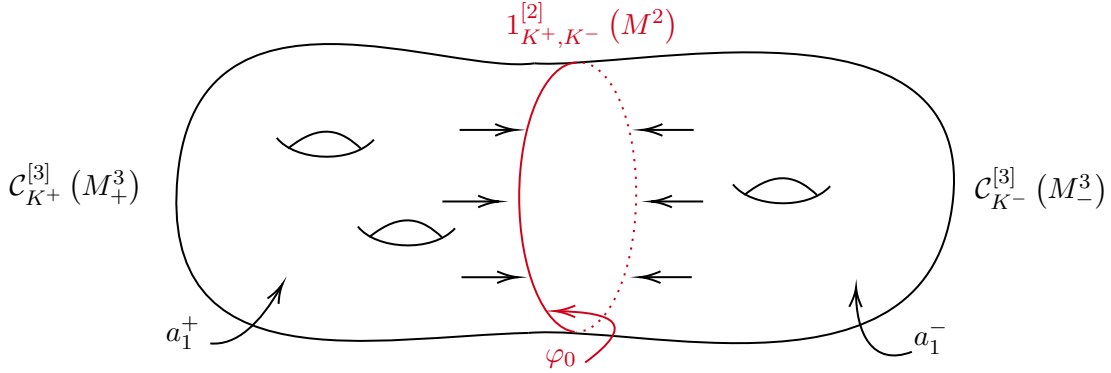
\begin{figure}
\centering
\tikzset{every picture/.style={line width=0.75pt}} 
\begin{tikzpicture}[x=0.75pt,y=0.75pt,yscale=-0.9,xscale=0.95]
\draw    (332.95,228.46) .. controls (252.66,230.44) and (137.67,259.19) .. (148.57,140.23) .. controls (159.47,21.28) and (290.33,84.72) .. (335.93,77.78) ;
\draw    (332.95,228.46) .. controls (390.45,223.5) and (545.09,272.08) .. (555.99,153.12) .. controls (566.9,34.16) and (386.48,81.75) .. (335.93,77.78) ;
\draw    (200.61,119.42) .. controls (210.03,132.3) and (239.77,134.29) .. (251.66,120.41) ;
\draw    (206.56,125.36) .. controls (223.91,110.49) and (234.81,115.45) .. (244.73,125.36) ;
\draw    (243.24,168.98) .. controls (252.66,181.87) and (282.4,183.85) .. (294.29,169.97) ;
\draw    (249.19,174.93) .. controls (266.53,160.06) and (277.44,165.02) .. (287.35,174.93) ;
\draw    (440.34,145.44) .. controls (449.76,158.33) and (479.5,160.31) .. (491.39,146.44) ;
\draw    (446.29,151.39) .. controls (463.64,136.52) and (474.54,141.48) .. (484.45,151.39) ;
\draw  [draw opacity=0] (357.16,227.87) .. controls (340.92,225.87) and (327.99,192.95) .. (327.99,152.62) .. controls (327.99,111.44) and (341.48,77.97) .. (358.21,77.3) -- (358.73,152.62) -- cycle ; \draw  [color={rgb, 255:red, 208; green, 2; blue, 27 }  ,draw opacity=1 ] (357.16,227.87) .. controls (340.92,225.87) and (327.99,192.95) .. (327.99,152.62) .. controls (327.99,111.44) and (341.48,77.97) .. (358.21,77.3) ;  
\draw  [draw opacity=0][dash pattern={on 0.84pt off 2.51pt}] (361.38,77.41) .. controls (377.59,79.64) and (390.04,112.75) .. (389.45,153.07) .. controls (388.86,194.25) and (374.89,227.52) .. (358.15,227.95) -- (358.73,152.62) -- cycle ; \draw  [color={rgb, 255:red, 208; green, 2; blue, 27 }  ,draw opacity=1 ][dash pattern={on 0.84pt off 2.51pt}] (361.38,77.41) .. controls (377.59,79.64) and (390.04,112.75) .. (389.45,153.07) .. controls (388.86,194.25) and (374.89,227.52) .. (358.15,227.95) ;  
\draw    (287.67,155) -- (315.33,155) ;
\draw [shift={(317.33,155)}, rotate = 180] [color={rgb, 255:red, 0; green, 0; blue, 0 }  ][line width=0.75]    (10.93,-3.29) .. controls (6.95,-1.4) and (3.31,-0.3) .. (0,0) .. controls (3.31,0.3) and (6.95,1.4) .. (10.93,3.29)   ;
\draw    (424.67,155) -- (399.33,155) ;
\draw [shift={(397.33,155)}, rotate = 360] [color={rgb, 255:red, 0; green, 0; blue, 0 }  ][line width=0.75]    (10.93,-3.29) .. controls (6.95,-1.4) and (3.31,-0.3) .. (0,0) .. controls (3.31,0.3) and (6.95,1.4) .. (10.93,3.29)   ;
\draw    (165.67,234.33) .. controls (179.32,235.63) and (191.7,223.62) .. (202.81,202.63) ;
\draw [shift={(203.67,201)}, rotate = 117.26] [color={rgb, 255:red, 0; green, 0; blue, 0 }  ][line width=0.75]    (10.93,-3.29) .. controls (6.95,-1.4) and (3.31,-0.3) .. (0,0) .. controls (3.31,0.3) and (6.95,1.4) .. (10.93,3.29)   ;
\draw    (534.33,239) .. controls (513.53,247.45) and (504.77,228.65) .. (504.97,204.84) ;
\draw [shift={(505,203)}, rotate = 91.55] [color={rgb, 255:red, 0; green, 0; blue, 0 }  ][line width=0.75]    (10.93,-3.29) .. controls (6.95,-1.4) and (3.31,-0.3) .. (0,0) .. controls (3.31,0.3) and (6.95,1.4) .. (10.93,3.29)   ;
\draw [color={rgb, 255:red, 208; green, 2; blue, 27 }  ,draw opacity=1 ]   (362.67,245) .. controls (390.43,223.2) and (378.02,217) .. (348.34,216.29) ;
\draw [shift={(346.5,216.25)}, rotate = 0.92] [color={rgb, 255:red, 208; green, 2; blue, 27 }  ,draw opacity=1 ][line width=0.75]    (10.93,-3.29) .. controls (6.95,-1.4) and (3.31,-0.3) .. (0,0) .. controls (3.31,0.3) and (6.95,1.4) .. (10.93,3.29)   ;
\draw    (297.67,197.33) -- (325.33,197.33) ;
\draw [shift={(327.33,197.33)}, rotate = 180] [color={rgb, 255:red, 0; green, 0; blue, 0 }  ][line width=0.75]    (10.93,-3.29) .. controls (6.95,-1.4) and (3.31,-0.3) .. (0,0) .. controls (3.31,0.3) and (6.95,1.4) .. (10.93,3.29)   ;
\draw    (296.67,115) -- (324.33,115) ;
\draw [shift={(326.33,115)}, rotate = 180] [color={rgb, 255:red, 0; green, 0; blue, 0 }  ][line width=0.75]    (10.93,-3.29) .. controls (6.95,-1.4) and (3.31,-0.3) .. (0,0) .. controls (3.31,0.3) and (6.95,1.4) .. (10.93,3.29)   ;
\draw    (418.67,197.3) -- (393.33,197.3) ;
\draw [shift={(391.33,197.3)}, rotate = 360] [color={rgb, 255:red, 0; green, 0; blue, 0 }  ][line width=0.75]    (10.93,-3.29) .. controls (6.95,-1.4) and (3.31,-0.3) .. (0,0) .. controls (3.31,0.3) and (6.95,1.4) .. (10.93,3.29)   ;
\draw    (419.67,115) -- (394.33,115) ;
\draw [shift={(392.33,115)}, rotate = 360] [color={rgb, 255:red, 0; green, 0; blue, 0 }  ][line width=0.75]    (10.93,-3.29) .. controls (6.95,-1.4) and (3.31,-0.3) .. (0,0) .. controls (3.31,0.3) and (6.95,1.4) .. (10.93,3.29)   ;

\draw (58.67,132.07) node [anchor=north west][inner sep=0.75pt]    {$\mathcal{C}_{K^{+}}^{[ 3]}\left( M_{+}^{3}\right)$};
\draw (566,134.07) node [anchor=north west][inner sep=0.75pt]    {$\mathcal{C}_{K^{-}}^{[ 3]}\left( M_{-}^{3}\right)$};
\draw (319.33,42.07) node [anchor=north west][inner sep=0.75pt]  [color={rgb, 255:red, 208; green, 2; blue, 27 }  ,opacity=1 ]  {$1_{K^{+} ,K^{-}}^{[ 2]}\left( M^{2}\right)$};
\draw (140.67,217.4) node [anchor=north west][inner sep=0.75pt]    {$a_{1}^{+}$};
\draw (533.73,219.69) node [anchor=north west][inner sep=0.75pt]    {$a_{1}^{-}$};
\draw (340,236.07) node [anchor=north west][inner sep=0.75pt]  [color={rgb, 255:red, 208; green, 2; blue, 27 }  ,opacity=1 ]  {$\varphi _{0}$};
\end{tikzpicture}
\caption{Two 1-condensation defects $\cC^{[3]}_{K^\pm}$, defined respectively over $M^3_\pm$, joined by an inflow interface $1_{K^+,K^-}^{[2]}(M^2)$ with $M^2 = \partial M^3_+ = -\partial M^3_-$. Each 1-condensation defect has a gauge field $a_1^\pm$ whose holonomies generate a $\widehat{\bZ}_{K^\pm}^{(1)}$ symmetry. The inflow interface matches the difference between the two flanking gauge field transformations, and identifies the common $\bZ_{\bar{K}}^{(1)}$ subgroup on both sides, where $\bar{K} = \gcd(K^+,K^-)$.}
\label{fig:interface_between_condensations}
\end{figure}

Given the plethora of condensation defects coexisting in the theory $\cT$, it is useful for us to also study interfaces mediating their worldvolume theories. For instance, interfaces are important when one wish to consider the fusion between two different condensation defects.

As we are going to see, not only do non-trivial interfaces exist naturally via anomaly inflow, but {\it every} condensation defect can be ``split'' into two such interfaces. Specifically, an $n$-condensation defect can be factorized into two gauging interfaces that mediate two $(5-n)$-dimensional subregions of spacetime. This will ultimately lead us to the notion of condensation completeness in our theory $\cT$.

\subsection{Interfaces between Condensation Defects}

Before analyzing the factorization of condensation defects, we first discuss the possible interfaces between them. In the following, we focus on two kinds of ``minimal'' interfaces, between a pair of condensation defects (of the same dimension): ``soft'' interfaces permeable to a select subset of symmetry operators, and ``hard'' interfaces that act as boundary conditions between two theories.

\paragraph{Soft interface:} Consider two 1-condensation defects $\cC^{[1]}_{K^+}(M^3_+)$ and $\cC^{[1]}_{K^-}(M^3_-)$ obtained respectively by 1-gauging a $\bZ_{K^+}^{(1)}$ and $\bZ_{K^-}^{(1)}$ symmetry in $\cT$. The supports of these operators are such that $\d M^3_+ = -\d M^3_- := M^2$, with the minus sign signifying a difference in orientation. See Figure \ref{fig:interface_between_condensations} for an illustration. The composite system has a 3d worldvolume action given by
\begin{equation}\label{eq:example_composite_action}
    S_{\rm 3d} = \int_{M^3_+}K^+ a_1^+\wedge dc_1^+ + a_1^+\wedge B_2^+ + \int_{M^3_-}K^- a_1^-\wedge dc_1^- + a_1^-\wedge B_2^-\,,
\end{equation}
where $a_1^\pm$ and $c_1^\pm$ are dynamical fields and $B_2^\pm$ are background data from the point of view of the worldvolume. If we consider the gauge transformation
\begin{equation}\label{eq:example_composite_gauge_transfo}
    a_1^\pm\to a_1^\pm + d\lambda_0\,,\qquad c_1^\pm \to c_1^\pm + d\omega_0^\pm\,,
\end{equation}
for scalar functions $\lambda_0, \omega_0^\pm$, one finds the variation in the action \eqref{eq:example_composite_action} to be
\begin{equation}
    \delta S_{\rm 3d} = \int_{M^2} \lambda_0\left(K^+ dc_1^+ + B_2^+ - K^- dc_1^- - B_2^-\right)\,.
\end{equation}

Let $\bar{K} = \gcd(K^+,K^-)$. By B\'ezout's identity, there exists a properly quantized gauge field $\bar{c}_1$ such that 
\begin{equation}
    \bar{K}d\bar{c}_1 = \left.(K^+dc_1^+ - K^-dc_1^-)\right|_{M^2}\,,
\end{equation}
Hence, a 2d interface theory on $M^2$, such as
\begin{equation}\label{eq:example_identity_interface_action}
    S_{\rm 2d} = \int_{M^2}\varphi_0\left(\bar{K}d\bar{c}_1 + B_2^+ - B_2^-\right)\,,
\end{equation}
preserves gauge invariance of the composite system, where $\varphi_0$ is a dynamical scalar field that transforms as $\varphi_0 \to\varphi_0 - \lambda_0$ under the variation \eqref{eq:example_composite_gauge_transfo}, while all other fields are background. In terms of the discrete cocycles $\cB_2^\pm$ associated with the continuum fields $B_2^\pm$, the scalar acts as a Lagrange multiplier enforcing the identification
\begin{equation}\label{eq:Lagrange_multiplier_identification}
    \cB_2^+|_{M^2} = \cB_2^-|_{M^2} \mod \bar{K}\,,
\end{equation}
and similarly for the dual fields $\cA_1^\pm|_{M^2}\text{ mod }\bar{K}$. Therefore, the interface identifies the defects generating the common subgroup ${\bZ}_{\bar{K}}^{(1)}\subset \widehat{\bZ}_{K^\pm}^{(1)}$ on the condensation defects $\cC^{[1]}_{K^\pm}(M^3_\pm)$.

The fate of the remaining subsets of the dual defects is subtler. If $\gcd(\bar{K},K^\pm/\bar{K}) = 1$, then the short exact sequence
\begin{equation}
    0 \to {\mathbb{Z}}_{\bar{K}} \xrightarrow{\times K^\pm/\bar{K}} \mathbb{Z}_{K^\pm} \xrightarrow{\text{mod} \, K^\pm/\bar{K}} {\mathbb{Z}}_{K^\pm/\bar{K}} \to 0
\end{equation}
splits, so the $\bZ_{\bar{K}}$ and $\bZ_{K^\pm/\bar{K}}$ components of $\mathcal{B}_2^\pm$ are independent, i.e.~$\mathcal{B}_2$ admits a decomposition 
\begin{equation}
    \cB_2^\pm = \frac{K^\pm}{\bar{K}}\,\cB_{2,{\bZ}_{\bar{K}}}^\pm + \bar{K}\cB_{2,{\bZ}_{K^\pm/\bar{K}}}^\pm \, .
\end{equation}
Consequently, we are only left with the option to impose the Dirichlet boundary condition,
\begin{equation}
    \mathcal{B}_2^\pm|_{M^2} \, \text{mod} \, K^\pm/\bar{K} = 0 \, ,
\end{equation}
such that the defects generating the $\mathbb{Z}_{K^\pm/\bar{K}}^{(1)}$ symmetry get trivialized on the interface.

On the other hand, if $\gcd(\bar{K},K^\pm/\bar{K}) \neq 1$, then the aforementioned short exact sequence no longer splits, in which case certain non-trivial combinations of defects generating the $\mathbb{Z}_{K^\pm/\bar{K}}^{(1)}$ subgroup may fuse into a defect belonging to the other $\mathbb{Z}_{\bar{K}}^{(1)}$ subgroup in $\cT$.\footnote{For instance, two charge-1 defects generating a $\mathbb{Z}_4$ symmetry fuse into a charge-1 defect generating a $\mathbb{Z}_2 \subset \mathbb{Z}_4$ symmetry.} Hence, these combinations of defects are identified with their counterparts from the opposite side of the interface.

The defects generating the residual $\overline{\bZ}_{N/K^\pm}^{(0)}$ symmetry on $\cC^{[1]}_{K^\pm}(M^3)$ follow parallel arguments, i.e.~defects generating the common subgroup of the symmetries can move freely across the interface at $M^2$, while others are trivialized. We will refer to the anomaly inflow interface described by \eqref{eq:example_identity_interface_action} as a ``soft interface,'' and denote it as
\begin{equation}
    1^{[2]}_{K^+,K^-}(M^2) = \int \mathcal{D}\varphi_0 \exp\bigg(2\pi i \int_{M^2}\varphi_0\left(\bar{K}d\bar{c}_1 + B_2^+ - B_2^-\right)\bigg) \, .
\end{equation}
The notation merely reflects the fact that the interface acts {\it almost} like an ``identity,'' and is ``minimal'' in the sense that is permeable to the subset of symmetry defects generating the common subgroup.\footnote{As with the $n$-condensation defects, the superscript in the square brackets indicates the codimension with respect to the $d$-dimensional spacetime.} Specifically, in the case where $K^+ = K^- = \bar{K}$, we recover the identity defect on $\cC^{[1]}_{K^\pm}(M^3)$.

While we focused above on an interface between two codimension-1 condensation defects, the construction is readily generalized to interfaces between two condensation defects of any codimension. Note that the notation has an implicit directionality, where $1^{[2]}_{K^+,K^-}(M^2)$ is defined such that it mediates from $\cC^{[1]}_{K^+}(M^3_+)$ on the ``left'' to $\cC^{[1]}_{K^-}(M^3_-)$ on the ``right.'' This direction and choice of left/right is arbitrary, but important to keep consistent when we discuss the fusion of these interfaces in Section \ref{sec:cond_complete} below.

\paragraph{Hard interface:} One can also construct a different interface by considering a slightly modified set up. We start with the same composite system \eqref{eq:example_composite_action}, but now consider the more general gauge transformation
\begin{equation}\label{eq:example_composite_gauge_transfo_general}
    a_1^\pm \to a_1^\pm + d\lambda_0^\pm\,,\qquad c_1^\pm \to c_1^\pm + d\omega_0^\pm\,,
\end{equation}
where we have allowed for different gauge parameters $\lambda_0^\pm$ on the two sides of the interface. Demanding this configuration to still be gauge invariant yields the inflow action,
\begin{equation}\label{eq:example_zero_interface_action}
    S_{\rm 2d} = \int_{M^2}\varphi^+_0\left(K^+dc_1^+ + B_2^+\right) - \varphi^-_0\left(K^-dc_1^- + B_2^-\right)\,,
\end{equation}
where $\varphi_0^\pm$ are dynamical scalar fields that transform as $\varphi_0^\pm\to \varphi_0^\pm - \lambda_0^\pm$ under \eqref{eq:example_composite_gauge_transfo_general}.

On this interface, the Lagrange multiplier fields $\varphi_0^\pm$ impose Dirichlet boundary conditions,
\begin{equation}
    (B_2^+ + K^+dc_1^+)|_{M^2} = 0\,,\qquad  (B_2^- + K^-dc_1^-)|_{M^2} = 0\,,
\end{equation}
which effectively trivializes {\it all} defects for both the $\widehat{\bZ}_{K^\pm}^{(1)}$ and $\overline{\bZ}_{N/K^\pm}^{(0)}$ symmetries when crossing the interface worldvolume. As this acts as a ``hard wall'' that kills all symmetry defects, we will refer to such an interface as the ``hard interface,'' and denote it $0^{[2]}_{K^+,K^-}$. Similarly to before, the construction is generalizes to interfaces between two condensation defects of any codimension.

More general interfaces between two given condensation defects can be obtained from either a soft or hard interface by decorating with additional topological defects, or by stacking with a symmetry-protected topological (SPT) order etc. The constructions of these interfaces are qualitatively straightforward and will not play important roles in the rest of this paper, so we are not going to pursue the details here.

\subsection{Condensation Completeness}\label{sec:cond_complete}

Condensation defects have two crucial properties which we are going to examine below. For concreteness, we will focus on the 1-condensation defect $\cC^{[1]}_K(M^3)$, obtained by 1-gauging a $\bZ_K^{(1)}\subseteq\bZ_N^{(1)}$ symmetry in $\cT$, along with the higher condensation defect $\cC^{[2]}_{\dot{K}}(M^2)$ obtained by 1-gauging a $\bZ_{\dot{K}}^{(1)}\subseteq\widehat{\bZ}_K^{(1)}$ symmetry in the $\cC^{[1]}_K(M^3)$ worldvolume theory. However, the resulting discussion is applicable to all other condensation defects in a straightforward manner. 
\paragraph{Idempotence:} Our first observation is that the condensation defect absorbs any symmetry defect used in its construction. Let $\cU_k(\Sigma^2)$ be a symmetry generator in the $\bZ_K^{(1)}$ subgroup with support $\Sigma^2 \in H_2(M^3;\bZ_K)$, then using the product defined in \eqref{eq:example_monoidal}, we have the fusion rule,
\begin{align}\label{eq:example_absorb_defect}
    \cU_k(\Sigma^2)\otimes \cC_K^{[1]}(M^3) &= \cU_k(\Sigma^2)\otimes\bigoplus_{\widetilde{\Sigma}^2\in H_2(M^3;\bZ_K)}\cU_1(\widetilde{\Sigma}^2)\nn\\
    &= \bigoplus_{\widetilde{\Sigma}^2\in H_2(M^3;\bZ_K)}\cU_1(\widetilde{\Sigma}^2)\nn\\
    &= \cC^{[1]}_K(M^3)\,,
\end{align}
where in the second line we see that the defect $\cU_k(\Sigma^2)$ simply rearranges the summands of the condensation defect.\footnote{Strictly speaking, to compute the monoidal product of these operators we should consider $\cU_k$ as dressing a trivial codimension-1 defect, i.e.~the fusion is given by $((\cU_k\acts 1^{[1]})\otimes \cC^{[1]}_K)(M^3)$. We will not be careful to distinguish such mathematical subtleties in what follows, though such structure can be reinstated if needed.} This fusion rule follows directly from the physical construction of $\cC^{[1]}_K(M^3)$ as a 1-gauging of the $\bZ_K^{(1)}$ symmetry. The absorption of $\cU_k$ by the condensation defect is the statement that the global symmetry generators become trivialized when the symmetry is gauged. 

Next we shall consider the more involved product coming from the parallel fusion $\cC^{[1]}_K(M^3) \otimes \cC^{[1]}_K(M^3)$ of a condensation defect with itself. From \eqref{eq:example_absorb_defect}, we can immediately deduce that this product must be proportional to $\cC^{[1]}_K(M^3)$. Carrying out the computation more explicitly using \eqref{eq:example_continuum}, such as in \cite{Roumpedakis:2022aik,Bah:2023ymy}, one finds that the coefficient of proportionality is a 3d decoupled TQFT $T_K(M^3)$. We denote such a fusion rule schematically as 
\begin{equation}\label{eq:example_idempotent}
    \cC^{[1]}_K\otimes\cC^{[1]}_K = |T_K| \, \cC^{[1]}_K\,,
\end{equation}
where $T_K(M^3)$ is the decoupled Dijkgraaf-Witten theory,
\begin{equation}
    \int \cD a_1'\cD c_1' \exp\left(2\pi i \int_{M^3}K a_1'\wedge dc_1'\right)\, .
\end{equation}
The exact interpretation of the coefficient $\left|T_K(M^3)\right|$ is subtle, and we will return to it later in Section \ref{sec:decoupled_TQFTs}. For now it is important to note that the fusion rule implies the existence of a 2-dimensional interface between the product $\cC_K^{[1]}\otimes\cC_K^{[1]}$ and the original defect $\cC_K^{[1]}$.

For our present purposes, the above tells us that fusing two condensation defects yields again a condensation defect, up to some coefficient. In this sense, condensation defects are {\it idempotent} operators, provided that the decoupled TQFT does not act on any bulk operators. We will give a precise definition of idempotence later in a categorical context.

We can repeat the argument for the higher condensation defect $\cC^{[2]}_{\dot{K}}$ living on the $\cC^{[1]}_K$ worldvolume and find that it too is idempotent, i.e.
\begin{equation}\label{eq:example_higher_idempotent}
    \cC^{[2]}_{\dot{K}}\otimes\cC^{[2]}_{\dot{K}} = |\fT_{\dot{K}}| \, \cC^{[2]}_{\dot{K}} \, ,
\end{equation}
where now $\fT_{\dot{K}}(M^2)$ is a decoupled 2-dimensional $\bZ_{\dot{K}}$ gauge theory, with continuum action given by
\begin{equation}
    \int\cD c_0'\cD a_0'\exp\left(2\pi i \int_{M^2}\dot{K} c_0' da_0'\right) \, .
\end{equation}
By the same token, one finds that all $n$-condensation defects in $\cT$ are idempotent for $1 \leq n \leq 4$, with fusion rules similar to \eqref{eq:example_idempotent} and \eqref{eq:example_higher_idempotent}.

\paragraph{Splitting:} We now claim that the 2-condensation defect $\cC^{[2]}_{\dot{K}}$ admits a natural splitting into two interfaces given by
\begin{equation}\label{eq:example_splitting}
    \cC^{[2]}_{\dot{K}}(M^2) = \iota(M^2) \otimes \rho(M^2) := 1^{[2]}_{K/\dot{K},K}(M^2) \otimes 1^{[2]}_{K,K/\dot{K}}(M^2) \, .
\end{equation} 
The proposed splitting can be inferred from the interface conditions imposed on the two sets of operators $\widehat{\cU}_{\hat{k}}(\gamma^1)$ and $\overline{\cU}_{\overline{n}}(\Sigma^2)$ as discussed following \eqref{eq:example_identity_interface_action}. Explicitly, the soft interface $1^{[2]}_{K,K/\dot{K}}(M^2)$ projects out the defects generating the subgroup $\bZ_{\dot{K}}^{(1)}\subseteq\widehat{\bZ}_K^{(1)}$ moving from left to right, while identifying those generating the common subgroup $\bZ_{K/\dot{K}}^{(1)}$.  Additionally, it projects out defects generating the subgroup $\bZ_{\dot{K}}^{(0)} \subseteq \overline{\bZ}_{N\dot{K}/K}^{(0)}$ when moving from the right to the left.

\begin{figure}
    \centering
\tikzset{every picture/.style={line width=0.75pt}} 
\begin{tikzpicture}[x=0.75pt,y=0.75pt,yscale=-1,xscale=1]
\draw  [draw opacity=0] (394,246.07) .. controls (393.96,246.07) and (393.92,246.07) .. (393.88,246.07) .. controls (381.52,246.07) and (371.5,234.37) .. (371.5,219.93) .. controls (371.5,205.49) and (381.52,193.79) .. (393.88,193.79) .. controls (394.33,193.79) and (394.78,193.81) .. (395.22,193.84) -- (393.88,219.93) -- cycle ; \draw  [color={rgb, 255:red, 208; green, 2; blue, 27 }  ,draw opacity=1 ] (394,246.07) .. controls (393.96,246.07) and (393.92,246.07) .. (393.88,246.07) .. controls (381.52,246.07) and (371.5,234.37) .. (371.5,219.93) .. controls (371.5,205.49) and (381.52,193.79) .. (393.88,193.79) .. controls (394.33,193.79) and (394.78,193.81) .. (395.22,193.84) ;  
\draw  [draw opacity=0] (268.42,246.07) .. controls (268.38,246.07) and (268.34,246.07) .. (268.3,246.07) .. controls (255.94,246.07) and (245.91,234.37) .. (245.91,219.93) .. controls (245.91,205.49) and (255.94,193.79) .. (268.3,193.79) .. controls (268.75,193.79) and (269.2,193.81) .. (269.64,193.84) -- (268.3,219.93) -- cycle ; \draw  [color={rgb, 255:red, 208; green, 2; blue, 27 }  ,draw opacity=1 ] (268.42,246.07) .. controls (268.38,246.07) and (268.34,246.07) .. (268.3,246.07) .. controls (255.94,246.07) and (245.91,234.37) .. (245.91,219.93) .. controls (245.91,205.49) and (255.94,193.79) .. (268.3,193.79) .. controls (268.75,193.79) and (269.2,193.81) .. (269.64,193.84) ;  
\draw    (395.22,193.84) .. controls (442.13,200.85) and (463.18,163.15) .. (506.14,181.56) .. controls (549.11,199.97) and (543.85,239.43) .. (503.51,256.09) .. controls (463.18,272.75) and (446.52,236.8) .. (395.22,246.45) ;
\draw    (457.48,213.57) .. controls (471.95,226.28) and (489.48,228.03) .. (501.76,214) ;
\draw    (463.62,217.95) .. controls (477.21,210.5) and (483.35,209.62) .. (495.62,219.27) ;
\draw    (270.16,246.62) .. controls (224.67,241) and (201.85,276.52) .. (159.1,257.6) .. controls (116.35,238.69) and (122.07,199.29) .. (162.6,183.1) .. controls (203.13,166.92) and (218.67,196.33) .. (270.77,194.01) ;
\draw    (163.86,216.73) .. controls (178.18,229.61) and (195.7,231.57) .. (208.13,217.69) ;
\draw    (169.95,221.19) .. controls (183.63,213.89) and (189.77,213.09) .. (201.94,222.88) ;

\draw    (269.64,193.84) -- (395.22,193.84) ;
\draw    (269.64,246.45) -- (395.22,246.45) ;
\draw  [draw opacity=0][dash pattern={on 0.84pt off 2.51pt}] (393.84,193.79) .. controls (393.88,193.79) and (393.92,193.79) .. (393.95,193.79) .. controls (406.32,193.83) and (416.31,205.56) .. (416.27,219.99) .. controls (416.23,234.43) and (406.17,246.11) .. (393.81,246.07) .. controls (393.36,246.07) and (392.91,246.05) .. (392.47,246.02) -- (393.88,219.93) -- cycle ; \draw  [color={rgb, 255:red, 208; green, 2; blue, 27 }  ,draw opacity=1 ][dash pattern={on 0.84pt off 2.51pt}] (393.84,193.79) .. controls (393.88,193.79) and (393.92,193.79) .. (393.95,193.79) .. controls (406.32,193.83) and (416.31,205.56) .. (416.27,219.99) .. controls (416.23,234.43) and (406.17,246.11) .. (393.81,246.07) .. controls (393.36,246.07) and (392.91,246.05) .. (392.47,246.02) ;  
\draw  [draw opacity=0][dash pattern={on 0.84pt off 2.51pt}] (268.25,193.79) .. controls (268.29,193.79) and (268.33,193.79) .. (268.37,193.79) .. controls (280.73,193.83) and (290.72,205.56) .. (290.68,219.99) .. controls (290.64,234.43) and (280.59,246.11) .. (268.23,246.07) .. controls (267.78,246.07) and (267.33,246.05) .. (266.89,246.02) -- (268.3,219.93) -- cycle ; \draw  [color={rgb, 255:red, 208; green, 2; blue, 27 }  ,draw opacity=1 ][dash pattern={on 0.84pt off 2.51pt}] (268.25,193.79) .. controls (268.29,193.79) and (268.33,193.79) .. (268.37,193.79) .. controls (280.73,193.83) and (290.72,205.56) .. (290.68,219.99) .. controls (290.64,234.43) and (280.59,246.11) .. (268.23,246.07) .. controls (267.78,246.07) and (267.33,246.05) .. (266.89,246.02) ;  
\draw    (334.44,49.01) .. controls (381,50.67) and (403.51,19.48) .. (446.48,37.89) .. controls (489.44,56.31) and (484.18,95.77) .. (443.85,112.43) .. controls (403.51,129.09) and (388.33,108) .. (333.82,101.62) ;
\draw    (397.81,69.9) .. controls (412.28,82.61) and (429.82,84.37) .. (442.09,70.34) ;
\draw    (403.95,74.28) .. controls (417.54,66.83) and (423.68,65.95) .. (435.95,75.6) ;
\draw    (333.82,101.62) .. controls (288.33,96) and (265.51,131.52) .. (222.76,112.6) .. controls (180.02,93.69) and (185.74,54.29) .. (226.27,38.1) .. controls (266.79,21.92) and (282.33,51.33) .. (334.44,49.01) ;
\draw    (227.53,71.73) .. controls (241.85,84.61) and (259.36,86.57) .. (271.8,72.69) ;
\draw    (233.61,76.19) .. controls (247.29,68.89) and (253.44,68.09) .. (265.6,77.88) ;

\draw  [draw opacity=0] (328.5,101.25) .. controls (328.46,101.25) and (328.42,101.25) .. (328.38,101.25) .. controls (316.02,101.25) and (306,89.55) .. (306,75.11) .. controls (306,60.67) and (316.02,48.97) .. (328.38,48.97) .. controls (328.83,48.97) and (329.28,48.98) .. (329.72,49.01) -- (328.38,75.11) -- cycle ; \draw  [color={rgb, 255:red, 144; green, 19; blue, 254 }  ,draw opacity=1 ] (328.5,101.25) .. controls (328.46,101.25) and (328.42,101.25) .. (328.38,101.25) .. controls (316.02,101.25) and (306,89.55) .. (306,75.11) .. controls (306,60.67) and (316.02,48.97) .. (328.38,48.97) .. controls (328.83,48.97) and (329.28,48.98) .. (329.72,49.01) ;  
\draw  [draw opacity=0][dash pattern={on 0.84pt off 2.51pt}] (332.44,49.01) .. controls (332.48,49.01) and (332.52,49.01) .. (332.56,49.01) .. controls (344.92,49.05) and (354.91,60.78) .. (354.87,75.22) .. controls (354.83,89.65) and (344.77,101.33) .. (332.41,101.3) .. controls (331.96,101.29) and (331.51,101.28) .. (331.07,101.25) -- (332.48,75.16) -- cycle ; \draw  [color={rgb, 255:red, 144; green, 19; blue, 254 }  ,draw opacity=1 ][dash pattern={on 0.84pt off 2.51pt}] (332.44,49.01) .. controls (332.48,49.01) and (332.52,49.01) .. (332.56,49.01) .. controls (344.92,49.05) and (354.91,60.78) .. (354.87,75.22) .. controls (354.83,89.65) and (344.77,101.33) .. (332.41,101.3) .. controls (331.96,101.29) and (331.51,101.28) .. (331.07,101.25) ;  
\draw [line width=0.75]    (332.83,122) -- (332.83,166.33)(329.83,122) -- (329.83,166.33) ;
\draw [shift={(331.33,173.33)}, rotate = 270] [color={rgb, 255:red, 0; green, 0; blue, 0 }  ][line width=0.75]    (10.93,-4.9) .. controls (6.95,-2.3) and (3.31,-0.67) .. (0,0) .. controls (3.31,0.67) and (6.95,2.3) .. (10.93,4.9)   ;
\draw [shift={(331.33,115)}, rotate = 90] [color={rgb, 255:red, 0; green, 0; blue, 0 }  ][line width=0.75]    (10.93,-4.9) .. controls (6.95,-2.3) and (3.31,-0.67) .. (0,0) .. controls (3.31,0.67) and (6.95,2.3) .. (10.93,4.9)   ;

\draw (136.67,20.4) node [anchor=north west][inner sep=0.75pt]    {$\mathcal{C}_{K}^{[ 1]}\left( M_{+}^{3}\right)$};
\draw (464,19.07) node [anchor=north west][inner sep=0.75pt]    {$\mathcal{C}_{K}^{[ 1]}\left( M_{-}^{3}\right)$};
\draw (298.67,17.73) node [anchor=north west][inner sep=0.75pt]  [color={rgb, 255:red, 144; green, 19; blue, 254 }  ,opacity=1 ]  {$\mathcal{C}_{\dot{K}}^{[ 2]}\left( M^{2}\right)$};
\draw (61,182.73) node [anchor=north west][inner sep=0.75pt]    {$\mathcal{C}_{K}^{[ 1]}\left( M_{+}^{3}\right)$};
\draw (535,180.07) node [anchor=north west][inner sep=0.75pt]    {$\mathcal{C}_{K}^{[ 1]}\left( M_{-}^{3}\right)$};
\draw (272.67,251.07) node [anchor=north west][inner sep=0.75pt]    {$\mathcal{C}_{K/\dot{K}}^{[ 1]}\left( M^{2} \times I\right)$};
\draw (243,164.4) node [anchor=north west][inner sep=0.75pt]  [color={rgb, 255:red, 208; green, 2; blue, 27 }  ,opacity=1 ]  {$1_{K,K/\dot{K}}^{[1]}$};
\draw (371.67,164.4) node [anchor=north west][inner sep=0.75pt]  [color={rgb, 255:red, 208; green, 2; blue, 27 }  ,opacity=1 ]  {$1_{K/\dot{K} ,K}^{[1]}$};
\end{tikzpicture}
    \caption{The ``cylinder gauging'' used in the factorization of $\cC_{\dot{K}}^{[2]}$. In the top diagram, the 2-condensation defect acts as an interface between two copies of $\cC_K^{[1]}$. In the bottom diagram, we have split $\cC^{[2]}_{\dot{K}}$ into two interfaces $1_{K,K/\dot{K}}^{[1]}$ and $1_{K/\dot{K},K}^{[1]}$ that respectively act from $\cC_K^{[1]}$ to $\cC_{K/\dot{K}}^{[1]}$ and vice versa.}
    \label{fig:cylinder_gauging}
\end{figure}
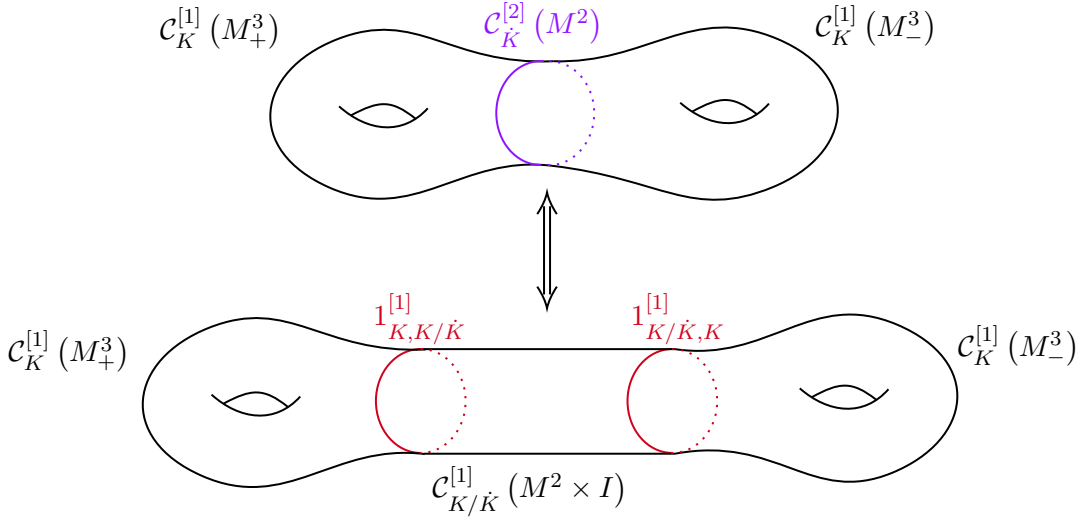

The splitting claim can be verified at the level of the worldvolume actions. By explicitly computing the fusion, one finds the following, up to a decoupled TQFT factor,
\begin{equation}\label{eq:example_iota_of_rho}
     \left(\iota\otimes\rho\right)(M^2) = \int\cD\varphi_0 \exp\left(2\pi i\int_{M^2}\varphi_0\left(\frac{K}{\dot{K}} \, d\overline{c}_1 + \left(B_{2,{\rm b}} - B_{2,{\rm i}}\right)\right)\right) \, .
\end{equation}
Here $B_{2,{\rm b}}$ and $B_{2,{\rm i}}$ respectively denote the background value of $B_2$ in the ambient $\cC^{[1]}_K(M^3)$ ``bulk'' theory and its value when restricted to $M^2$ ``interface.'' Using the discussion surrounding \eqref{eq:example_identity_interface_action}, this worldvolume action trivializes the generators of a $\bZ_{\dot{K}}^{(1)}\subseteq\widehat{\bZ}_K^{(1)}$ symmetry while allowing the remaining $\bZ_{K/\dot{K}}$ generators to pass through unchanged. Additionally, defects generating the residual symmetry $\overline{\bZ}_{N/K}^{(0)}$ can pass freely through the $\iota\otimes\rho$ interface. We thus see that, when regarded as a defect inserted within the $\cC^{[1]}_K(M^3)$ worldvolume theory, the composition $\iota\otimes\rho$ correctly recovers the behavior of the higher condensation defect $\cC^{[2]}_{\dot{K}}(M^2)$.

There is a clear geometric interpretation of what we mean by ``splitting.'' Recall that $\rho = 1^{[2]}_{K,K/\dot{K}}(M^2)$ and $\iota = 1^{[2]}_{K/\dot{K},K}(M^2)$ are anomaly inflow interfaces (with opposite orientations) separating two 3d ``bulk theories'' $\cC^{[1]}_K$ and $\cC^{[1]}_{K/\dot{K}}$. The factorization $\cC^{[2]}_{\dot{K}}(M^2) = \iota(M^2) \otimes \rho(M^2)$ can then be realized by the setup in Figure \ref{fig:cylinder_gauging}, where $M^2$ partitions $M^3$ into two disjoint pieces $M^3_\pm$, such that $M^3_+ \sqcup M^3_- = M^3$ and $\d M^3_+ = -\d M^3_- = M^2$.

Importantly, on both ends of the upper diagram, we have the same 3d ``bulk theory'' $\cC^{[1]}_K$, so $\cC^{[2]}_{\dot{K}}(M^2)$ acts as a {\it defect insertion} in this theory. In the lower diagram, however, we nucleate a cylinder $M^2 \times I$ sandwiched between $M^3_\pm$, within which a different ``bulk theory'' $\cC^{[1]}_{K/\dot{K}}$ lives, i.e.~we gauge the subgroup $\bZ_{\dot{K}}^{(1)}\subseteq\widehat{\bZ}_K^{(1)}$ along this cylinder. It is important to note that the manifold being sandwiched has to be the cylinder $M^2 \times I$, so as to preserve the topological data on $M^2$. Hence, the original 2-condensation defect $\cC^{[2]}_{\dot{K}}(M^2)$ reincarnates here as two {\it interfaces} between different theories, identified respectively with $1^{[2]}_{K,K/\dot{K}}(M^2)$ and $1^{[2]}_{K/\dot{K},K}(M^2)$.\footnote{Such a construction can be understood as a variation of the technique of half-space gauging used in the study of non-invertible duality defects \cite{Kaidi:2021xfk,Choi:2021kmx,Choi:2022jqy,Choi:2022zal}.}

The geometric realization above offers us a precise relation between higher gauging and ordinary gauging. Two ``bulk theories,'' differing from each other by an ordinary gauging (of some subgroup), are mediated by a condensation defect, in the sense that it splits into a pair of interfaces separating the bulk theories as described above.

Note that if we reverse the order of composition of interfaces, i.e.~$\left(\rho\otimes\iota\right)(M^2)$, we will find an action nearly identical to \eqref{eq:example_iota_of_rho}. However, due to the ordering, the resulting defect should now be interpreted as an insertion in the $\cC^{[1]}_{K/\dot{K}}$ theory, thus changing the interpretations of $\overline{c}_1$, $B_{2,b}$, and $B_{2,i}$. In such a setup the $\bZ_{\dot{K}}^{(1)}$ symmetry is {\it already} trivialized in the ambient space, and so $\rho\otimes\iota$ acts as the trivial defect in this theory, i.e.
\begin{equation}
    \left(\rho\otimes\iota\right)(M^2) = 1^{[2]}_{K/\dot{K}}(M^2) \, .
\end{equation}

The procedure above holds not only for the chosen example of the $\cC^{[2]}_{\dot{K}}$ defect, but also, for example, for the dual $\cC^{[2]}_{\dot{N}}(M^2)$ defect on the $\cC^{[1]}_K(M^3)$ worldvolume theory. In fact, one can find that {\it any} condensation defect in $\cT$, such as those in Figure \ref{fig:condensation_defects_diagram}, can be split into appropriate $\rho$ and $\iota$ interfaces with the aforementioned properties. Precisely, if $\cC^{[n]}_G(M^{4-n})$ denotes an arbitrary $n$-condensation defect in $\cT$ obtained by higher-gauging some group $G$, then it is idempotent, and there exist interfaces $\rho(M^{4-n})$ and $\iota(M^{4-n})$ such that
\begin{equation}\label{eq:example_splitting_general}
    \left(\iota \otimes\rho\right)(M^{4-n}) = \cC^{[n]}_G(M^{4-n}) \, , \qquad \left(\rho\otimes\iota\right)(M^{4-n}) = 1^{[n]}_G(M^{4-n}) \, ,
\end{equation}
where $1^{[n]}_G$ denotes the identity operator on $\cC^{[n-1]}_G(M^{5-n})$. This amounts physically to the statement that any defect constructed from a higher-gauging procedure can be reproduced by considering a pair of parallel gauging interfaces and dimensionally reducing along the interval sandwiched between them.

A crucial application is the case of $n=1$, where the 1-condensation defect $\cC^{[1]}_K(M^3)$ in the 4d theory $\cT$ can be split into two 3d interfaces $\rho(M^3)$ and $\iota(M^3)$. Here the cylinder $M^3\times I$ between $\rho$ and $\iota$ supports the theory $\cT/\bZ_K^{(1)}$, which describes a different global form of $\cT$ obtained by gauging the subgroup $\bZ_K^{(1)} \subseteq \bZ_N^{(1)}$ in the original 4d theory. In this sense, we see that the existence of 1-condensation defects $\cC^{[1]}_K(M^3)$, which are intrinsically defined within $\cT$, secretly encodes information from distinct theories $\cT/\bZ_K^{(1)}$. We will return to this point in the next section.

\section{Higher-categorical Structures}\label{sec:categorical_structures}

In Section \ref{sec:4d_example}, we already showed that the collection of symmetry defects, including the various $n$-condensation defects, in the 4d QFT $\mathcal{T}$ are naturally $\mathbb{C}$-linear, additive, and monoidal on purely physical grounds. The same conclusion applies to generic $d$-dimensional QFTs as well. Consequently, we would like to assemble a higher category $\mathfrak{C}$ encoding the data of the symmetry of $\mathcal{T}$.

To be more specific, $\mathfrak{C}$ will be a $d$-category where a certain sub-$(d-1)$-category incorporates all the symmetry defects of $\mathcal{T}$ and their fusion rules, such that it forms a fusion $(d-1)$-category in the sense of \cite{Johnson-Freyd:2020usu}.\footnote{Strictly speaking, the definition we are using in this paper is that for a {\it separable weak} (multi)fusion $(d-1)$-category. For our purposes, we will refer to it loosely as a fusion $(d-1)$-category.} The main focus in the rest of this paper is to argue that $\mathfrak{C}$ is indeed also Karoubi-complete, and to explore its physical implications. In order to do so, we will first need to introduce some (higher) categorical notions and terminologies.

As a warm-up, let us briefly review the inductive definition of a higher category. A {\it 0-category} is simply a {\it set} of objects. For our purposes, we always want this set to be finite. A {\it 1-category} contains the following data:
\begin{itemize}
    \item a collection of objects (or 0-morphisms),
    \item for any two objects $X$ and $Y$, a collection $\Hom(X,Y)$ of 1-morphisms $f: X \to Y$, such that $\Hom(X,Y)$ forms a set, i.e.~a 0-category,
    \item for any two maps $f\in \Hom(X,Y)$ and $g\in \Hom(Y,Z)$, there is a well-defined composition $g\circ f\in \Hom(X,Z)$,
    \item for any object $X$, there is an identity 1-morphism $1_X \in \Hom(X,X)$.
\end{itemize}
The collection $\Hom(X,X)$ of {\it endomorphisms} of an object $X$, i.e.~1-morphisms $f: X \to X$, is conveniently denoted instead as ${\rm End}(X)$. Moreover, the composition must be defined such that it is associative and unital, with the appropriate identity 1-morphisms acting as units. In what follows, we refer to such data (and higher generalizations) as the {\it coherence data} of a category.

To define a 2-category, we need the following data:
\begin{itemize}
    \item a collection of objects and their 1-morphisms as above,
    \item for any two 1-morphisms $f_1,f_2 \in \Hom(X,Y)$, a collection $\Hom(f_1,f_2)$ of 2-morphisms $\alpha: f_1 \to f_2$ satisfying suitable coherence relations, such that $\Hom(f_1,f_2)$ forms a 1-category.
\end{itemize}
Furthermore, the corresponding coherence data may be weakened to hold only up to invertible 2-morphisms.\footnote{In this work, we do not demand any strictness on our categories, so that any morphisms need only hold up to a higher isomorphism. As such any time the word ``category'' is used it should be implicitly understood to be weak.}

An {\it $n$-category} $\fC$ is then inductively defined, at least formally, to consist of objects, 1-morphisms, 2-morphisms, \dots, up to $n$-morphisms, such that $\Hom(X,Y)$ for any objects $X$ and $Y$ forms an $(n-1)$-category. We will henceforth use the notation $\fC_m$ to denote the collection of all $m$-morphisms, $0 \leq m \leq n$, in an $n$-category.

\subsection{The Symmetry $d$-category}

\begin{figure}
    \centering
\tikzset{every picture/.style={line width=0.75pt}} 
\begin{tikzpicture}[x=0.75pt,y=0.75pt,yscale=-1,xscale=1]
\draw    (100,71) .. controls (145.21,39.65) and (184.21,40.64) .. (228.65,70.1) ;
\draw [shift={(230,71)}, rotate = 213.98] [color={rgb, 255:red, 0; green, 0; blue, 0 }  ][line width=0.75]    (10.93,-3.29) .. controls (6.95,-1.4) and (3.31,-0.3) .. (0,0) .. controls (3.31,0.3) and (6.95,1.4) .. (10.93,3.29)   ;
\draw    (101,81) .. controls (143.24,113.67) and (184.17,114) .. (229.62,81.98) ;
\draw [shift={(231,81)}, rotate = 144.34] [color={rgb, 255:red, 0; green, 0; blue, 0 }  ][line width=0.75]    (10.93,-3.29) .. controls (6.95,-1.4) and (3.31,-0.3) .. (0,0) .. controls (3.31,0.3) and (6.95,1.4) .. (10.93,3.29)   ;
\draw    (165.5,58) -- (165.5,90)(162.5,58) -- (162.5,90) ;
\draw [shift={(164,98)}, rotate = 270] [color={rgb, 255:red, 0; green, 0; blue, 0 }  ][line width=0.75]    (10.93,-3.29) .. controls (6.95,-1.4) and (3.31,-0.3) .. (0,0) .. controls (3.31,0.3) and (6.95,1.4) .. (10.93,3.29)   ;
\draw   (377.28,142.57) -- (377.28,33.64) -- (453.01,8) -- (453.01,116.93) -- cycle ;
\draw    (377.28,88.11) -- (453.01,62.47) ;
\draw    (377.28,88.11) -- (415.14,75.29) ;
\draw [shift={(415.14,75.29)}, rotate = 341.3] [color={rgb, 255:red, 0; green, 0; blue, 0 }  ][fill={rgb, 255:red, 0; green, 0; blue, 0 }  ][line width=0.75]      (0, 0) circle [x radius= 3.35, y radius= 3.35]   ;

\draw (72,67.4) node [anchor=north west][inner sep=0.75pt]    {$X$};
\draw (242,67.4) node [anchor=north west][inner sep=0.75pt]    {$Y$};
\draw (156,23.4) node [anchor=north west][inner sep=0.75pt]    {$f_{1}$};
\draw (156,112.4) node [anchor=north west][inner sep=0.75pt]    {$f_{2}$};
\draw (145,66.4) node [anchor=north west][inner sep=0.75pt]    {$\alpha $};
\draw (379.28,34.53) node [anchor=north west][inner sep=0.75pt]    {$X$};
\draw (378.67,121.92) node [anchor=north west][inner sep=0.75pt]    {$Y$};
\draw (378.61,62.42) node [anchor=north west][inner sep=0.75pt]    {$f_{1}$};
\draw (433.67,44.26) node [anchor=north west][inner sep=0.75pt]    {$f_{2}$};
\draw (417.14,76.18) node [anchor=north west][inner sep=0.75pt]    {$\alpha $};
\end{tikzpicture}
    \caption{The relation between categorical and geometric descriptions of defects in our category of symmetry defects $\fC$. On the left we depict 1- and 2-morphisms graphically, while on the right we depict the realization in spacetime. Here $X$ and $Y$ are objects (codimension-1 operators), $f_1$ and $f_2$ are 1-morphisms (codimension-2 operators), and $\alpha$ is a 2-morphism (codimension-3 operator).}
    \label{fig:morphisms_as_interfaces}
\end{figure}
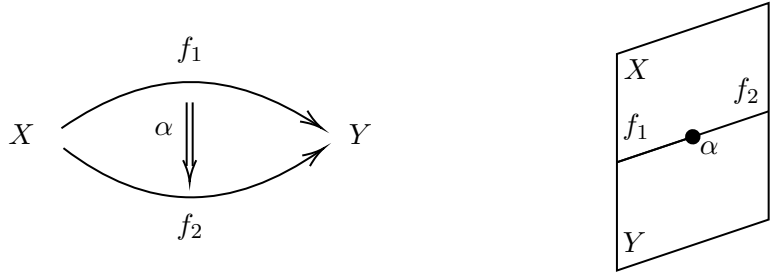

To construct an $n$-category $\fC$ which encapsulates all the topological defects of various codimensions, we adopt the geometric interpretation that a 1-morphism $f: X \to Y$ can be realized as an interface between two objects $X$ and $Y$. Likewise, a 2-morphism $\alpha: f_1 \to f_2$ can be realized as an interface between two 1-morphisms $f_1,f_2: X \to Y$, etc., see Figure \ref{fig:morphisms_as_interfaces}. This provides a natural grading of the morphisms in $\fC$ in terms of the codimensions of the corresponding defects. In particular, point defects in $\cT$ should be assigned as $n$-morphisms in $\fC$, since there are geometrically no interfaces between points, just like there are no $(n+1)$-morphisms between $n$-morphisms in an $n$-category.

By induction, a codimension-$m$ defect in a $d$-dimensional QFT $\cT$ corresponds to an $(n-d+m)$-morphism in $\fC$. The next question is how the integer $n$ should be related to the dimension $d$ of spacetime. If we care exclusively about the defects in a fixed theory $\cT$, then it is sufficient to set $n=d-1$, such that codimension-1 defects are 0-morphisms, i.e.~objects, in $\fC$. Such a convention is used in part of the physics literature \cite{Moore:1988qv, Bhardwaj:2017xup, Bhardwaj:2022yxj, Copetti:2023mcq, Thorngren:2019iar}.

On the other hand, in the 4d QFT example we studied in the previous sections, we found that theories with different global forms, i.e.~$\mathcal{T}/\bZ_K^{(1)}$ for $K$ any divisor of $N$, naturally make an appearance when studying the splittability of 1-condensation defects. Since this is an important property establishing the relation between condensation defects and gauging, we shall encode such information by setting $n=d$ instead, such that the objects in our $d$-category $\fC$ are the collection of global forms of $\cT$ in the sense of \cite{Aharony:2013hda}.\footnote{Said more explicitly, the objects of $\fC$ only include different phases that can be obtained from $\cT$ via discrete gauging of a (non-trivial) global symmetry in $\cT$. In this way we do {\it not} include theories obtained by stacking $\cT$ with arbitrary $d$-dimensional TQFTs. This latter construction is explored more in \cite{Bhardwaj:2024xcx}.} In short, the collections of $m$-morphisms in $\fC$, with $0 \leq m \leq d$, are schematically given by
\begin{itemize}
    \item $\fC_0 = \{\text{theories related to $\cT$ by (ordinary) gauging}\}$,
    \item $\fC_1 = \{\text{codimension-1 defects (in all theories) and interfaces between theories}\}$,
    \item $\fC_2 = \{\text{codimension-2 defects and interfaces between codimension-1 defects}\}$,
    \item $\fC_m = \{\text{codimension-$m$ defects and interfaces between codimension-$(m-1)$ defects}\}$,
    \item $\fC_d = \{\text{point defects and junctions between line defects}\}$.
\end{itemize}

Restricting to our 4d example, we have
\begin{equation}
    \fC_0 = \left\{\cT/\bZ_K^{(1)} \right|\left. \bZ_K^{(1)} \subseteq \bZ_N^{(1)}\right\} \, .
\end{equation}
The 1-morphisms of $\fC$ consists of all codimension-1 topological defects. Non-trivial 1-morphisms in $\Hom(\cT,\cT')$ are codimension-1 gauging interfaces $\rho$ (or $\iota$), such as the soft interface $1_{\cT,\cT'}^{[1]}$, between the theories $\cT$ and $\cT'$. Non-trivial 1-morphisms in $\text{End}(\cT)$ are 1-condensation defects $\cC_K^{[1]}$, viewed as topological defects in the theory $\cT$. Note that if $\cT$ had a 0-form symmetry, then $\text{End}(\cT)$ would also include the corresponding codimension-1 symmetry generators. Similarly, $\text{End}(\cT')$ consists of all codimension-1 topological defects in the theory $\cT' = \cT/\bZ_K^{(1)}$ for any $K$.

Let us dwell for a moment on the codimension-1 interfaces present in $\Hom(\cT,\cT')$. Importantly the gauging interfaces here are {\it topological}, and thus can be described as a gapped interface theory between $\cT$ and $\cT'$. In other words, the 1-morphisms from $\cT$ to $\cT'$ in our category $\fC$ correspond to gapped boundary interfaces between the respective theories. If there are no such interfaces, then $\Hom(\cT,\cT')$ is empty.

To further manifest the categorical nature of symmetry defects, recall from Section \ref{subsec:condensation_defects} that on the 1-condensation defect $\cC_K^{[1]}$, there are a pair of global symmetries, $\widehat{\bZ}_K^{(1)} \times \overline{\bZ}_{N/K}^{(0)}$, whose symmetry generators respectively have codimensions 2 and 1 with respect to the 3-dimensional worldvolume. Importantly, the defects $\widehat{\cU}_{\hat{k}}$ in \eqref{eq:quantum_symmetry_step1} generating the Pontryagin dual symmetry $\widehat{\bZ}_K^{(1)}$ are localized to the worldvolume of $\cC_K^{[1]}$, and do not exist independently in the bulk theory $\cT$.

In terms of categorical data, since the identity defect $1_\cT^{[1]}$ and the 1-condensation defect $\cC_K^{[1]}$ are distinct 1-morphisms in $\fC$, it follows that $\text{End}(1_\cT^{[1]})$ and $\text{End}(\cC_K^{[1]})$ are distinct 2-categories in their own rights. The localization of the line defect $\widehat{\cU}_{\hat{k}}$ to $\cC_K^{[1]}$ is simply the statement that $\widehat{\cU}_{\hat{k}}$ is a 1-morphism in $\text{End}(\cC_K^{[1]})$ but not $\text{End}(1_\cT^{[1]})$.\footnote{To clarify, $\widehat{\cU}_{\hat{k}}$ is a 1-morphism in $\text{End}(\cC_K^{[1]})$ as a 2-category, but a 3-morphism in $\fC$ as a 4-category.}

Similar remarks apply to all other symmetry defects that are localized to some given $n$-condensation defect. We thus see that a higher category conveniently provides a hierarchical structure of ``bins'' for us to sort the otherwise tremendous amount of topological defects in our theory $\cT$.

\subsection{Monoidal Structure}\label{sec:monoidal_structure}

In our example in Section \ref{sec:4d_example}, we argued intuitively how the category ${\rm End}(\cT)$ of symmetry defects in $\cT$ is monoidal through the OPE \eqref{eq:example_monoidal}. While this does define a product $\otimes_m$ for $m$-morphisms in $\fC$, or codimension-$m$ defects, for $1\leq m \leq n$, this is not enough to say that ${\rm End}(\cT)$ is monoidal. Additional data is needed to label a general $n$-category as being monoidal. Part of this data is a list of coherence conditions, such as associativity and unitality, that is suitably categorified. We will describe this data somewhat inductively by first describing monoidal 1-categories, then monoidal 2-categories, before generalizing the discussion to monoidal $n$-categories.

Let $\mathfrak{D}$ be a 1-category\footnote{We denote an arbitrary category with $\mathfrak{D}$ so as to not confuse with our category $\fC$ of symmetry defects. One can view $\mathfrak{D}$ as describing ${\rm End}(\cT)$ in a physical setting.} with a functor $\otimes: \mathfrak{D}\times\mathfrak{D}\to\mathfrak{D}$. We say $\mathfrak{D}$ is a monoidal category if also equipped with the following data:
\begin{itemize}
    \item a distinguished object $1^{[1]}\in \mathfrak{D}_0$ called the {\it monoidal unit},
    \item a natural isomorphism $a_{X,Y,Z}:(X\otimes_0 Y)\otimes_0 Z\xrightarrow{\raisebox{-0.65ex}{$\sim$}} X\otimes_0 (Y\otimes_0 Z)$ called the {\it associator},
    \item natural isomorphisms, $L_X:1^{[1]}\otimes_0 X\xrightarrow{\raisebox{-0.65ex}{$\sim$}} X$ and $R_X: X\otimes_0 1^{[1]} \xrightarrow{\raisebox{-0.65ex}{$\sim$}} X$, called left and right {\it unitors} respectively,
\end{itemize}
such that $a, L$, and $R$ satisfy relations known as the pentagon and triangle identities. The necessary coherence relations are presented in Appendix \ref{app:categorical_notions}. Note that ${\rm End}(X)$ is immediately a monoid for any $X$ in $\mathfrak{D}$, where the operation $\otimes_1$ is simply composition of the morphisms. If $X$ is taken to be the monoidal unit $1^{[1]}$, one can further show that ${\rm End}(1^{[1]})$ has the structure of a {\it commutative} monoid.\footnote{A proof of this can be found in, for example, Proposition 2.2.10 of \cite{etingof2015tensor}.}

\begin{figure}
    \centering
\tikzset{every picture/.style={line width=0.75pt}} 
\begin{tikzpicture}[x=0.75pt,y=0.75pt,yscale=-1,xscale=1]
\draw  [fill={rgb, 255:red, 189; green, 16; blue, 224 }  ,fill opacity=0.35 ] (331,34.58) -- (352.94,34.58) -- (352.94,86.53) -- (331,86.53) -- cycle ;
\draw  [fill={rgb, 255:red, 74; green, 144; blue, 226 }  ,fill opacity=0.55 ] (353.17,34.58) -- (375.11,34.58) -- (375.11,86.53) -- (353.17,86.53) -- cycle ;
\draw [line width=2.25]    (248.5,49.58) -- (331,49.58) ;
\draw [line width=2.25]    (248.5,71.25) -- (331,71.25) ;
\draw [line width=2.25]  [dash pattern={on 2.53pt off 3.02pt}]  (374.83,49.58) -- (457.33,49.58) ;
\draw [line width=2.25]  [dash pattern={on 2.53pt off 3.02pt}]  (374.83,71.25) -- (457.33,71.25) ;
\draw  [fill={rgb, 255:red, 74; green, 144; blue, 226 }  ,fill opacity=0.55 ] (331,121.75) -- (352.94,121.75) -- (352.94,173.69) -- (331,173.69) -- cycle ;
\draw  [fill={rgb, 255:red, 189; green, 16; blue, 224 }  ,fill opacity=0.35 ] (353.17,121.75) -- (375.11,121.75) -- (375.11,173.69) -- (353.17,173.69) -- cycle ;
\draw [line width=2.25]    (248.5,136.75) -- (331,136.75) ;
\draw [line width=2.25]    (248.5,158.42) -- (331,158.42) ;
\draw [line width=2.25]  [dash pattern={on 2.53pt off 3.02pt}]  (374.83,136.75) -- (457.33,136.75) ;
\draw [line width=2.25]  [dash pattern={on 2.53pt off 3.02pt}]  (374.83,158.42) -- (457.33,158.42) ;
\draw    (154,84.83) -- (154,118.42)(151,84.83) -- (151,118.42) ;
\draw [shift={(152.5,126.42)}, rotate = 270] [color={rgb, 255:red, 0; green, 0; blue, 0 }  ][line width=0.75]    (10.93,-3.29) .. controls (6.95,-1.4) and (3.31,-0.3) .. (0,0) .. controls (3.31,0.3) and (6.95,1.4) .. (10.93,3.29)   ;

\draw (190.92,52.65) node [anchor=north west][inner sep=0.75pt]    {$X\otimes _{0} Y$};
\draw (190.92,139.82) node [anchor=north west][inner sep=0.75pt]    {$X\otimes _{0} Y$};
\draw (462.42,52.65) node [anchor=north west][inner sep=0.75pt]    {$X'\otimes _{0} Y'$};
\draw (462.42,139.82) node [anchor=north west][inner sep=0.75pt]    {$X'\otimes _{0} Y'$};
\draw (131.58,51.82) node [anchor=north west][inner sep=0.75pt]    {$f\otimes _{0} g:$};
\draw (131.58,138.98) node [anchor=north west][inner sep=0.75pt]    {$g\otimes _{0} f:$};
\draw (156.33,96.23) node [anchor=north west][inner sep=0.75pt]    {$\phi _{f,g}$};
\draw (270,3) node [anchor=north west][inner sep=0.75pt]    {$\textcolor{DiagramMagenta}{\left( 1_{X}^{[ 2]} \otimes_{0} g\right)} \otimes_{1} \textcolor{DiagramBlue}{\left( f\otimes _{0} 1_{Y}^{[ 2]}\right)}$};
\draw (270,178) node [anchor=north west][inner sep=0.75pt]    {$\textcolor{DiagramBlue}{\left( f\otimes _{0} 1_{Y}^{[ 2]}\right)} \otimes_{1}\textcolor{DiagramMagenta}{\left( 1_{X}^{[ 2]} \otimes _{0} g\right)}$};
\end{tikzpicture}
    \caption{Depiction of the interchanger 2-isomorphism, where we are using the shorthand $\phi_{f,g}:f\otimes_0 g \xRightarrow{} g\otimes_0 f$ to represent the full expression in \eqref{eq:interchanger}. The solid lines represent $X\otimes_0 Y$ while the dashed lines represent $X'\otimes_0Y'$; the interfaces between these depict the two orders of 1-morphisms exchanged by the interchanger $\phi_{f,g}$. Importantly, this is distinct from the statement that $\otimes_1$ is commutative, as reversing the order of composition $f\otimes_1 g\Rightarrow g\otimes_1 f$ will generally be ill-defined as domains and codomains are moved around.}
    \label{fig:interchanger}
\end{figure}
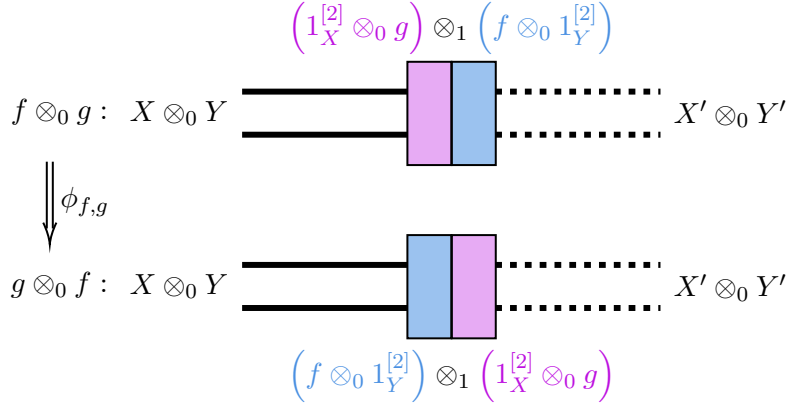

Moving up to a monoidal 2-category, which we continue to refer to as $\mathfrak{D}$, the pentagon and triangle equations need only hold up to some higher isomorphism. Additionally there is extra data in {\it interchanger} morphisms as follows \cite{2018arXiv181211933D}. Let $f:X\to X'$ and $g:Y\to Y'$ be two 1-morphisms in $\mathfrak{D}$, then there must be an interchanger 2-isomorphism
\begin{equation}\label{eq:interchanger}
    \phi_{f,g}: \left(f\otimes_0 1_{Y'}^{[2]}\right)\otimes_1\left(1_X^{[2]}\otimes_0 g\right) \xRightarrow{\raisebox{-0.65ex}{$\sim$}} \left(1_{X'}^{[2]}\otimes_0 g\right)\otimes_1\left(f\otimes_0 1_Y^{[2]}\right)
\end{equation}
that satisfies certain coherence relations such as compatibility with the tensor product $\otimes_0$ on objects. The 2-morphism $\phi_{f,g}$ is depicted in Figure \ref{fig:interchanger}. This morphism can be intuitively understood as the ``commuting'' of $f$ and $g$ as interfaces from $X\otimes_0 Y$ to $X'\otimes_0 Y'$. Above, the operation $\otimes_1$ should be understood as composition of the 1-morphisms, where $\otimes_0$ is the tensor product of the original 2-category.

If we now consider ${\rm End}(X)$ for any object $X$ in a 2-category, this will be a monoidal 1-category where the tensor product $\otimes_1$ is again given by composition. In the special case of the monoidal unit, ${\rm End}(1^{[1]})$ is further a {\it braided} monoidal 1-category, meaning there exists an isomorphism,\footnote{One can prove this with a categorification of the proof mentioned in the previous footnote.}
\begin{equation}\label{eq:braiding_morphism}
    c_{f,g}:f\otimes_1 g \xRightarrow{\raisebox{-0.65ex}{$\sim$}} g\otimes_1 f\,,
\end{equation}
induced by the interchanger morphism \eqref{eq:interchanger}, for any $f,g\in {\rm End}(1^{[1]})$. This braiding data must itself satisfy a relation known as the hexagon identity, presented in Appendix \ref{app:categorical_notions}.

We can continue to define a general monoidal $n$-category by including both the data of a product $\otimes_m$ for every $0\leq m<n$ and a list of coherence data at each step, such that each $\otimes_m$ is compatible with each $\otimes_{m'}$ for $m' > m$. A result of this coherence data is that each level of a monoidal $n$-category is ``more monoidal'' than the last. For example, in the case of a monoidal 1-category, as discussed above, the 0-category ${\rm End}(1^{[1]})$ is a {\it commutative} monoid, also known as a 2-monoidal 0-category. We can then say that a monoidal 1-category with a single object (the unit) is equivalent to a commutative monoid.

For a monoidal 2-category, ${\rm End}(1^{[1]})$ is a {\it braided} monoidal 1-category, thus we can say a monoidal 2-category with a single object is equivalent to a braided monoidal 1-category, or a 2-monoidal 1-category. At each level, there is more and more coherence data needed for the consistency of the category. The physical interpretation of this additional monoidal structure is that the operators at higher levels in the category correspond to operators of higher codimension. As a result, the higher one goes in the category, the more transverse dimensions in which the operators can move.

Importantly, at each level of a monoidal $n$-category, the relevant coherence data is not uniquely specified. There can be inequivalent choices of such data that still satisfy the necessary conditions and ensure consistency of the monoidal structure. For example, let our category $\mathfrak{D}$ be the monoidal 1-category ${\rm Vec}_G$, the category of $G$-graded complex vector spaces. The monoidal product of ${\rm Vec}_G$ is inherited from the group multiplication of $G$. In order to fully define this category as a monoidal category, we need an associator morphism $a$ that satisfies the necessary coherence conditions. However, such a morphism is not unique; it is only specified up to a 3-cocycle $\omega \in H^3(BG;U(1))$ \cite{etingof2015tensor}. One must then {\it choose} a value for $\omega$ for a full definition of ${\rm Vec}_G^{\omega}$.

In general, all the higher coherence data of a monoidal $n$-category can be similarly modified and thus must be specified in order to have a consistent monoidal structure. We will return to this discussion in Section \ref{sec:anomalous_symmetries}, where we will relate this higher coherence data to the anomalies of a QFT built from ${\rm End}(\cT)$. The example we studied in the earlier sections is non-anomalous, and hence correspondeds to a trivial choice of cocycle for the relevant coherence data.

Previously, we have motivated our category ${\rm End}(\cT)$ through the symmetry defects present in a physical QFT, and have further investigated the monoidal structure through the properties of the OPE \eqref{eq:example_monoidal}. An interesting aside is that if we assumed that $\fC$ were a higher category from the beginning, then any monoidal structure would be automatic. This is due to the Delooping Hypothesis, which states that a $k$-monoidal $n$-category is equivalent to a $(k-j)$-monoidal $(n+j)$-category for any $0\leq j\leq k$, where every $j'$-morphism is equivalent to the identity for $j'<j$ \cite{baez2007lecturesncategoriescohomology}. Applying this to our desired category, let us consider the 0-monoidal $d$-subcategory of $\fC$ comprised of $\cT$ and its endomorphisms. It follows from the Delooping Hypothesis that ${\rm End}(\cT)$ is a (1-)monoidal $(d-1)$-category, which is exactly what we were looking for. The physical manifestation of this data is that we can define an OPE \eqref{eq:example_monoidal} for each $\otimes_m$ in ${\rm End}(\cT)$, and furthermore this OPE is subject to anomalies as alluded to before. We thus see that an alternative (although somewhat circular) approach to establishing the monoidality of ${\rm End}(\cT)$ would be to posit that $\fC$ is a higher category in its own right, then the Delooping Hypothesis ensures all the relevant data for a coherent monoidal structure.

\subsection{Categorical Condensation}\label{sec:categorical_condensation}

To establish the Karoubi completeness of our symmetry $d$-category on firmer ground, we shall review the (higher) categorical definition of a ``condensation.'' Following \cite{Gaiotto:2019xmp,Johnson-Freyd:2020usu}, an object $X$ in an $n$-category $\mathfrak{C}$ {\it condenses} onto another object $Y$, denoted as $X \cond Y$, if there exist a pair of 1-morphisms, $\rho_1: X \rightleftarrows Y :\iota_1$, and an {\it idempotent} 1-morphism, $e_1 \cong \iota_1 \circ \rho_1: X \to X$, such that
\begin{equation}
	e_1 \circ e_1 \cond e_1 \, , \qquad \rho_1 \circ \iota_1 \cond 1_Y\label{eq:condensation_monad_data}
\end{equation}
are themselves condensations respectively in the endomorphism $(n-1)$-categories $\text{End}(X)$ and $\text{End}(Y)$, where $1_Y$ is the identity 1-morphism of $Y$. The naming of $X\cond Y$ as a ``condensation'' is motivated physically, as $Y$ can be obtained from $X$ by proliferating defects described by $e$ on the worldvolume of the latter. In this way one ``condenses'' the defect $e$ on $X$ to obtain $Y$.

Unpacking the definition, this means that there exist two pairs of 2-morphisms, $\rho_2: e_1 \circ e_1 \rightleftarrows e_1 :\iota_2$ and $\rho_2': \rho_1 \circ \iota_1 \rightleftarrows 1_Y :\iota_2'$, as well as two idempotent 2-morphisms, $e_2 \cong \iota_2 \circ \rho_2: e_1 \circ e_1 \to e_1 \circ e_1$ and $e_2' \cong \iota_2' \circ \rho_2': \rho_1 \circ \iota_1 \to \rho_1 \circ \iota_1$, such that
\begin{equation}
	e_2 \circ e_2 \cond e_2 \, , \qquad \rho_2 \circ \iota_2 \cond 1_{e_1} \, , \qquad e_2' \circ e_2' \cond e_2' \, , \qquad \rho_2' \circ \iota_2' \cond 1_{1_Y}\label{eq:next_level_condensation_monad_data}
\end{equation}
are condensations respectively in the $(n-2)$-categories $\text{End}(e_1 \circ e_1)$, $\text{End}(e_1)$, $\text{End}(\rho_1 \circ \iota_1)$, and $\text{End}(1_Y)$.\footnote{As remarked before, the composition between 2-morphisms should be understood to be distinct from that between 1-morphisms. More precisely, in 2-categorical terms, here we are using the {\it vertical composition} of 2-morphisms which composes along 1-morphisms, rather than the {\it horizontal composition} which composes along objects.} One then iterates the procedure above all the way to the level of $n$-morphisms, whose endomorphisms form sets.

In the terminology of \cite{Gaiotto:2019xmp}, a condensation $X \cond Y$ between objects in an $n$-category is an {\it $n$-condensation}. A $0$-condensation is then simply an equality between $n$-morphisms, since there are no $(n+1)$-morphisms by construction. Although we will not use this terminology in the rest of the paper, we would like to point out that, a priori, claiming that a pair of morphisms, $\rho: X \rightleftarrows Y :\iota$, satisfies the definition of an $n$-condensation requires checking a total of roughly $\sum_{m=1}^n 2^{m-1}$ relations as outlined earlier, thus making it seem like a formidable task in general. Nonetheless, for cases of interest in our context, we typically only need to check a small subset of the relations, and the rest follows essentially for free.

\paragraph{Example 1 (isomorphisms):} To illustrate the idea, let us consider a trivial example of a categorical condensation, i.e.~an isomorphism. By definition, an isomorphism of an object $X$ in $\mathfrak{C}$ is a 1-morphism $f: X \to X$ admitting an inverse 1-morphism $f^{-1}: X \to X$, such that $f \circ f^{-1} = f^{-1} \circ f = 1_X$. Let us set $\rho_1 = f$, $\iota_1 = f^{-1}$, and $e_1 = 1_X$. They obviously satisfy
\begin{equation}\label{eq:identity_as_condensation}
    e_1 \circ e_1 = e_1 \cond e_1 \, , \qquad \rho_1 \circ \iota_1 = 1_X \cond 1_X \, ,
\end{equation}
where the corresponding $\rho_2, \iota_2, \rho_2', \iota_2'$ can all be chosen to be the identity 2-morphism $1_{1_X}$, such that they automatically satisfy \eqref{eq:next_level_condensation_monad_data}. Likewise, $\rho_3, \iota_3, \dots$ can all be chosen to be the identity 3-morphism $1_{1_{1_X}}$, so on and so forth. We thus conclude that $f: X \rightleftarrows X :f^{-1}$ is indeed a condensation, i.e.~$X$ condenses onto itself via any isomorphism.

\paragraph{Example 2 (direct sums):} A slightly more non-trivial example is as follows. The direct sum $X_1 \oplus X_2$ of two objects $X_1$ and $X_2$ in $\mathfrak{C}$ is defined by the existence of two projections 1-morphisms $\rho_i: X_1 \oplus X_2 \to X_i$ and two inclusion 1-morphisms $\iota_i: X_i \to X_1 \oplus X_2$ for $i=1,2$, such that
\begin{equation}
    \rho_i \circ \iota_j \cong \delta_{ij} 1_{X_i} \, , \qquad (\iota_1 \circ \rho_1) \oplus (\iota_2 \circ \rho_2) \cong 1_{X_1 \oplus X_2} \, .
\end{equation}
Let $e_i \coloneqq \iota_i \circ \rho_i$, then one can check that
\begin{equation}
    \begin{gathered}
        e_i \circ e_i = \iota_i \circ \rho_i \circ \iota_i \circ \rho_i \cong \iota_i \circ 1_{X_i} \circ \rho_i = \iota_i \circ \rho_i = e_i \cond e_i \, ,\\
        \rho_i \circ \iota_i \cong 1_{X_i} \cond 1_{X_i} \, .\label{eq:projection_as_condensation}
    \end{gathered}
\end{equation}
In other words, the projection $X_1 \oplus X_2 \cond X_i$ is a condensation for both $i=1,2$. One obviously recovers $X_i \oplus 0 \cond X_i$ as a special case.

\paragraph{Karoubi completeness:} We say that an $n$-category is {\it Karoubi-complete} (also known as {\it condensation-complete}) if all its idempotent $m$-morphisms $e_m$ split for all $1 \leq m \leq n$, i.e.~$e_m \cong \iota_m \circ \rho_m$ for some pair $(\rho_m,\iota_m)$, and they satisfy the recursive relations elucidated above.\footnote{More precisely, such a higher categorification of an idempotent is known as a {\it condensation monad} \cite{Gaiotto:2019xmp,Johnson-Freyd:2020usu}, but we will refer to it as an ``idempotent'' for simplicity.} Karoubi completeness is necessary (but not sufficient) for an $n$-category to be a fusion $n$-category \cite{Johnson-Freyd:2020usu}. Therefore, if we expect a physical theory to have its symmetries encoded by a fusion $n$-category, then it must be Karoubi-complete.  Importantly, the morphisms involved in each level are not necessarily unique. For a category to be Karoubi-complete we need only existence of $\iota_m$ and $\rho_m$ for each $m$, not uniqueness.

In Section \ref{sec:cond_complete}, we saw how the 3-category ${\rm End}(\cT)$, and furthermore the 4-category $\fC$, was Karoubi-complete from purely physical grounds. The condensation defects $\cC^{[4-m]}_G(M^m)$ are the idempotent $m$-morphisms in $\fC$, and the interfaces $\iota$ and $\rho$ that obeyed \eqref{eq:example_splitting_general} demonstrate how the condensation defects split as required. As the condensation defects are the only idempotent topological defects in $\fC$, we thus see that both $\fC$ and ${\rm End}(\cT)$ are Karoubi-complete. In the next section, we will generalize our example to more general $d$-dimensional QFTs, and analyze their Karoubi completeness from a categorical standpoint.

\subsection{Full Dualizability}\label{sec:full_dualizability}

So far we have discussed most of the necessary qualities for an $n$-category to be considered fusion, and have seen how they arise naturally in physical QFTs. There is one more property needed for a fusion $n$-category as per the proposed definition of \cite{Johnson-Freyd:2020usu}: it must be fully dualizable in an appropriate sense. As this requirement is rather technical, we leave the mathematical details of this statement to Appendix \ref{app:categorical_notions}, and focus here on the conceptual picture and manifestation in physical settings. The requirement that ${\rm End}(\cT)$ is fully dualizable can be understood through three intuitive properties. We will discuss each of these three in turn.

The first condition for full dualizability of ${\rm End}(\cT)$ is that every defect is {\it semisimple}, i.e.~expressible as a finite sum of {\it simple} defects.\footnote{A given $m$-dimensional defect $\cO(M^m)$ is {\it simple} or {\it indecomposable} if there is no way to write it as a sum $\cO(M^m) = (\cO_1\oplus\cO_2)(M^m)$ of other $m$-dimensional defects.} In particular this applies to products of defects under the OPE. In the QFTs we have discussed, one can confirm that this condition is satisfied, as all fusion is either group-like or idempotent.

The second condition is that there is a finite number of defects at each categorical level. To be more precise, we need only a finite number of simple defects so that direct sums of operators are still allowed. As we are concerned with QFTs whose symmetry groups are finite, this condition is automatically satisfied since we only have a finite number of symmetry generators. Additionally, we only have finitely many $m$-condensation defects for each $m$, as there are only a finite number of subgroups to be gauged for a given symmetry group.

The third condition is that each defect has a dual is similar but distinct from the statement that each defect has an inverse. Being more precise, if $f:X\to Y$ is an $m$-morphism in a monoidal $n$-category $\mathfrak{D}$, we say $f$ has a dual $f^\vee : Y\to X$ if there exist the following $(m+1)$-morphisms,\footnote{The data stated is here is, strictly speaking, the data for a {\it left} dual of $f$, and there are similar conditions to define a right dual. We will not take care to distinguish between left and right duals in the present paper, as in all the examples we are concerned with the left and right duals coincide. As such we refer to $f^\vee$ as ``the'' dual of $f$.} 
\begin{equation}\label{eq:dual_maps}
    d_f:f^\vee\otimes f\to 1^{[m]}_X\qquad b_f:1^{[m]}_Y \to f\otimes f^\vee\,,
\end{equation}
that satisfy certain coherence relations detailed in Appendix \ref{app:categorical_notions}. This should not be confused with saying $f^\vee$ is an inverse to $f$, as the maps in \eqref{eq:dual_maps} are not isomorphisms in general. When considering the fusion of a defect with its dual, what is required is that there is an interface to the trivial defect of appropriate dimension. Said differently, the fusion should admit topological boundary conditions. In the example of Sections \ref{sec:4d_example} and \ref{sec:karoubi_completeness}, this followed almost by definition of the defects. For example, the dual of $\cU_n(\Sigma^2)$ is given by $\cU_{N-n}(\Sigma^2)$, as can be read off from the fusion rule,
\begin{equation}
    (\cU_n\otimes\cU_{N-n})(\Sigma^2) = \cU_N(\Sigma^2) = 1^{[2]}(\Sigma^2)
    \,.
\end{equation}
By the same token, as we will demonstrate more generally in Section \ref{sec:gen_anomaly_free}, the existence of maps $\cC_G^{[m]}\otimes\cC_G^{[m]}\to \cC_G^{[m]}\to 1^{[m]}$ leads one to conclude that the $m$-condensation defects are their own dual.\footnote{As a side remark, the notion of duals applies equally well to non-invertible defects, e.g.~duality defects \cite{Choi:2021kmx,Kaidi:2021xfk} and chiral symmetry defects \cite{Choi:2022jqy,Cordova:2022ieu}, in higher dimensional QFTs. In these cases, the non-invertible defect $\cD$ has a fusion rule of the form $\cD \otimes \overline{\cD} \cong \cC^{[1]}$ with $\cC^{[1]}$ a 1-condensation defect, whereas $\overline{\cD}$ is typically an orientation reversal of $\cD$. Therefore, $\cD$ and $\overline{\cD}$ are dual to each other, {\it not} because the condensation defect $\cC^{[1]}$ ``contains'' a codimension-2 (or higher) trivial defect $1^{[2]}$, but because there exists a categorical condensation $\cC^{[1]} \cond 1^{[1]}$.} This, combined with the two properties discussed in the previous paragraphs, give the necessary conditions that ${\rm End}(\cT)$ is fully dualizable.

\subsection{Decoupled TQFT Fusion Coefficients}\label{sec:decoupled_TQFTs}

In Section \ref{sec:cond_complete}, we saw the presence of ``fusion coefficients'' described by decoupled TQFTs such as in \eqref{eq:example_idempotent}. We previously made sense of similar coefficients by saying they should count something, as in \eqref{eq:example_non-neg_coeffs}, and therefore are non-negative integers. However, the decoupled TQFTs do not a priori lend themselves to such an interpretation as integers.

Here we analyze in what precise sense the TQFTs are ``coefficients,'' discuss the values these coefficients take, and lay some groundwork to discuss idempotent defects in greater generality in the next section. Importantly, the fact that fusion coefficients are inherently subtler than just being integers is strictly a higher-categorical phenomenon. Before investigating this claim in general spacetime dimensions, let us first review some known results of 2-dimensional theories.

If $\cT$ is only 2-dimensional, then the only symmetry defects described in ${\rm End}(\cT)$ are topological lines and points. As in \eqref{eq:example_monoidal}, we may compute a fusion of two line operators, $\cO_a(M^1)$ and $\cO_b(M^1)$, in order to obtain a new line operator $(\cO_a\otimes\cO_b)(M^1)$. Suppose the product $(\cO_a\otimes\cO_b)(M^1)$ can itself be expressed as a linear combination of other line operators $\cO_c(M^1)$, we say that there is a {\it fusion rule} as follows,
\begin{equation}\label{eq:general_2d_fusion_rule}
    (\cO_a\otimes\cO_b)(M^1) \cong \bigoplus_i N_{a,b}^{c}\cO_c(M^1) \, ,
\end{equation}
where $N_{a,b}^{c}$ counts the number of times each line $\cO_c$ appears in the decomposition, exactly as in the discussion surrounding \eqref{eq:example_non-neg_coeffs}. More precisely, it counts for each $c$ the number of inequivalent pairs of projection and inclusion interfaces $\rho_{c,n}:\cO_a\otimes\cO_b \leftrightarrows \cO_c : \iota_{c,n}$, where $n = 1,\dots,N_{a,b}^{c}$.

Recall that, for a 2-dimensional theory, ${\rm End}(\cT)$ constitutes a 1-category. By definition, the collection of 1-morphisms, $\Hom(\cO_a\otimes\cO_b,\cO_c)$, is a {\it set} whose elements form a basis for the vector space of maps between $\cO_a\otimes\cO_b$ and $\cO_c$. The fusion coefficients
\begin{equation}
    N_{a,b}^{c} = {\rm dim}(\Hom(\cO_a\otimes\cO_b,\cO_c)) = {\rm dim}(\Hom(\cO_c,\cO_a\otimes\cO_b))
\end{equation}
measure precisely the dimension of (finite-dimensional) vector spaces, and so they are necessarily non-negative integers in the case of a 2-dimensional QFT \cite{2002math......3060E,etingof2015tensor, Moore:1988qv}.

If we now consider the case of a $d$-dimensional QFT, then $\cO_a(M^{d-1})$ and $\cO_b(M^{d-1})$ are objects in the $(d-1)$-category ${\rm End}(\cT)$. The fusion rule analogous to \eqref{eq:general_2d_fusion_rule} can then be written as
\begin{equation}
    (\cO_a\otimes\cO_b)(M^{d-1}) \cong \bigoplus_i \big|\fT_{a,b}^{c}(M^{d-1})\big|\,\cO_c(M^{d-1})\,,
\end{equation}
where now the fusion coefficients $\big|\fT_{a,b}^{c}\big|$ are to be associated with some notion of dimension for a $(d-2)$-category $\Hom(\cO_a\otimes\cO_b, \cO_c)$. This $(d-2)$-category,
\begin{equation}
    \fT_{a,b}^{c} \coloneqq \Hom(\cO_a\otimes\cO_b, \cO_c) \, , 
\end{equation}
encodes all at once the collection of $(d-2)$-dimensional topological interfaces between $\cO_a\otimes\cO_b$ and $\cO_c$ (comprising the objects of $\fT_{a,b}^c$), the $(d-3)$-dimensional topological interfaces between these $(d-2)$-dimensional ones (comprising the 1-morphisms of $\fT_{a,b}^c$), and so on all the way down to topological local operators (comprising the $(d-2)$-morphisms of $\fT_{a,b}^c$). In other words, it describes a $(d-1)$-dimensional topological field theory in its own right, decoupled from the degrees of freedom on the other ``bulk'' operators $\cO_a$.

In many physical examples, such as Dijkgraaf-Witten theories \cite{Dijkgraaf:1989pz}, the topological operators comprising $\fT^c_{a,b}(M^{d-1})$ form a $(d-2)$-groupoid: a $(d-2)$-category whose $m$-morphisms are invertible for all $1\leq m \leq d-2$. There exists a well-defined notion of its {\it homotopy cardinality} \cite{quinn1995lectures, 2000math......4133B, 2009arXiv0908.4305B}, given by
\begin{equation}\label{eq:homotopy_cardinality}
    \big|\fT_{a,b}^c\big| = \sum_{x\in\pi_0(\fT_{a,b}^c)}\prod_{k=1}^{d-2}\left|\pi_k (\fT_{a,b}^c,x)\right|^{(-1)^k} = \sum_{x\in\pi_0(\fT_{a,b}^c)} \frac{\big|\pi_2 (\fT_{a,b}^c,x)\big|\big|\pi_4 (\fT_{a,b}^c,x)\big|\cdots}{\big|\pi_1 (\fT_{a,b}^c,x)\big|\big|\pi_3 (\fT_{a,b}^c,x)\big|\cdots}\,,
\end{equation}
which is typically a (non-negative) rational number rather than an integer.

Let us consider two special cases of the formula above. When $d=2$, $\fT_{a,b}^c$ is a Hom-{\it set}, so $\pi_k$ is automatically trivial for all $k>0$. The formula \eqref{eq:homotopy_cardinality} then reduces to the usual sum over distinct elements of a set, i.e.~its cardinality. Similarly, when $\fT_{a,b}^c$ is described by an Eilenberg-MacLane space $K(G,m)$ for some finite group $G$,\footnote{Such a space is defined as a topological space such that its only non-trivial homotopy group is $\pi_m(K(G,m)) = G$ \cite{Eilenberg:1945abc, Eilenberg:1950abc, Eilenberg:1954abc}.} then \eqref{eq:homotopy_cardinality} yields that $|K(G,m)| = |G|^{(-1)^m}$. Note that when $d=1$, the homotopy cardinality also agrees with the definition in \cite{2018arXiv181211933D} for the dimension of a finite pre-semisimple 2-category.

At first sight, the idea that the fusion coefficient is a rational number, rather than an integer, is perhaps an unfamiliar one. However, it should not be surprising that as one goes beyond the realm of 1-categories, there are more ``substructures'' (i.e.~invertible higher morphisms) that provide redundancies and could modify the na\"ive counting of fusion channels. Such a concept is not new in physics, as theories containing higher-form gauge symmetries also have gauge transformations {\it of} gauge transformations that must properly be accounted for in the partition function. Roughly speaking, the formula \eqref{eq:homotopy_cardinality} is computing the ``effective size'' of the TQFT $\fT_{a,b}^c$ by taking into account redundancies coming from gauge transformations, gauge transformations of gauge transformations, etc. 

An explicit way to see this accounting of redundancies is that the coefficient $\big|\fT_{a,b}^{c}(M^{d-1})\big|$ evaluates the path integral for this decoupled $(d-1)$-dimensional theory on the worldvolume $M^{d-1}$. Indeed, one computes the partition function of a finite-group $p$-form gauge theory in the absence of any anomaly as
\begin{equation}\label{eq:finite-group_partition_function}
    Z[M^{d-1},G] = \frac{|H^p(M^{d-1};G)||H^{p-2}(M^{d-1};G)|\cdots}{|H^{p-1}(M^{d-1};G)|\cdots}\,,
\end{equation}
where each factor counts the allowed gauge transformation of specified form degree \cite{Horowitz:1989sol,Gaiotto:2014kfa, Freed:2022qnc}. By evaluating \eqref{eq:homotopy_cardinality} for $\fT_{a,b}^c(M^{d-1})$ given by the $(d-2)$-category of symmetry operators for a $p$-form symmetry on $M^{d-1}$, one can explicitly recover \eqref{eq:finite-group_partition_function} directly. This is briefly reviewed in Appendix \ref{app:homotopy_cardinality}, and discussed more thoroughly in \cite{Freed:2009qp, Freed:2022qnc}. The equivalence between these two expressions is somewhat non-trivial, and is important in the quantization of TQFTs.

\section{Anomaly-free Finite Abelian Symmetries}\label{sec:gen_anomaly_free}

In Section \ref{sec:4d_example}, we saw how the characteristics of a higher fusion category, reviewed in Section \ref{sec:categorical_structures}, arise naturally in a specific 4-dimensional QFT with a $\bZ_N^{(1)}$ symmetry. In this section we generalize our QFT $\cT$ to be $d$-dimensional, defined on a spacetime $W^d$, with a global symmetry group $G^{(p)}$. To this end, let us suppose that $G^{(p)}$ has a finite Abelian subgroup\footnote{By the fundamental theorem of finite Abelian groups we can write such any such $A^{(p)}$ as
\begin{equation}\label{eq:fund_thm_finite_ab_groups}
    A^{(p)} \cong \bigoplus_i \bZ_{p_i^{n_i}}\,,
\end{equation}
for some (not necessarily distinct) prime numbers $p_i$ and integers $n_i$. We generally consider the group $A^{(p)}$ in the abstract, but one can use the expression \eqref{eq:fund_thm_finite_ab_groups} for explicit computations as was done in Section \ref{sec:4d_example}.}
\begin{equation}
    A^{(p)}\subseteq G^{(p)}\,,
\end{equation}
such that $A^{(p)}$ has a trivial self-anomaly. The exact value of $p$ is irrelevant. We shall denote the corresponding symmetry generators as $\cU_a(\Sigma^{d-p-1})$ for $a\in A^{(p)}$ and $\Sigma^{d-p-1}\subset W^d$ some submanifold of spacetime. We further assume that $A^{(p)}$ does not participate in non-trivial (higher) coherence relations with the rest of $\fC$, so that it is not part of any higher group.

More generally, one could consider multiple anomaly-free subgroups $A^{(p)}, A'^{(p)},\dots \subseteq G^{(p)}$ that have mixed anomalies between them. In such a case the following discussion applies to each group individually, but it must be modified collectively. An example of such a case is 4d Yang-Mills theory with gauge group $SU(N)/\bZ_k$ for $k|N$. As long as $k\neq 1,N$, such a theory will have two 1-form symmetries, an ``electric'' $\bZ_{N/k}^{(1)}$ and a ``magnetic'' $\widehat{\bZ}_k^{(1)}$, that are both free of self-anomalies but have a mixed anomaly between them. We will return to such anomalies in Section \ref{sec:anomalous_symmetries}.

When considering the characteristics of a higher fusion category, we have discussed above how $(d-1)$-category ${\rm End}(\cT)$ immediately has the properties of monoidality, $\bC$-linearity, additivity, and full dualizability. In order to show that the $d$-category $\fC$ is also Karoubi-complete, we must demonstrate that all idempotent $m$-morphisms split, in the sense reviewed in Section \ref{sec:categorical_condensation}, for all $1\leq m \leq d$.

We claim that the non-trivial idempotent $m$-morphisms in our category $\fC$ are furnished by $m$-condensation defects $\cC^{[m]}_H(M^{d-m})$ for groups $H$ to be specified below. Starting with codimension-1, we first verify that the 1-condensation defects are idempotent in the appropriate sense before discussing their splitting into gauging interfaces. This will be a generalization of the example in Section \ref{sec:karoubi_completeness}, presented in the more categorical context of the previous section. The construction and physical interpretation of the associated data will be further expanded on.

\subsection{1-gauging and 1-condensation defects}

Each 1-condensation defect is constructed by the (untwisted) 1-gauging of a subgroup, $K^{(p)}\subseteq A^{(p)}$, defined explicitly by summing over insertions of the $K^{(p)}$ generators on a codimension-1 submanifold,
\begin{equation}\label{eq:general_1-condensation_defect}
    \cC^{[1]}_K(M^{d-1}) = \bigoplus_{\substack{\Sigma^{d-p-1}\in H_{d-p-1}(M^{d-1};K)\\s\in S}}\cU_s(\Sigma^{d-p-1}) \, .
\end{equation}
Since a generic (finite Abelian) group $K^{(p)}$ may have more than one generating element, we have to sum over them in $S$, a minimal generating set of $K^{(p)}$, so that the entirety of $K^{(p)}$ is included in the above sum. For example, if $K^{(p)} = \bZ_2\times\bZ_2$, then we could take $S =\{(1,0),(0,1)\}$ as the generating set. The resulting $\cC^{[1]}_K(M^{d-1})$ will have the fusion rule
\begin{equation}\label{eq:idempotent_fusion}
    (\cC^{[1]}_K\otimes\cC^{[1]}_K)(M^{d-1}) \cong \big|\fT_K(M^{d-1})\big|\,\cC^{[1]}_K(M^{d-1})\,, 
\end{equation}
where $\fT_K(M^{d-1})$ is some decoupled TQFT. This is a direct generalization of \eqref{eq:example_idempotent}, and can be understood similarly. When the two defects fuse, the diagonal combination of both factors yields a new copy of $\cC^{[1]}_K(M^{d-1})$, while the remaining degrees decouple and form the TQFT coefficient. 

Given the fusion rule \eqref{eq:idempotent_fusion}, we can immediately see that the 1-condensation defect $\cC_K^{[1]}(M^{d-1})$ acts as an idempotent 1-morphism in $\fC$. Recall from the discussion of Section \ref{sec:decoupled_TQFTs} that the coefficient $\big|\fT_K(M^{d-1})\big|$ is to be understood categorically as the ``effective size'' of the $(d-2)$-category $\Hom\big(\cC^{[1]}_K\otimes\cC_K^{[1]},\cC_K^{[1]}\big)$. As long as this coefficient is nonzero, we see that this category is non-empty, and so there is a 2-morphism in $\fC$ from $\cC^{[1]}_K\otimes\cC^{[1]}_K$ to $\cC_K^{[1]}$ acting as a higher-categorical ``projection'' map. Said differently, there is a codimension-2 interface between the fusion $\cC_K^{[1]}\otimes\cC_K^{[1]}$ and a single copy of $\cC_K^{[1]}$. Using the result of Example 2 in Section \ref{sec:categorical_condensation}, we thus find
\begin{equation}\label{eq:e_of_e}
    \cC_K^{[1]}\otimes\cC_K^{[1]} \cond \cC_K^{[1]}
\end{equation}
to be a categorical condensation. In particular, it follows that we do not have to explicitly check all the recursive relations needed to define a categorical condensation for every subgroup $K^{(p)}$, and they are indeed idempotent 1-morphisms in $\fC$.

Our goal is to show that the every idempotent 1-morphism $\cC_K^{[1]}$ itself takes part in a categorical condensation between some pair of objects in our $d$-category $\fC$. This can be done by studying the splitting property of the 1-condensation defects $\cC_K^{[1]}$. To this end, let us apply the cylinder gauging argument in Section \ref{sec:cond_complete}. More precisely, consider two 1-morphisms
\begin{equation}
    \rho_K: \cT \leftrightarrows \cT/K^{(p)}: \iota_K
\end{equation}
corresponding to gauging/ungauging interfaces for the subgroup $K^{(p)}$. As hinted at by the notation, $\rho_K$ acts to project out the subset of codimension-$(p+1)$ defects $\cU_k$ for all $k\in K^{(p)}$, whereas $\iota_K$ acts as to include the set $\overline{\cU}_{\overline{a}}$ for $\overline{a} \in A^{(p)}/K^{(p)}$ as a subset of the defects $\cU_a$.\footnote{In the notation of Section \ref{sec:karoubi_completeness}, the 1-morphisms $\rho_K$ and $\iota_K$ would be $1^{[0]}_{\cT,\cT/K^{(p)}}(M^{d-1})$ and $1^{[0]}_{\cT/K^{(p)},\cT}(M^{d-1})$ respectively. We use $\rho_K$ and $\iota_K$ here to keep the notation more compact, and highlight the similarities to projection/inclusion maps as well.}

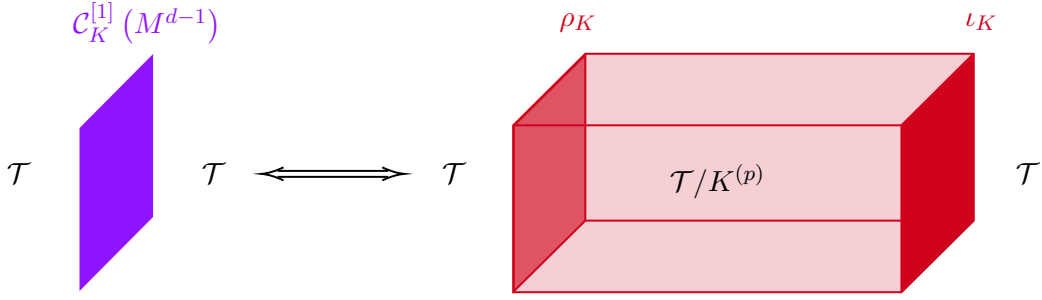
\begin{figure}
    \centering
\tikzset{every picture/.style={line width=0.75pt}} 
\begin{tikzpicture}[x=0.75pt,y=0.75pt,yscale=-1,xscale=1]

\draw  [color={rgb, 255:red, 208; green, 2; blue, 27 }  ,draw opacity=1 ][fill={rgb, 255:red, 208; green, 2; blue, 27 }  ,fill opacity=0.19 ] (383.34,88.22) -- (419.32,52.24) -- (613.1,52.24) -- (613.1,136.19) -- (577.12,172.17) -- (383.34,172.17) -- cycle ; \draw  [color={rgb, 255:red, 208; green, 2; blue, 27 }  ,draw opacity=1 ] (613.1,52.24) -- (577.12,88.22) -- (383.34,88.22) ; \draw  [color={rgb, 255:red, 208; green, 2; blue, 27 }  ,draw opacity=1 ] (577.12,88.22) -- (577.12,172.17) ;
\draw  [color={rgb, 255:red, 208; green, 2; blue, 27 }  ,draw opacity=1 ][fill={rgb, 255:red, 208; green, 2; blue, 27 }  ,fill opacity=1 ] (613.1,53.52) -- (613.1,136.19) -- (577.38,171.62) -- (577.38,88.95) -- cycle ;
\draw [color={rgb, 255:red, 208; green, 2; blue, 27 }  ,draw opacity=1 ]   (419.32,136.16) -- (611.83,136.16) ;
\draw  [color={rgb, 255:red, 208; green, 2; blue, 27 }  ,draw opacity=1 ][fill={rgb, 255:red, 208; green, 2; blue, 27 }  ,fill opacity=0.6 ] (419.32,52.87) -- (419.32,136.16) -- (383.6,171.75) -- (383.6,88.46) -- cycle ;
\draw  [color={rgb, 255:red, 144; green, 19; blue, 254 }  ,draw opacity=1 ][fill={rgb, 255:red, 144; green, 19; blue, 254 }  ,fill opacity=1 ] (203.06,53.57) -- (203.06,134.12) -- (167.34,170.34) -- (167.34,89.8) -- cycle ;
\draw    (265.74,111.15) -- (320.9,111.15)(265.74,114.15) -- (320.9,114.15) ;
\draw [shift={(328.9,112.65)}, rotate = 180] [color={rgb, 255:red, 0; green, 0; blue, 0 }  ][line width=0.75]    (10.93,-3.29) .. controls (6.95,-1.4) and (3.31,-0.3) .. (0,0) .. controls (3.31,0.3) and (6.95,1.4) .. (10.93,3.29)   ;
\draw [shift={(257.74,112.65)}, rotate = 0] [color={rgb, 255:red, 0; green, 0; blue, 0 }  ][line width=0.75]    (10.93,-3.29) .. controls (6.95,-1.4) and (3.31,-0.3) .. (0,0) .. controls (3.31,0.3) and (6.95,1.4) .. (10.93,3.29)   ;

\draw (129.03,105.92) node [anchor=north west][inner sep=0.75pt]    {$\mathcal{T}$};
\draw (226.38,105.92) node [anchor=north west][inner sep=0.75pt]    {$\mathcal{T}$};
\draw (162.19,24.69) node [anchor=north west][inner sep=0.75pt]  [font=\normalsize,color={rgb, 255:red, 144; green, 19; blue, 254 }  ,opacity=1 ]  {$\mathcal{C}_{K}^{[ 1]}\left( M^{d-1}\right)$};
\draw (346.3,105.91) node [anchor=north west][inner sep=0.75pt]    {$\mathcal{T}$};
\draw (632.75,105.91) node [anchor=north west][inner sep=0.75pt]    {$\mathcal{T}$};
\draw (460.08,106.31) node [anchor=north west][inner sep=0.75pt]    {$\mathcal{T} /K^{( p)}$};
\draw (404.9,30.48) node [anchor=north west][inner sep=0.75pt]  [color={rgb, 255:red, 208; green, 2; blue, 27 }  ,opacity=1 ]  {$\rho _{K}$};
\draw (607.37,30.48) node [anchor=north west][inner sep=0.75pt]  [color={rgb, 255:red, 208; green, 2; blue, 27 }  ,opacity=1 ]  {$\iota _{K}$};
\end{tikzpicture}
    \caption{The splitting of the 1-condensation defect $\cC_K^{[1]}(M^{d-1})$. On the left, the defect acts as a codimension-1 interface from the theory $\cT$ to itself. On the right, we split $\cC_K^{[1]}$ into $\rho_K$ and $\iota_K$, which respectively act as interfaces from $\cT$ to $\cT/K^{(p)}$ and vice versa.}
    \label{fig:cylinder_spacetime_gauging}
\end{figure}

From the cylinder gauging, we have the following fusion,
\begin{equation}\label{eq:iota_of_rho}
    \iota_K \otimes \rho_K \cong \cC^{[1]}_K: \cT\to\cT \, .
\end{equation}
On the contrary, the fusion $\rho_K\otimes\iota_K:\cT/K^{(p)}\to\cT/K^{(p)}$ acts as an identity interface in the gauged theory, giving the following condensation,
\begin{equation}\label{eq:rho_of_iota}
    \rho_K \otimes\iota_K \cond 1_{\cT/K^{(p)}}\,,
\end{equation}
using the fact that identity morphisms are trivial condensations as discussed in Example 1 of Section \ref{sec:categorical_condensation}. The factorization $\cC_K^{[1]} = \rho_K\otimes\iota_K$ is depicted in Figure \ref{fig:cylinder_spacetime_gauging}.

The two maps \eqref{eq:iota_of_rho} and \eqref{eq:rho_of_iota} provide us with the splitting of $\cC_K^{[1]}$ that we were looking for. Combined with the confirmation in \eqref{eq:e_of_e} that the 1-condensation defects are idempotent, we have thus found condensations, 
\begin{equation}\label{eq:T_to_T/K}
    \cT \cond \cT/K^{(p)}\,,
\end{equation}
via the idempotent 1-morphism $\cC_K^{[1]}$. Physically, the statement of this categorical condensation is that the theory $\cT/K^{(p)}$ can be obtained from $\cT$ by proliferating a network of $\cC^{[1]}_K$ defects on $W^d$. Recall, however, that $\cC_K^{[1]}$ is already described by a network of $K^{(p)}$ generators, so \eqref{eq:T_to_T/K} is the familiar statement that summing over all insertions of $K^{(p)}$ symmetry defects on $W^d$ is the same as gauging the $K^{(p)}$ symmetry in all of spacetime.

\subsection{Condensations within condensations}

We have seen that the 1-condensation defects $\cC^{[1]}_K$ describe idempotent 1-morphisms in $\fC$, and subsequently split into gauging/ungauging interfaces. However, as we saw in the 4d example in previous sections there is still a rich hierarchy of $(n>1)$-condensation defects living on the $\cC_K^{[1]}(M^{d-1})$ worldvolume theories. We now turn to these defects and see how they fit into the Karoubi completeness of $\fC$.

First, let us recall that when regarded as a $(d-1)$-dimensional theory, $\cC_K^{[1]}(M^{d-1})$ generally has two global symmetries.
\begin{itemize} 
    \item A $(d-p-2)$-form symmetry $\widehat{K}^{(d-p-2)}$ arising as the Pontryagin dual symmetry of the now-gauged $K^{(p)}$ symmetry. Since $K^{(p)}$ is a finite Abelian group and hence can be decomposed as \eqref{eq:fund_thm_finite_ab_groups}, we have 
    \begin{equation}
        \widehat{K}^{(d-p-2)} \cong K^{(p)}
    \end{equation}
    as groups. The defects generating this symmetry will be denoted as $\widehat{\cU}_{\hat{k}}(\Sigma^p)$, and correspond to $(d-p)$-morphisms in our $d$-category $\fC$
    \item For $p\geq 1$, there is a residual $(p-1)$-form symmetry $\overline{A}^{(p-1)} \coloneqq (A/K)^{(p-1)}$, where the degree shift is due to working in one lower dimension. The defects generating this residual theory will be denoted as $\overline{\cU}_{\overline{a}}(\Sigma^{d-p-1})$, and correspond to $(p+1)$-morphisms in $\fC$.
\end{itemize}

As an aside, there is a mixed anomaly between the two global symmetries $\widehat{K}^{(d-p-2)}$ and $\overline{A}^{(p-1)}$ whenever the following group extension does not split,
\begin{equation}
    0\to K^{(p-1)}\to A^{(p-1)}\to\overline{A}^{(p-1)}\to 0 \, .
\end{equation}
This condition is equivalent to the statement that the extension class, a 2-cocycle valued in group cohomology $H^2(\overline{A},K)\cong {\rm Ext}(\overline{A},K)$, is non-trivial. In such a case, moving a $\widehat{K}^{(d-p-2)}$ symmetry generator across a trivalent junction of $\overline{A}^{(p-1)}$ generators creates a phase determined by the extension class \cite{Tachikawa:2017gyf}. Categorically, this is captured by extra braiding data in $\fC$, as the two operators are seen to be charged under one another.

Like in Section \ref{sec:cond_complete}, we can construct higher condensation defects by 1-gauging subgroups of these global symmetries.
\begin{itemize}
    \item If $p\leq d-2$, we can 1-gauge a subgroup of the quantum symmetry $L^{(d-p-2)}\subseteq \widehat{K}^{(d-p-2)}$ on a codimension-1 submanifold $M^{d-2}\subset M^{d-1}$ of $\cC^{[1]}_K$. This would give a 2-condensation defect as follows,
    \begin{equation}\label{eq:general_2-condensation_defect}
        \cC^{[2]}_L(M^{d-2}) \coloneqq \bigoplus_{\substack{\Sigma^p\in H_p(M^{d-2};L) \\ s\in S}} \widehat{\cU}_s(\Sigma^p) \, .
    \end{equation}
    \item If $p\geq 1$, we may also 1-gauge a subgroup of the residual symmetry $B^{(p-1)}\subseteq\overline{A}^{(p-1)}$ in the same way to find a 2-condensation defect $\cC_B^{[2]}(M^{d-2})$.
\end{itemize}
Using arguments completely analogous to those in the previous subsection, we find that both of these 2-condensation defects are idempotent, i.e.
\begin{equation}
    \cC^{[2]}_L \otimes \cC^{[2]}_L \cond \cC_L^{[2]}\,,\qquad \cC_B^{[2]}\otimes \cC^{[2]}_B \cond \cC_B^{[2]}
\end{equation}
are both condensations in the endomorphism $(d-2)$-category ${\rm End}(\cC_K^{[1]})$. Physically this is again the statement that the fusion reproduces the same 2-condensation defect up to some decoupled TQFT.

Furthermore, both 2-condensation defects admit splittings into codimension-1 gauging/ungauging interfaces on $\cC_K^{[1]}$. More explicitly, there exist 2-morphisms $\rho_L: \cC_K^{[1]} \leftrightarrows \cC_{K/L}^{[1]}:\iota_L$ such that $\cC^{[2]}_L \cong \iota_L\otimes\rho_L$, and
\begin{equation}
    \rho_L\otimes\iota_L \cond 1_{\cC_{K/L}^{[1]}}
\end{equation}
is a condensation in the endomorphism $(d-2)$-category ${\rm End}(\cC_{K/L}^{[1]})$. Similarly to the 1-condensation defect, the maps $\rho_L$ and $\iota_L$ admit interpretations as projection and inclusion maps respectively. The 2-morphism $\rho_L$ projects out $\widehat{\cU}_{\ell}$ defects for $\ell\in L^{(d-p-2)}$, while $\iota_L$ acts to include the $\overline{\cU}_{\overline{k}}$ defects for $\overline{k} \in (K/L)^{(d-p-1)}$, regarded as a subset of the defects $\overline{\cU}_{\overline{a}}$. By the earlier arguments, this tells us that 
\begin{equation}
    \cC^{[1]}_K\cond \cC_{K/L}^{[1]}
\end{equation}
is a categorical condensation via the idempotent 2-morphism $\cC^{[2]}_L$. Note that in the limiting case where $L^{(d-p-2)} = \widehat{K}^{(d-p-2)}$, we have a condensation $\cC^{[1]}_K \cond \cC_{\{*\}}^{[1]} = 1_{\cT}$. This should be interpreted physically as the statement that condensation defects can end \cite{Copetti:2023mcq}.

Likewise, we can see that the 2-condensation defect $\cC^{[2]}_B$ splits via 2-morphisms $\rho_B: \cC_K^{[1]}\leftrightarrows\cC^{[1]}_{\check{B}} : \iota_B$ such that $\cC_B^{[2]}\cong \iota_B\otimes\rho_B$, and
\begin{equation}
    \rho_B\otimes\iota_B \cond 1_{\cC_{\check{B}}^{[1]}}
\end{equation}
is a condensation in the endomorphism $(d-2)$-category ${\rm End}(\cC_{\check{B}}^{[1]})$. Here, the group $\check{B}^{(p-1)} \subset A^{(p-1)}$ is defined via the group extension
\begin{equation}\label{eq:group_extension}
    0\to K^{(p-1)}\to \check{B}^{(p-1)} \to B^{(p-1)}\to 0\,.
\end{equation}
Our choice of defining $\check{B}$ through a group extension, rather than denoting the 1-condensation by $\cC_{\overline{A}/B}^{[1]}$, is intentional so that we may stay within the same duality frame when discussing all the 1-condensation defects; this is merely a notational choice. Generally, we will discuss the gauging of (subgroups of) the Pontryagin dual symmetry on a given defect through quotients, and gaugings of (subgroups of) the residual symmetry on a given defect through group extensions. The 2-morphism $\rho_B$ projects out the $\cU_b$ defects for $b\in B^{(p-1)}$, while the 2-morphism $\iota_B$ acts to include the defects $\cU_{\tilde{a}}$ for $\tilde{a}\in (A/\check{B})^{(p-1)}$ as a subset of the $\overline{\cU}_{\overline{a}}$.\footnote{Here we have used the third isomorphism theorem to rewrite $\overline{A}/B$ as $A/\check{B}$, utilizing that $K\subseteq \check{B}\subseteq A$. Additionally, since $A$ is Abelian, all subgroups above are normal subgroups.}

Let us stress that for both $\cC_L^{[2]}$ and $\cC_B^{[2]}$, the splittings are generally not unique. In the former condensation $\cC_K^{[1]}\cond\cC_{K/L}^{[1]}$, one can possibly compute the quotient $(\widehat{K}/L)^{(d-p-2)}$ in multiple inequivalent ways. For example, in the case $\widehat{K}^{(d-p-2)} = \bZ_4\times\bZ_2$ and $L^{(d-p-2)} = \bZ_2$, the quotient $(\widehat{K}/L)^{(d-p-2)}$ could be $\bZ_2\times\bZ_2$ or $\bZ_4$. In the latter condensation $\cC_K^{[1]}\cond \cC_{\check{B}}^{[1]}$, there can similarly exist multiple inequivalent choices for the group extension $\check{B}^{(p-1)}$. Such a choice is specified by a 2-cocycle in $H^2(B,K) \cong {\rm Ext}(B,K)$, which is trivial if and only if the extension \eqref{eq:group_extension} splits, such that $\check{B}^{(p-1)} \cong K^{(p-1)}\times B^{(p-1)}$.

One can iterate the previous construction at the next level to build 3-condensation defects $\cC_H^{[3]} \cong \iota_H\otimes \rho_H$, for suitable subgroups $H$, as endomorphisms of the 2-condensation defects described above. As expected, these will be idempotent and split in completely analogous ways. In fact, one can iterate all the way up to $d$-condensation defects giving condensations $\cC_H^{[d-1]} \cond \cC_{H'}^{[d-1]}$ for some $H'\subseteq H$. The procedure then truncates as a $d$-condensation defect can only condense (in a categorical manner) onto itself.

An important property of the discussed categorical condensations is that they can be composed (see Appendix \ref{app:categorical_notions}). This corresponds physically to the ability for one to sequentially gauge nested subgroups of a symmetry, which can be expressed as a sequence of categorical condensations,
\begin{equation}
    \cdots \cond \cC_{\check{B}}^{[1]}\cond \cC_K^{[1]}\cond \cC^{[1]}_{K/L}\cond \cdots\,, 
\end{equation}
for some $L\subseteq \widehat{K} \cong K \subseteq \check{B}$. Using a dual set of 2-condensation defects, we also get the reverse order,
\begin{equation}
    \cdots \cond \cC_{K/L}^{[1]}\cond \cC_K^{[1]}\cond \cC^{[1]}_{\check{B}}\cond \cdots \, .
\end{equation}
More generally, one can always construct non-trivial closed orbits that run through $\cC_H^{[1]}$ for all possible subgroups $H$, including the trivial subgroup $H = \{*\}$.\footnote{If we allow for twisted gauging, then there can exist multiple disconnected components where two given condensation defects are not related by any sequence of categorical condensations.} This applies equally to condensation defects at all codimensions, or equivalently, all categorical levels of $\fC$.

\subsection{$q$-gauging and $q$-condensation defects}

So far, we have discussed how the physical notion of 1-gauging produces 1-condensation defect $\cC_K^{[1]}$, which in turn gives rise to categorical condensations $\cT \cond \cT/K^{(p)}$ in the $d$-category $\fC$. On top of that, we have illustrated how this notion is not restricted only to the level of objects in $\fC$, since the procedure can be carried out iteratively. Namely, by performing a 1-gauging with respect to the condensation defect worldvolume obtained by the previous step, one finds 2-condensation defects, then 3-condensation defects, etc. In what follows, we show that a special case of our construction reduces to more general $q$-gauging for $1\leq q\leq p+1$. Hence, our construction provides a universal framework to encode all possible $q$-gaugings that one can perform within a given family of theories (and the contained subtheories) in the language of categorical condensations.

As is now the theme, it suffices to demonstrate explicitly the case of $q=2$, then the rest follows inductively. To obtain a 2-condensation defect from a direct 2-gauging of a subgroup $K^{(p)}\subseteq A^{(p)}$, we sum over insertions of generators on a codimension-2 submanifold $M^{d-2}\subset W^d$, i.e.
\begin{equation}
    \cC_K^{[2]}(M^{d-2}) \coloneqq \bigoplus_{\substack{\Sigma^{d-p-1}\in H_{d-p-1}(M^{d-2};K) \\ s\in S}} \cU_s(\Sigma^{d-p-1}) \, .
\end{equation}
Following previous arguments, it is straightforward to show that $\cC_K^{[2]}\otimes\cC_K^{[2]}\cond \cC_K^{[2]}$ is a condensation, such that $\cC_K^{[2]}$ is an idempotent 2-morphism in $\fC$. Equivalently, it can be seen as a 1-morphism in the $(d-1)$-category ${\rm End}(1_\cT^{[1]})$. It remains to be seen how this defect splits.

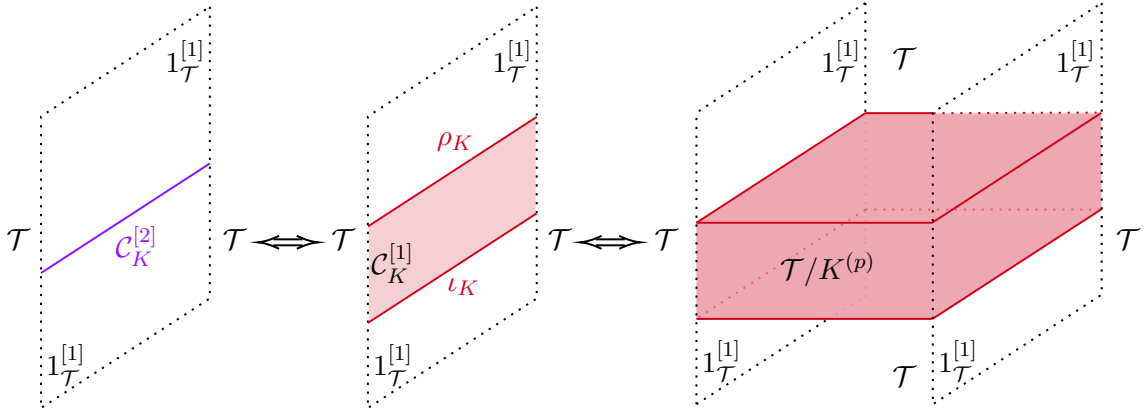
\begin{figure}
    \centering
\tikzset{every picture/.style={line width=0.75pt}} 
\begin{tikzpicture}[x=0.75pt,y=0.75pt,yscale=-1,xscale=0.96]
\draw  [dash pattern={on 0.84pt off 2.51pt}]  (363.09,213.85) -- (431.69,170.86) ;
\draw  [color={rgb, 255:red, 255; green, 255; blue, 255 }  ,draw opacity=1 ][fill={rgb, 255:red, 208; green, 2; blue, 27 }  ,fill opacity=0.19 ] (573.8,67.4) -- (450.9,67.4) -- (363.09,122.53) -- (363.09,122.53) -- (486,122.53) -- cycle ;
\draw  [draw opacity=0][fill={rgb, 255:red, 208; green, 2; blue, 27 }  ,fill opacity=0.19 ] (573.46,67.13) -- (573.8,115.9) -- (450.9,115.9) -- (450.9,115.9) -- (450.9,67.4) -- cycle ;
\draw  [draw opacity=0][fill={rgb, 255:red, 208; green, 2; blue, 27 }  ,fill opacity=0.19 ] (450.9,67.4) -- (450.9,115.9) -- (363.09,171.02) -- (363.09,171.02) -- (363.09,122.53) -- cycle ;
\draw  [draw opacity=0][fill={rgb, 255:red, 208; green, 2; blue, 27 }  ,fill opacity=0.19 ] (450.9,115.9) -- (573.8,115.9) -- (486,171.02) -- (486,171.02) -- (363.09,171.02) -- cycle ;
\draw  [draw opacity=0][fill={rgb, 255:red, 208; green, 2; blue, 27 }  ,fill opacity=0.19 ] (485.82,122.32) -- (485.82,170.82) -- (363.09,170.96) -- (363.09,170.96) -- (363.09,122.46) -- cycle ;
\draw  [color={rgb, 255:red, 255; green, 255; blue, 255 }  ,draw opacity=1 ][fill={rgb, 255:red, 208; green, 2; blue, 27 }  ,fill opacity=0.19 ] (280.05,69.28) -- (280.05,117.78) -- (192.25,172.91) -- (192.25,172.91) -- (192.25,124.41) -- cycle ;
\draw  [dash pattern={on 0.84pt off 2.51pt}] (110.08,13.88) -- (110.08,160.58) -- (22.27,215.73) -- (22.27,69.04) -- cycle ;
\draw [color={rgb, 255:red, 144; green, 19; blue, 254 }  ,draw opacity=1 ]   (22.27,147.82) -- (110.08,92.69) ;
\draw  [dash pattern={on 0.84pt off 2.51pt}] (280.05,13.88) -- (280.05,160.58) -- (192.25,215.73) -- (192.25,69.04) -- cycle ;
\draw [color={rgb, 255:red, 208; green, 2; blue, 27 }  ,draw opacity=1 ]   (192.25,124.41) -- (280.05,69.28) ;
\draw [color={rgb, 255:red, 208; green, 2; blue, 27 }  ,draw opacity=1 ]   (192.25,172.91) -- (280.05,117.78) ;
\draw    (143.15,130.88) -- (158.21,130.88)(143.15,133.88) -- (158.21,133.88) ;
\draw [shift={(166.21,132.38)}, rotate = 180] [color={rgb, 255:red, 0; green, 0; blue, 0 }  ][line width=0.75]    (10.93,-3.29) .. controls (6.95,-1.4) and (3.31,-0.3) .. (0,0) .. controls (3.31,0.3) and (6.95,1.4) .. (10.93,3.29)   ;
\draw [shift={(135.15,132.38)}, rotate = 0] [color={rgb, 255:red, 0; green, 0; blue, 0 }  ][line width=0.75]    (10.93,-3.29) .. controls (6.95,-1.4) and (3.31,-0.3) .. (0,0) .. controls (3.31,0.3) and (6.95,1.4) .. (10.93,3.29)   ;
\draw [color={rgb, 255:red, 208; green, 2; blue, 27 }  ,draw opacity=1 ]   (363.09,122.53) -- (450.9,67.4) ;
\draw  [draw opacity=0][fill={rgb, 255:red, 208; green, 2; blue, 27 }  ,fill opacity=0.19 ] (573.63,67.26) -- (573.63,115.76) -- (485.82,170.89) -- (485.82,170.89) -- (485.82,122.39) -- cycle ;
\draw [color={rgb, 255:red, 208; green, 2; blue, 27 }  ,draw opacity=1 ]   (485.82,122.39) -- (573.63,67.26) ;
\draw [color={rgb, 255:red, 208; green, 2; blue, 27 }  ,draw opacity=1 ]   (485.82,170.89) -- (573.63,115.76) ;
\draw [color={rgb, 255:red, 208; green, 2; blue, 27 }  ,draw opacity=1 ]   (363.09,122.53) -- (486,122.53) ;
\draw [color={rgb, 255:red, 208; green, 2; blue, 27 }  ,draw opacity=1 ]   (363.09,170.89) -- (485.82,170.89) ;
\draw [color={rgb, 255:red, 208; green, 2; blue, 27 }  ,draw opacity=1 ]   (450.9,67.4) -- (486.29,67.4) ;
\draw [color={rgb, 255:red, 208; green, 2; blue, 27 }  ,draw opacity=1 ] [dash pattern={on 0.84pt off 2.51pt}]  (486.29,67.4) -- (573.46,67.13) ;
\draw [color={rgb, 255:red, 208; green, 2; blue, 27 }  ,draw opacity=0.42 ] [dash pattern={on 0.84pt off 2.51pt}]  (450.9,115.9) -- (363.09,170.89) ;
\draw [color={rgb, 255:red, 208; green, 2; blue, 27 }  ,draw opacity=0.42 ] [dash pattern={on 0.84pt off 2.51pt}]  (450.9,115.9) -- (573.8,115.9) ;
\draw    (310.8,130.88) -- (325.86,130.88)(310.8,133.88) -- (325.86,133.88) ;
\draw [shift={(333.86,132.38)}, rotate = 180] [color={rgb, 255:red, 0; green, 0; blue, 0 }  ][line width=0.75]    (10.93,-3.29) .. controls (6.95,-1.4) and (3.31,-0.3) .. (0,0) .. controls (3.31,0.3) and (6.95,1.4) .. (10.93,3.29)   ;
\draw [shift={(302.8,132.38)}, rotate = 0] [color={rgb, 255:red, 0; green, 0; blue, 0 }  ][line width=0.75]    (10.93,-3.29) .. controls (6.95,-1.4) and (3.31,-0.3) .. (0,0) .. controls (3.31,0.3) and (6.95,1.4) .. (10.93,3.29)   ;
\draw  [dash pattern={on 0.84pt off 2.51pt}]  (363.09,213.85) -- (363.09,67.13) ;
\draw  [dash pattern={on 0.84pt off 2.51pt}]  (450.9,12) -- (363.09,67.13) ;
\draw  [dash pattern={on 0.84pt off 2.51pt}]  (450.9,12) -- (450.9,67.4) ;
\draw  [dash pattern={on 0.84pt off 2.51pt}] (573.8,12) -- (573.8,158.69) -- (486,213.85) -- (486,67.15) -- cycle ;
\draw [color={rgb, 255:red, 0; green, 0; blue, 0 }  ,draw opacity=0.08 ] [dash pattern={on 0.84pt off 2.51pt}]  (450.9,160) -- (450.9,131.71) ;
\draw [color={rgb, 255:red, 0; green, 0; blue, 0 }  ,draw opacity=0.08 ] [dash pattern={on 0.84pt off 2.51pt}]  (450.9,67.4) -- (450.9,131.71) ;
\draw [color={rgb, 255:red, 0; green, 0; blue, 0 }  ,draw opacity=0.08 ] [dash pattern={on 0.84pt off 2.51pt}]  (431.69,170.86) -- (450.9,160) ;

\draw (2.74,124.18) node [anchor=north west][inner sep=0.75pt]    {$\mathcal{T}$};
\draw (115.19,124.18) node [anchor=north west][inner sep=0.75pt]    {$\mathcal{T}$};
\draw (171.85,124.18) node [anchor=north west][inner sep=0.75pt]    {$\mathcal{T}$};
\draw (284.29,124.4) node [anchor=north west][inner sep=0.75pt]    {$\mathcal{T}$};
\draw (85.52,29.4) node [anchor=north west][inner sep=0.75pt]    {$1_{\mathcal{T}}^{[ 1]}$};
\draw (23.33,182.14) node [anchor=north west][inner sep=0.75pt]    {$1_{\mathcal{T}}^{[ 1]}$};
\draw (255.52,29.4) node [anchor=north west][inner sep=0.75pt]    {$1_{\mathcal{T}}^{[ 1]}$};
\draw (193.43,182.14) node [anchor=north west][inner sep=0.75pt]    {$1_{\mathcal{T}}^{[ 1]}$};
\draw (59.59,122.7) node [anchor=north west][inner sep=0.75pt]  [color={rgb, 255:red, 144; green, 19; blue, 254 }  ,opacity=1 ]  {$\mathcal{C}_{K}^{[ 2]}$};
\draw (192.63,131.24) node [anchor=north west][inner sep=0.75pt]    {$\mathcal{C}_{K}^{[ 1]}$};
\draw (226.55,75.65) node [anchor=north west][inner sep=0.75pt]  [color={rgb, 255:red, 208; green, 2; blue, 27 }  ,opacity=1 ]  {$\rho _{K}$};
\draw (231.83,148.88) node [anchor=north west][inner sep=0.75pt]  [color={rgb, 255:red, 208; green, 2; blue, 27 }  ,opacity=1 ]  {$\iota _{K}$};
\draw (426.43,29.4) node [anchor=north west][inner sep=0.75pt]    {$1_{\mathcal{T}}^{[ 1]}$};
\draw (364.27,180.26) node [anchor=north west][inner sep=0.75pt]    {$1_{\mathcal{T}}^{[ 1]}$};
\draw (548.46,29.4) node [anchor=north west][inner sep=0.75pt]    {$1_{\mathcal{T}}^{[ 1]}$};
\draw (487.18,180.26) node [anchor=north west][inner sep=0.75pt]    {$1_{\mathcal{T}}^{[ 1]}$};
\draw (339.79,124.4) node [anchor=north west][inner sep=0.75pt]    {$\mathcal{T}$};
\draw (579.79,124.4) node [anchor=north west][inner sep=0.75pt]    {$\mathcal{T}$};
\draw (405.13,135.83) node [anchor=north west][inner sep=0.75pt]    {$\mathcal{T} /K^{( p)}$};
\draw (463.5,33.4) node [anchor=north west][inner sep=0.75pt]    {$\mathcal{T}$};
\draw (463.5,193.4) node [anchor=north west][inner sep=0.75pt]    {$\mathcal{T}$};
\end{tikzpicture}
    \caption{The splitting of the 2-condensation defect $\cC_K^{[2]}$ when viewed as a standalone codimension-2 defect. On the left, $\cC_K^{[2]}$ can be seen as either a codimension-2 interface from $\cT$ to $\cT$ or as a codimension-1 interface from $1_\cT^{[1]}$ to $1_\cT^{[1]}$. In the middle, we use the latter interpretation to split $\cC_K^{[2]}$ into interfaces $\rho_K$ and $\iota_K$ mediating from $1_\cT^{[1]}$ to the 1-condensation defect $\cC_K^{[1]}$ and back. On the right, we further split both the $1_\cT^{[1]}$ and $\cC_K^{[1]}$ defects into their respective components.}
    \label{fig:2-condensation_splitting}
\end{figure}

We know that, when regarded as a $(d-1)$-dimensional theory, the transparent defect $1_\cT^{[1]} = \cC_{\{*\}}^{[1]}$ has on its worldvolume a residual $\overline{A}^{(p-1)} \coloneqq A^{(p)}/\{*\} \cong A^{(p)}$ symmetry. Therefore, the 2-condensation defect $\cC_K^{[2]}$ should be viewed as some 1-gauging of the residual symmetry in $1_\cT^{[1]}$. According to our prescription, the 2-condensation defect $\cC_K^{[2]}$ should be mediating a condensation $\cC_{\{*\}}^{[1]} \cond \cC_{\check{K}}^{[1]}$ for some group $\check{K}^{(p-1)}$, defined by the extension\footnote{Here we are distinguishing between $\{*\}$ and 0 in notation in order to emphasize the role of the group extension.}
\begin{equation}
    0\to \{*\}^{(p-1)}\to \check{K}^{(p-1)}\to K^{(p-1)}\to 0\,,
\end{equation}
i.e.~$\check{K}^{(p-1)}\cong K^{(p-1)}$. We see that our 2-condensation defect splits into codimension-2 gauging/ungauging interfaces between the trivial 1-condensation defect $1_\cT^{[1]}$ and the non-trivial 1-condensation defect $\cC_K^{[1]}$. This configuration is depicted in Figure \ref{fig:2-condensation_splitting}. 

The argument above shows that a 2-gauging induces a categorical condensation not at the level of objects, but at the level of 1-morphisms in $\fC$. In general, a $q$-gauging induces a categorical condensation at the level of $(q-1)$-morphisms in $\fC$ described by $\cC_H^{[q-1]}$. Hence, we observe that the physical notion of $q$-gauging for any $1 \leq q \leq p+1$, as introduced by \cite{Roumpedakis:2022aik}, fits naturally in the structure of a Karoubi-complete higher category, and is moreover a special case of a much richer zoo of configurations. In this sense, if we replace the transparent defect $1_\cT^{[1]} = \cC_{\{*\}}^{[1]}$ in the previous example by non-trivial condensation defects, and repeat the same exercise of splitting the various idempotents involved, then we can regard the resultant procedure as a generalized notion of higher-gauging.

\section{Anomalous Finite Abelian Symmetries}\label{sec:anomalous_symmetries}

Previously, we have always specified that all the symmetries we were dealing with were strictly anomaly-free. Nonetheless, we regularly saw hints of how the presence of anomalies could modify the discussion and introduce subtleties. In the current section, we will address additional details arising when anomalies are present. We begin with a general discussion of (higher) anomalies and their relation to the categorical structure of $\fC$, and follow with three examples relating to its Karoubi completeness.

Consider again a general $d$-dimensional Euclidean QFT $\cT$ with a global $p$-form symmetry $G^{(p)}$, containing some finite Abelian subgroup $A^{(p)} \subseteq G^{(p)}$. Now, we will invert our previous assumption and take the entirety of $A^{(p)}$ to be self-anomalous. This is not restrictive; if there were some anomaly-free subgroup $H^{(p)}\subseteq A^{(p)}$, one could repeat the analysis of the previous sections on $H^{(p)}$ and apply the subsequent discussion on the remaining quotient $A^{(p)}/H^{(p)}$.

\subsection{From 0-gauging to $q$-gauging}

\paragraph{Categorical rephrasing:} Before discussing how a non-trivial anomaly affects whether $\fC$ is Karoubi-complete, let us recall how this data is described categorically. We will focus on a specific global form for our QFT, so that we are studying the $(d-1)$-category ${\rm End}(\cT)$ describing the symmetry operators of $\cT$. For simplicity, let us first consider the case that the entire symmetry structure is concentrated in one level via $A^{(p)}$, so that we are not yet incorporating condensation defects. In other words, the {\it only} non-trivial morphisms in ${\rm End}(\cT)$ are focused in degree $p$ and labeled by elements of $A$. For any physical QFT, one also has a $U(1)^{(d-1)}$ symmetry whose action is given by insertion of a unit complex number, as correlation functions are unaffected by an overall phase \cite{Tachikawa:2017gyf}. We then have $(d-1)$-morphisms in ${\rm End}(\cT)$ labeled by elements of $U(1)$.

These two pieces of data combine to tell us that ${\rm End}(\cT)$ has the structure of a $(d-1)$-group concentrated in the $
p^{\rm th}$ and $(d-1)^{\rm th}$ levels.\footnote{We take here the definition that an $n$-group is an $n$-category with a single object such that all $m$-morphisms are invertible for $1\leq m\leq n$. In the physics literature this encompasses many examples of ``trivial'' $n$-groups, where ``non-trivial'' $n$-groups correspond to there being non-trivial Postnikov invariants in the categorical data.} Such a structure is completely classified by its Postnikov invariant, given by a $(d+1)$-cocycle
\begin{equation}\label{eq:postnikov_class}
    \omega \in H^{d+1}(B^{p+1}A; U(1))\,,
\end{equation}
that acts as a ``twist'' on the higher coherence data of the monoidal product of ${\rm End}(\cT)$ \cite{breen1998monoidalcategoriesmultiextensions,breen1998braidedncategoriessigmastructures,baez2007lecturesncategoriescohomology}. Said differently, when computing the fusion of multiple operators in ${\rm End}(\cT)$, there is an extra phase, described by $\omega$, that affects how this multiplication works. This is exactly the familiar classification of 0-anomalies for group-like symmetries, but phrased in terms of the higher coherence data for a higher category. Let us emphasize that the data encoded by $\omega$ does not introduce any {\it extra} morphisms to our $d$-category $\fC$, but only constrains the behavior of {\it existing} morphisms.

An example that illustrates the discussion above is ordinary quantum mechanics, i.e.~a QFT with $d=1$. For a given symmetry $A^{(0)}$, it is well-known that Hilbert space states may transform in a projective representation of $A^{(0)}$. Let us label the symmetry generators as $\cU_a$ for $a\in A^{(0)}$. If a state $\ket{\psi}$ transforms projectively, then there is a 2-cocycle in group cohomology,
\begin{equation}
    \omega \in H^2_{\rm grp}(A;U(1))
\end{equation}
that modifies the multiplication law of symmetry operators as
\begin{equation}
    \cU_a\cU_{a'}\ket{\psi} = \omega(a,a')\,\cU_{aa'}\ket{\psi}\,.
\end{equation}
Using the equivalence between group cohomology and simplicial/singular cohomology, \begin{equation}
    H^2_{\rm grp}(G;U(1)) \cong H^2(BG;U(1))\,,
\end{equation}
as well as that quantum mechanics is a 1-dimensional quantum theory, we see that the projective phase $\omega$ is an example of the Postnikov class \eqref{eq:postnikov_class}.

For more general theories, the cocycle $\omega$ above determines whether or not the symmetry group $A^{(p)}$ can be gauged in all of spacetime. However, as we are presently interested in condensation defects, we may also ask if $A^{(p)}$ can be gauged only on codimension-$q$ submanifolds of spacetime or if it is instead $q$-anomalous, for $q \leq p+1$. By studying the symmetry operators defined over a codimension-$q$ submanifold of our QFT $\cT$, we are effectively considering a QFT that lives in $(d-q)$ spacetime dimensions, instead of the original $d$ dimensions.

To reflect this, our symmetry structure should be described by the category ${\rm End}(\cT)$ truncated {\it from the bottom} to only contain levels $q$ and above. The truncated category can be viewed as a $(d-q-1)$-group concentrated in the $(p-q)^{\rm th}$ and $(d-q-1)^{\rm th}$ levels. As before, such a higher group is classified by its Postnikov invariant,
\begin{equation}\label{eq:truncated_postnikov_class}
    \omega^{\rm tr} \in H^{d-q+1}(B^{p-q+1}A;U(1))\,,
\end{equation}
which extends the previous classification of 0-anomalies for group-like symmetries to general $q$-anomalies that obstruct higher gauging. The coherence data described by $\omega^{\rm tr}$ then determines which condensation defects can and cannot be consistently constructed, and furthermore how they behave in $\fC$.

\paragraph{Inducing $q$-anomalies:} However, the data of a $q$-anomaly is not fully independent data in a physical theory, and in fact is induced by the data of a $q'$-anomaly for $q' \leq q$. This can be seen via the transgression homomorphism on cohomology\footnote{We would like to thank Dalton Sakthivadivel for discussions on this point. A similar statement can be found in footnote 36 of \cite{Freed:2022qnc}.} 
\begin{equation}\label{eq:transgression_map_1}
    \tau_1: H^{d+1}(X;U(1)) \to H^{d}(\Omega X;U(1))\,,\qquad\omega\mapsto\int_{S^1}{\rm ev}_1^*(\omega)\,,
\end{equation}
where $\Omega X := {\rm Hom}(S^1,X)$ is the loop space of $X$ and ${\rm ev}_1^*$ is the pullback of the evaluation map to cohomology.\footnote{In more detail, the evaluation map ${\rm ev}_1: S^1\times \Omega X \to X$ takes a map $f:S^1 \to X$, a point $x\in S^1$, and evaluates them as ${\rm ev}_1(x,f) = f(x)$.} Specifically, we are considering the diagram,
\begin{equation}
    \begin{tikzcd}
        \Omega X & S^1\times\Omega X \arrow[r, "{\rm ev}_1"] \arrow[l] & X\,,
    \end{tikzcd}
\end{equation}
and moving to cohomology to get the following set of maps,
\begin{equation}
    \begin{tikzcd}
        H^{d}(\Omega X;U(1)) & H^{d+1}(S^1\times\Omega X;U(1)) \arrow[l, "\int_{S^1}"'] & H^{d+1}(X;U(1)) \arrow[l, "{\rm ev}_1^*"'] \, .
    \end{tikzcd}
\end{equation}
The right map is given by the standard pullback of ${\rm ev}_1$, whereas the left map comes from integrating out the $S^1$ fiber. This gives the shift in degree necessary for the $\tau_1$ map in \eqref{eq:transgression_map_1}.

Through an iterated application of $\tau_1$, one finds the following homomorphism,
\begin{equation}
    \tau_q: H^{d+1}(X;U(1)) \to H^{d-q+1}(\Omega^q X;U(1))\,,\qquad \omega\mapsto \underbrace{\int\cdots\int}_q{\rm ev}_q^*(\omega) \, .
\end{equation}
Specializing to the case where $X = B^{p+1}A$, then since $\Omega BY\simeq Y$ for $Y$ a deloopable space, we find that 0-anomalies and $q$-anomalies are related via
\begin{equation}\label{eq:induced_anomaly_map}
    \tau_q: H^{d+1}(B^{p+1}A;U(1)) \to H^{d-q+1}(B^{p-q+1}A;U(1))\,,\qquad \omega\mapsto \omega^{\rm tr} \, .
\end{equation}
The presence of this homomorphism demonstrates that a non-trivial 0-anomaly may induce a non-trivial $q$-anomaly that prevents the $A^{(p)}$ symmetry from being $q$-gauged.

Furthermore, the only possible $q$-anomalies are given by the image of $\tau_q$ rather than the entire codomain $H^{d-q+1}(B^{p-q+1}A;U(1))$. The reason is as follows. Suppose there were another cocycle $\omega^{\rm tr}_0$ such that $\tau_q(\omega) = \omega^{\rm tr} \neq \omega^{\rm tr}_0$, then it would lead to an ambiguity affecting the fusion rule of two defects. Consistency of the fusion in the $d$-category $\fC$ demands that $\omega^{\rm tr}_0$ and $\omega^{\rm tr}$ must coincide, so that higher $q$-anomalies may only take values in the image of \eqref{eq:induced_anomaly_map}. If the 0-anomaly for a given symmetry vanishes, we see that all higher anomalies similarly vanish using that $\tau_q$ is a homomorphism for all $q$, thereby implying the statement of \cite{Roumpedakis:2022aik} that a $q$-gaugeable symmetry is also $q'$-gaugeable for all $q' \geq q$.

Physically, the map $\tau_q$ can be understood by forgetting the braiding of operators in the ambient $q$ directions transverse to the chosen $M^{d-q}$.\footnote{This is a generalization of the map discussed in \cite{Roumpedakis:2022aik}, wherein a non-trivial R-symbol for anyons in a 3d TQFT could induce a non-trivial F-symbol and thus obstruct 1-gauging of the associated topological defect lines.} Although this braiding is forgotten, there may still be phases arising from the fusion of defects, or even additional braiding {\it within} $M^{d-q}$ if the dimensions allow it. However, such data must be consistent with how the defects interact in the rest of spacetime, as $M^{d-q}$ is merely a subregion of $W^d$ with no additional structure to it. In other words, an inability to define gauging with respect to $M^{d-q} \subset W^d$ necessarily means that we also cannot define gauging consistently on $W^d$. Consequently, the possible $q$-anomalies are not completely independent from the data of lower anomalies. This motivates why the higher anomalies are only valued in the subgroup ${\rm im}(\tau_q)\subseteq H^{d-q+1}(B^{p-q+1}A;U(1))$, as this is the subgroup sensitive to the lower anomalies.

\paragraph{Higher groups:} The previous discussion may also be applied to a more complicated symmetry structure. Let us consider now the case in which $\cT$ has two non-trivial Abelian symmetries, $A_1^{(p)}$ and $A_2^{(r)}$, where without loss of generality we will take $p\leq r$. In this case, the symmetry structure is constructed sequentially, so that potential anomalies may arise in how each level acts on the previous. In particular, there may be a Postnikov invariant classifying the extension of $A_1$ by $A_2$ to construct a non-trivial higher group $\mathcal{G}$, described via the following short exact sequence,
\begin{equation}\label{eq:higher_group}
    0\to A_2^{(r)}\to \mathcal{G}\to A_1^{(p)}\to 0 \, .
\end{equation}
The different forms $\mathcal{G}$ can take, and hence the possible extensions, are completely classified by a cocycle\footnote{More generally, there may be a non-trivial action of $A_1^{(p)}$ on $A_2^{(r)}$ labeled by some representation $\rho$. In such cases the Postnikov invariant will instead be valued in {\it twisted} cohomology $H^{r+2}_\rho(B^{p+1}A_1; A_2)$.}
\begin{equation}\label{eq:postnikov_non-trivial}
    \sigma \in H^{r+2}(B^{p+1}A_1; A_2)\,.
\end{equation}
If $\sigma$ is trivial, then the total symmetry group is a product of $A_1^{(p)}$ and $A_2^{(r)}$ with no mixing. This cocycle provides another potential obstruction to gauging the $A_1^{(p)}$ symmetry, and is sometimes referred to as an ``obstruction to symmetry fractionalization'' in the condensed matter literature \cite{Barkeshli:2014cna}.\footnote{The $A_2^{(r)}$ symmetry is unaffected by the presence of a non-trivial $\sigma$, and so is blind to the presence of a higher group. However, it may still have a $q$-anomaly coming from a $\omega^{\rm tr}$ cocycle as above.} While not an anomaly in the traditional sense,\footnote{For discussions on the relation of Postnikov invariants for non-trivial 2-groups to traditional anomalies, as well as their realization in explicit theories, see for example \cite{Cordova:2018cvg,Benini:2018reh}.} we will refer to such data as an anomaly for the $A_1^{(p)}$ symmetry. 

It should be noted that the cocycle $\sigma$ is independent of the data carried by $\omega$ above. Indeed, regardless of the $\sigma$ chosen, and thus the exact form of the extension \eqref{eq:higher_group}, the total symmetry structure $\mathcal{G}$ may still have an anomaly measured by a $(d+1)$-cocycle $\omega \in H^{d+1}(B\mathcal{G};U(1))$. Such cohomology groups can be constructed using the Leray-Serre spectral sequence, regarding the classifying space $B\mathcal{G}$ as a fibration,
\begin{equation}
    B^{r+1}A_2 \hookrightarrow B\mathcal{G} \to B^{p+1}A_1 \, ,
\end{equation}
as discussed more in \cite{Tachikawa:2017gyf,Yu:2020twi}.

Despite being independent coherence data, $\sigma$ and $\omega$ behave similarly when considering higher gauging. Specifically, the cocycle $\sigma$ can also induce more general $q$-anomalies. In the truncated category, there is a cocycle that modifies the coherence data given by
\begin{equation}
    \sigma^{\rm tr} \in H^{r-q+2}(B^{p-q+1}A_1;A_2)\,.
\end{equation} 
In the assumptions of the previous sections, we have implicitly taken both $\omega$ and $\sigma$, and thus $\omega^{\rm tr}$ and $\sigma^{\rm tr}$, to be trivial whenever relevant. 

If desired, one could construct more non-trivial higher groups by iterating this process. In particular, one could start with $\mathcal{G}$ and consider an extension by yet another group $A_3^{(m)}$ for some $m>r$. This extension will again be classified by a Postnikov invariant valued in an appropriate cohomology group, and will construct a higher group $\widetilde{\mathcal{G}}$ with anomalies specified by the Postnikov data. Repeating this, each step will introduce new cocycles that affect the anomaly structure of the associated QFT.

\subsection{Examples}

Let us now see how the presence of non-trivial $q$-anomalies can affect the previous discussion on condensation completeness. We look at three examples of varying complexity, and see how the higher condensation defect trees of Figure \ref{fig:condensation_defects_diagram} are modified.

\paragraph{Non-trivial $q$-anomaly:} The simplest case to consider is when a given $p$-form symmetry $A^{(p)}$ has a non-trivial $q$-anomaly, given by a cocycle $\omega^{\rm tr}$. By definition, we cannot construct a codimension-$q$ condensation defect by summing over generators of $A^{(p)}$ on codimension-$q$ submanifolds $M^{d-q}\subset W^d$. The cocycle $\omega^{\rm tr}$ would introduce phases to such a sum, and make any expression ill-defined due to ambiguities in the operator insertions.

We thus have the process terminate before it even begins. If $\omega^{\rm tr}$ is non-trivial, we cannot even construct a well-defined $q$-condensation defect $\cC_A^{[q]}(M^{d-q})$ to begin with, so there is nothing to consider as an idempotent map, and furthermore no requirement of splitting by Karoubi completeness. In this case, a $q$-anomaly has simplified the discussion to almost nothing.

\paragraph{Trivial 1-anomaly, non-trivial self 0-anomaly:} Our next example arises when a symmetry can only be gauged on a subset of spacetime, but cannot be gauged everywhere. For a concrete realization, we can consider $U(1)_{pN}$ Chern-Simons theory in $d=3$. This theory is well-known to have an anomalous $\bZ_N^{(1)}$ symmetry, where the 0-anomaly can be seen from the fact that the Wilson lines are charged under themselves. Said differently, the line operators braid with one another, and produce a phase characterized by the integer $p$ \cite{Hsin:2018vcg}. This integer corresponds to the 0-anomaly, and the corresponding phase is encoded in a cocycle
\begin{equation}
    \omega \in H^{4}(B^2\bZ_N; U(1)) \cong \begin{cases}
        \bZ_{2N}\,, & N \text{ even}\\
        \bZ_N\,, & N \text{ odd}\,.
    \end{cases}
\end{equation}
We thus see that one cannot gauge the $\bZ_N^{(1)}$ in all of spacetime in the $U(1)_{pN}$ theory. It should be noted that for spin theories, the anomaly is always valued in $\bZ_N$ regardless of if $N$ is even or odd.\footnote{This was discussed in \cite{Hsin:2018vcg}, and can be understood as coming from accounting for a transparent spin-half line in the theory.} As previously stated, we are considering spin theories in the present paper, and thus will utilize the result for odd $N$ in what follows.

While the $\bZ_N^{(1)}$ symmetry cannot be 0-gauged on account of the above 0-anomaly, we can still ask if the symmetry can be 1-gauged. From the above discussion, the possible cocycles encoding a 1-anomaly are given by
\begin{equation}
    \omega^{\rm tr} \in H^{3}(B\bZ_N; U(1)) \cong \bZ_N,\,
\end{equation}
and moreover induced by the 0-anomaly by forgetting the braiding data, i.e.~the 1-anomaly is given by the image of the map
\begin{equation}
    H^{4}(B^2\bZ_N; U(1)) \to H^{3}(B\bZ_N; U(1))\,,\qquad \omega\mapsto\omega^{\rm tr}\,.
\end{equation}
For odd $N$, however, or for spin theories, this map is trivial and so is any potential 1-anomaly. As a result, we can create a 1-condensation defect $\cC_K^{[1]}(M^2)$ by summing over insertions of lines on $M^2$ generating a $\bZ_K\subseteq \bZ_N$ symmetry.

If one were to proceed from here to construct a splitting $\cC_K^{[1]}\cong \iota_K\otimes\rho_K$, they would immediately run into a problem. In the cylinder gauging construction, the actual cylinder is required to support a codimension-0 subregion of spacetime where the $\bZ_K^{(1)}\subseteq\bZ_N^{(1)}$ is gauged. However, in the $U(1)_{pN}$ theory, this symmetry is 0-anomalous and so the cylinder theory appears to be ill-defined! As it turns out, this train of logic was too quick, as we did not stop to check if the 1-condensation defect $\cC_K^{[1]}$ was idempotent and so proceeding with splitting the defect may be moot. Indeed, by computing the fusion of $\cC_K^{[1]}$ with itself, one finds\footnote{Explicit details of this fusion can be found in Appendix A of \cite{Roumpedakis:2022aik}.}
\begin{equation}\label{eq:anomalous_idempotent}
    (\cC_K^{[1]}\otimes\cC_K^{[1]})(M^2) \cong \big|\fT_{L}(M^2)\big|\,\cC^{[1]}_{L}(M^2)\,,
\end{equation}
where $L \coloneqq\gcd(K,pN/K)$. We thus see that for generic $N$ and $K$, the 1-condensation defect $\cC_K^{[1]}$ is not idempotent and thus does not need to split in order to satisfy Karoubi completeness of $\fC$. In fact, from the above we see that $\cC_K^{[1]}$ is idempotent for generic $N$ and $K$ {\it if and only if} the 0-anomaly parameterized by $p$ vanishes. In this way we see that the 0-anomaly seems to act as an obstruction to the defect being idempotent.

Note that there are special values of $p$, $K$, and $N$ such that $\cC_K^{[1]}$ is still idempotent in \eqref{eq:anomalous_idempotent}. In particular, if $pN = nK^2$ for some integer $n$ one finds $L = K$, the fusion
\begin{equation}
    (\cC^{[1]}_K\otimes\cC_K^{[1]})(M^2) \cong \big|\fT_K(M^2)\big|\,\cC_K^{[1]}(M^2)
\end{equation}
tells us that for these choices of $p,K$, and $N$, the 1-condensation defect $\cC_K^{[1]}$ is idempotent, and hence should split in order for the 3-category $\fC$ to be Karoubi complete. However, for such values the subgroup $\bZ_K^{(1)}\subseteq \bZ_N^{(1)}$ is anomaly-free, and so can be 0-gauged without any issue. Therefore the procedure of Section \ref{sec:gen_anomaly_free}, including the cylinder gauging, can proceed without issue for this 1-condensation defect.

\paragraph{Trivial 1-anomaly, non-trivial mixed 0-anomaly:}

Our final example is a slight generalization of the previous one, where we take a 3d QFT $\cT$ with a $\bZ_N^{(1)}\times \bZ_M^{(1)}$ symmetry. The possible 0-anomaly of this symmetry group is captured by the cocycle\footnote{Here we are using the results for odd $N$ and $M$ as we did in the previous example. For even values of $N$ or $M$, the corresponding cohomology group is isomorphic to $\bZ_{2N}\oplus\bZ_{\gcd(N,M)}\oplus\bZ_{2M}$.}
\begin{equation}
    \omega \in H^4(B^2(\bZ_N\times\bZ_M);U(1)) \cong \bZ_N \oplus \bZ_{\gcd(N,M)}\oplus\bZ_M\,.
\end{equation}
The middle $\bZ_{\gcd(N,M)}$ factor captures a mixed anomaly between the two symmetry groups, while the two flanking factors describe any self-anomalies; the middle describes how the generating lines are charged under {\it each other} while the outer factors describe how the lines are charged under {\it themselves}. Let us take $\bZ_N^{(1)}$ and $\bZ_M^{(1)}$ to be individually anomaly-free, but for there to be a mixed anomaly between them characterized by an integer valued in $\bZ_{\gcd(N,M)}$. This set-up describes, for example, 3d $BF$-theory in the case where $N=M$.

The 1-anomaly for this theory is given by the induced cocycle
\begin{equation}
    \omega^{\rm tr}\in H^3(B(\bZ_N\times\bZ_M);U(1)) \cong \bZ_N \oplus 0 \oplus \bZ_M\,,
\end{equation}
where again the two flanking factors describe possible self-anomalies for the $\bZ_N^{(1)}$ and $\bZ_M^{(1)}$ symmetries. As we have assumed these self-anomalies to be trivial, we see that the 1-anomaly vanishes for the whole symmetry group irrespective of $N$ and $M$. As a result, we can construct a 1-condensation defect $\cC^{[1]}_{\bZ_N\times\bZ_M}(M^2)$ by gauging the whole symmetry group on a codimension-1 submanifold of spacetime. 

There is one more subtlety present in the construction of $\cC^{[1]}_{\bZ_N\times\bZ_M}(M^2)$: there are multiple inequivalent ways to gauge the $\bZ_N^{(1)}\times\bZ_M^{(1)}$ symmetry labeled by an invertible field theory realizing the given symmetry. It is more accurate to say there is a {\it family} of condensation defects $\cC^{[1]}_{\bZ_N\times\bZ_M,\,\alpha}(M^2)$, where the label
\begin{equation}\label{eq:SPT_discrete_torsion}
    \alpha \in H^{2}(B(\bZ_N\times\bZ_M);U(1)) \cong \bZ_{\gcd(N,M)}
\end{equation}
characterizes the possible invertible field theories, also known as Symmetry Protected Topological (SPT) or discrete torsion phases.

To simplify the discussion, and thus emphasize the impact on Karoubi completeness, let us further specialize to the case where $N = M = p$ for $p$ an odd prime number. This case was studied, including the various 1-condensation defects and their fusion, in \cite{Roumpedakis:2022aik}. There the authors found that only two condensations $\cC_{\bZ_p\times\bZ_p,\,\alpha}(M^2)$ are idempotent:
\begin{equation}\label{eq:discrete_torsion_idempotents}
    \cC_{\bZ_p\times\bZ_p,\,0}^{[1]}\qquad\text{and}\qquad \cC_{\bZ_p\times\bZ_p,\,p-1}^{[1]}\,.
\end{equation}
In all other cases, the 1-condensation defects were found to produce a different defect under fusion and thus fail to be idempotent. Accordingly, for the symmetry category $\fC$ to be Karoubi-complete, we only need to discuss the splitting for the two 1-condensation defects \eqref{eq:discrete_torsion_idempotents}. 

However, these two 1-condensation defects in particular can be written as a product of two others. If we write the symmetry group as $\bZ_{p_1}^{(1)}\times\bZ_{p_2}^{(1)}$ to keep track of the two copies of $\bZ_p$, we can write the idempotent 1-condensation defects as follows,
\begin{equation}
    \cC_{\bZ_p\times\bZ_p,\,0}^{[1]} = \cC_{p_1}^{[1]}\otimes\cC_{p_2}^{[1]},\qquad \cC^{[1]}_{\bZ_p\times\bZ_p,\,p-1} = \cC^{[1]}_{p_2}\otimes\cC_{p_1}^{[1]} \, .
\end{equation}
which shows that the two idempotent 1-condensation defects can be constructed by sequentially 1-gauging the two subgroups $\bZ_{p_1}$ and $\bZ_{p_2}$. The fact that the order matters for which is gauged first can be understood from the mixed 0-anomaly between the subgroups. As the generating defects are charged under one another, the corresponding 1-condensation defects will not directly commute.

Once we have written the 1-condensation defects of interest as products of $\cC_{p_1}^{[1]}$ and $\cC_{p_2}^{[1]}$, the splitting is identical to that in Sections \ref{sec:4d_example} and \ref{sec:karoubi_completeness}, i.e.~the constituent 1-condensation defects are themselves idempotent, and we can split them as usual. This allows us to split the composite 1-condensation defects as
\begin{equation}
    \begin{gathered}
        \cC_{\bZ_p\times\bZ_p,\,0}^{[1]} = \iota_{p_1}\otimes \rho_{p_1}'\,,\qquad \rho_{p_1}' \coloneqq \rho_{p_1}\otimes \cC_{p_2}^{[1]}\,,\\
        \hspace{-9pt}\cC_{\bZ_p\times\bZ_p,\,p-1}^{[1]} = \iota_{p_1}'\otimes\rho_{p_1}\,,\qquad \iota_{p_1}' \coloneqq \cC_{p_2}^{[1]}\otimes\iota_{p_1}\,,
    \end{gathered}
\end{equation}
where $\rho_i$ and $\iota_i$ are respectively the gauging and ungauging interfaces of Section \ref{sec:karoubi_completeness}.\footnote{It should be noted that this choice of splitting is not unique, and we have only chosen one such possibility. However, as discussed in Section \ref{sec:categorical_structures}, there is no expectation that the splitting interfaces are unique, so this is not an obstruction to Karoubi completeness.} This further justifies the claim that the the two 1-condensation defects \eqref{eq:discrete_torsion_idempotents} can be constructed via a sequential 1-gauging, as we could similarly split the remaining factor of $\cC_{p_2}^{[1]}$ into another set of gauging and ungauging interfaces. We therefore see the only idempotent 1-condensation defects can indeed be split without issue, and the symmetry category $\fC$ is still Karoubi-complete even in the case of a mixed anomaly.

\paragraph{More general constructions:} We have so far given three specific examples for how the presence of non-trivial anomalies can affect the $m$-condensation defects in a theory, and hence how Karoubi completeness manifests in the associated $d$-category $\fC$. Let us now take a step back and outline a general procedure for such an analysis. 

The first step in studying the effect of anomalies it to have the anomalies in the first place. Either by fiat or by computation, one can find the 0-anomalies $\omega$ and $\sigma$ and thus their remnants as $q$-anomalies $\omega^{\rm tr}$ and $\sigma^{\rm tr}$. If the latter are trivial, then $m$-condensation defects of appropriate dimension can be constructed out of the relevant symmetry operators. This also requires the possible data of SPTs as labeled by $\alpha$ in \eqref{eq:SPT_discrete_torsion}. Once the $m$-condensation defects are constructed, the question becomes which, if any, $m$-morphisms of $\fC$ are idempotent. Any that are should then be able to be factored into gauging/ungauging interfaces. These may only come from products of other $m$-condensation defects as in the final example above, in which case the possible splittings are inherited from the individual factors. If any of these steps fail, the process truncates without any bearing on $\fC$ being Karoubi-complete.

\section{Discussion and Outlook}\label{sec:discussion}

It is important for us to emphasize that what we have done in this work is not meant to be a rigorous mathematical proof that all QFTs have symmetry structures described by higher fusion categories.\footnote{In fact, there are known counterexamples which are not described by fusion categories \cite{Gaiotto:2019xmp}.} Rather, we have illustrated that the defining features of a (separable weak) higher fusion category \cite{Gaiotto:2019xmp,Johnson-Freyd:2020usu} have natural physical origins. Particularly, the $n$-condensation defects for $1 \leq n \leq d$ in a $d$-dimensional QFT are necessarily arranged in a hierarchical manner that respects the nested structure of a higher category, due to the fact that Pontryagin dual/quantum symmetries are localized to the worldvolume in which the original symmetry is (higher) gauged.

Among the properties of a higher fusion category, the notion of Karoubi completeness provides a crucial organizing principle for us to investigate condensation defects, which constitute a cornerstone in the modern study of generalized symmetries in QFTs. It reveals the precise relation between $n$-condensation defects (which have codimension $n$) and gauging with respect to a codimension-$(n-1)$ submanifold of spacetime. For anomaly-free (finite Abelian) symmetries, an $n$-condensation defect is idempotent (in a suitably categorified sense), and so it always factorizes into a pair of gauging interfaces sandwiching a ``bulk theory'' of dimension $d-n+1$. The spectrum of operators present in such a bulk theory matches with that in the original $n$-condensation defect. For anomalous (finite Abelian) symmetries, an $n$-condensation defect may or may not be idempotent. We argued that the condition for it to be idempotent, hence splittable, corresponds precisely to the anomaly being trivializable, such that a corresponding ``bulk'' exists.
    
In this paper, although we only focused on finite Abelian (group-like) symmetries, we saw that there is already a rich interplay between the symmetry generators, charged operators, and condensation defects. An immediate question one may ask is whether our analysis of higher condensation defects can be generalized to finite non-Abelian symmetries. In this case, the quantum symmetry one obtains by (higher) gauging a group $G$ is no longer the Pontryagin dual $\widehat{G} = \text{Hom}(G,U(1))$, which is still Abelian by definition. Instead, it should be replaced with a generalization known as the Tannaka dual, i.e.~the category $\text{Rep}(G)$ of $G$-representations, or potentially higher-categorical analogues depending on the dimensions involved (e.g.~see \cite{Bartsch:2022mpm,Bartsch:2022ytj,Bartsch:2023pzl,Bartsch:2023wvv}). Assuming Karoubi completeness always holds as a general property of the symmetry $d$-category $\fC$ for a $d$-dimensional QFT, we expect the corresponding ``non-Abelian'' condensation defects to be splittable into gauging interfaces sandwiching a ``bulk theory'' with $\text{Rep}(G)$ symmetry. On top of that, analogously to the construction of the ``dual'' higher condensation defects in Section \ref{subsec:condensation_defects} by 1-gauging the Pontryagin-dual symmetry $\widehat{G}$, here one will need to study the gauging of the Tannaka-dual symmetry $\text{Rep}(G)$ in a higher-categorical setting for generic spacetime dimensions.

More generally, it will be interesting to apply the same philosophy to study non-invertible symmetries which are not group-like in nature. However, to the best of our knowledge, it is somewhat unclear how to unambiguously define the notions of ``gauging'' and ``anomaly'' of a non-invertible symmetry in generic spacetime dimensions (see \cite{Kaidi:2023maf,Choi:2023xjw,Sun:2023xxv,Cordova:2023bja,Antinucci:2023ezl,Choi:2023vgk,Antinucci:2024ltv,DelZotto:2025yoy} for some partial results). We anticipate that a good definition of condensation defects for non-invertible symmetries should nevertheless have appropriate idempotence and splitting properties that are compatible with these notions.

Another key aspect in the study of generalized symmetries is the construction of the Symmetry Topological Field Theory (SymTFT) \cite{Apruzzi:2021nmk,Freed:2022qnc}, which lives in a $(d+1)$-dimensional bulk. In terms of the fusion category encoding the symmetries of our $d$-dimensional QFT, the SymTFT roughly corresponds to the notion of its Drinfeld center \cite{2002math......3060E}, or rather its higher-categorical equivalent (e.g.~see \cite{Johnson-Freyd:2020usu}). Simply put, the $(d+1)$-dimensional SymTFT bulk is supported on the manifold $W^d \times I$, whose boundary consists of two copies of the spacetime manifold $W^d$. These two copies respectively accommodate the gapped and gapless degrees of freedom, such that upon the interval compactification of the SymTFT, one recovers the full QFT. For finite group-like symmetries, a choice of global form of the QFT amounts to a choice of boundary condition on the gapped boundary of the SymTFT. As we learned in this paper, if we insert a 1-condensation defect in the QFT $\cT$, then it splits into a pair of gauging interfaces sandwiching a gauged theory in the middle, say, $\cT/K^{(p)}$. From the perspective of the SymTFT bulk, such a configuration has to result from a hybrid choice of boundary conditions imposed on the gapped boundary. Moreover, one may even insert condensation defects within condensation defects, which unfold into a non-trivial, nested, network of sub-theories of various codimensions. It will then be instructive to systematically characterize the corresponding boundary conditions, incorporating the entire hierarchy of higher condensation defects, and establish concrete physical realizations in terms of the SymTFT.

A related question is the holographic realization of higher condensation defects. As pointed out by \cite{Apruzzi:2022rei,GarciaEtxebarria:2022vzq}, followed by a series of subsequent work, non-invertible symmetry defects, including condensation defects, have a holographic dual description in terms of D-branes in string theory. Particularly, it was shown in \cite{Bah:2023ymy} that 1-condensation defects can be understood as the consequence of pairs of D-branes and anti-D-branes undergoing tachyon condensation \cite{Sen:1998ii,Sen:1998sm,Sen:1999mg}, at least in the examples studied therein. In this scenario, the top-level D$p$-brane charge cancels between the D$p$-brane and the $\overline{\text{D}p}$-brane, but the lower-level D$(p-2)$-brane charge generically does not vanish, and is interpreted quantum mechanically as a superposition of D$(p-2)$-branes localized within the worlvolume of the $\text{D}p$-$\overline{\text{D}p}$-brane pair, which is the defining characteristic of a 1-condensation defect. Furthermore, the dual gauge field on the 1-condensation defect, whose holonomies are the symmetry generators of the Pontryagin dual symmetry, admits a natural stringy interpretation as the Chan-Paton gauge field on the D-branes. It is therefore possible for one to reconstruct the hierarchy of higher condensation defects in terms of D-brane physics. Doing so will hopefully also shed light on the precise relation between two parallel mathematical descriptions, namely, generalized symmetries in QFTs using higher fusion categories, and D-branes in string theory using K-theory.


\acknowledgments

We are grateful to Riccardo Argurio, Lakshya Bhardwaj, Victor Carmona, Iñaki García-Etxebarria, Andrea Grigoletto, Theo Johnson-Freyd, Tian Lan, Jamie Pearson, Daniel Roggenkamp, Konstantinos Roumpedakis, Dalton Sakthivadivel for interesting conversations and correspondence. IB, EL, and TW are supported in part by the Simons Collaboration on Global Categorical Symmetries and also by the NSF grant PHY-2412361.


\appendix

\addtocontents{toc}{\protect\setcounter{tocdepth}{1}}

\section{Some More Categorical Notions}\label{app:categorical_notions}

In this appendix, we supply some basic categorical description of the defining properties of a separable weak higher (multi)fusion category that was not discussed in the main text.\footnote{A multifusion $n$-category generally has non-trivial topological point operators in the bulk, which would be the case in the presence of a $(d-1)$-form symmetry. Meanwhile, a fusion $n$-category is a multifusion $n$-category with a single bulk point operator which is the identity. In this paper, we do not carefully distinguish between these two notions, provided the context is clear. Note that the $d$-condensation defects that we constructed earlier do not fall under this criterion, because they are localized to some submanifold rather than spacetime (i.e.~bulk) itself.} For more technical details, the reader is referred to \cite{Johnson-Freyd:2020usu} and \cite{2018arXiv181211933D}.

\subsection{Fusion Category Adjectives}

\paragraph{$\bC$-linearity:} For a general $n$-category to be $\bC$-linear, every set of $n$-morphisms must form a complex vector space and composition of $n$-morphisms must be linear over $\bC$. In the category of symmetry operators in $\cT$, and moreover in the entire $d$-category $\fC$, the top-level morphisms describe topological operators supported on codimension-$d$ manifolds, i.e.~points, in spacetime. We have already seen how such operators form a complex vector space in the discussion surrounding \eqref{eq:linearity_linearcombo}. 

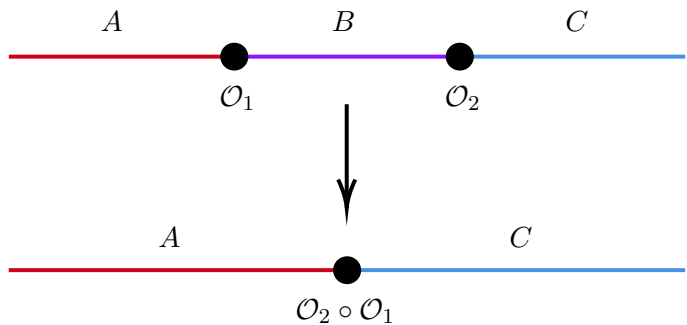
\begin{figure}
    \centering
\tikzset{every picture/.style={line width=0.75pt}} 

\begin{tikzpicture}[x=0.75pt,y=0.75pt,yscale=-1,xscale=1]

\draw [color={rgb, 255:red, 208; green, 2; blue, 27 }  ,draw opacity=1 ][line width=1.5]    (161.34,82.24) -- (273.89,82.24) ;
\draw [color={rgb, 255:red, 144; green, 19; blue, 254 }  ,draw opacity=1 ][line width=1.5]    (273.89,82.24) -- (386.45,82.24) ;
\draw [color={rgb, 255:red, 74; green, 144; blue, 226 }  ,draw opacity=1 ][line width=1.5]    (386.45,82.24) -- (499,82.24) ;
\draw  [fill={rgb, 255:red, 0; green, 0; blue, 0 }  ,fill opacity=1 ] (267.39,82.24) .. controls (267.39,78.65) and (270.3,75.74) .. (273.89,75.74) .. controls (277.48,75.74) and (280.39,78.65) .. (280.39,82.24) .. controls (280.39,85.83) and (277.48,88.74) .. (273.89,88.74) .. controls (270.3,88.74) and (267.39,85.83) .. (267.39,82.24) -- cycle ;
\draw  [fill={rgb, 255:red, 0; green, 0; blue, 0 }  ,fill opacity=1 ] (379.95,82.24) .. controls (379.95,78.65) and (382.86,75.74) .. (386.45,75.74) .. controls (390.04,75.74) and (392.95,78.65) .. (392.95,82.24) .. controls (392.95,85.83) and (390.04,88.74) .. (386.45,88.74) .. controls (382.86,88.74) and (379.95,85.83) .. (379.95,82.24) -- cycle ;
\draw [color={rgb, 255:red, 208; green, 2; blue, 27 }  ,draw opacity=1 ][line width=1.5]    (161.3,189.63) -- (332.35,189.63) ;
\draw [color={rgb, 255:red, 74; green, 144; blue, 226 }  ,draw opacity=1 ][line width=1.5]    (326.67,189.63) -- (498.96,189.63) ;
\draw  [fill={rgb, 255:red, 0; green, 0; blue, 0 }  ,fill opacity=1 ] (323.71,189.67) .. controls (323.71,186.08) and (326.62,183.17) .. (330.21,183.17) .. controls (333.8,183.17) and (336.71,186.08) .. (336.71,189.67) .. controls (336.71,193.26) and (333.8,196.17) .. (330.21,196.17) .. controls (326.62,196.17) and (323.71,193.26) .. (323.71,189.67) -- cycle ;
\draw [line width=1.5]    (330.01,106.18) -- (330.01,155.29) ;
\draw [shift={(330.01,158.29)}, rotate = 270] [color={rgb, 255:red, 0; green, 0; blue, 0 }  ][line width=1.5]    (14.21,-4.28) .. controls (9.04,-1.82) and (4.3,-0.39) .. (0,0) .. controls (4.3,0.39) and (9.04,1.82) .. (14.21,4.28)   ;

\draw (205.76,57.78) node [anchor=north west][inner sep=0.75pt]    {$A$};
\draw (321.07,57.78) node [anchor=north west][inner sep=0.75pt]    {$B$};
\draw (437.78,57.78) node [anchor=north west][inner sep=0.75pt]    {$C$};
\draw (234.52,166.64) node [anchor=north west][inner sep=0.75pt]    {$A$};
\draw (409.77,166.64) node [anchor=north west][inner sep=0.75pt]    {$C$};
\draw (264.96,95.44) node [anchor=north west][inner sep=0.75pt]    {$\mathcal{O}_{1}$};
\draw (377.51,95.44) node [anchor=north west][inner sep=0.75pt]    {$\mathcal{O}_{2}$};
\draw (302.96,202.79) node [anchor=north west][inner sep=0.75pt]    {$\mathcal{O}_{2} \circ \mathcal{O}_{1}$};
\end{tikzpicture}
    \caption{The OPE of local operators giving rise to composition when viewed categorically. Here we have two local operators which we can view as morphisms $A\xrightarrow{\cO_1} B$ and $C\xrightarrow{\cO_2}$. When fused together, they yield a composition $A\xrightarrow{\cO_2\circ\cO_1}C$ from the OPE \eqref{eq:example_monoidal}.}
    \label{fig:local_operator_OPE}
\end{figure}

Now we need only to discuss the linearity of compositions, which can be understood through the OPE \eqref{eq:example_monoidal} applied to local operators. Let us consider an example with two operators $\cO_{1,2}$ such as in Figure \ref{fig:local_operator_OPE}. There, $\cO_1(x)$ acts as an interface between region $A$ and region $B$, while $\cO_2(y)$ does the same between region $B$ and region $C$. This can be framed categorically as the statement that, as $d$-morphisms, $\cO_1\in\Hom(A,B)$ and $\cO_2\in\Hom(B,C)$. By the OPE \eqref{eq:example_monoidal}, we can describe the product as a $d$-morphism $(\cO_2\circ\cO_1)\in\Hom(A,C)$ resulting from a composition of the previous ones,
\begin{equation}
    \circ: \Hom(B,C)\times\Hom(A,B)\to \Hom(A,C)\,,\qquad (\cO_2,\cO_1)\mapsto\cO_2\circ\cO_1\,,
\end{equation}
that is linear in both arguments. Combined with previous statement that the topological local operators form a complex vector space, we have found that ${\rm End}(\cT)$, and moreover $\fC$, is $\bC$-linear.

\paragraph{Additivity:} For a general $n$-category to be additive, the definition is recursive. A 1-category $\mathfrak{D}$ is additive if the following conditions hold.
\begin{itemize}
    \item It has a {\it zero object}: an object $0$ such that ${\rm End}(0)$ is the zero vector space.
    \item It has {\it direct sums}: for $X_{1,2}$ any two objects in $\mathfrak{D}$, an object $X_1\oplus X_2$ together with maps $\iota_i:X_i\leftrightarrows X_1\oplus X_2:\rho_i$ such that
    \begin{itemize}
        \item $\rho_i\circ \iota_i = {\rm id}_{X_i}$ for $i=1,2$,
        \item $\rho_j\circ\iota_i = 0$ for $i\neq j$,
        \item $\iota_1\circ\rho_1 + \iota_2\circ\rho_2 = {\rm id}_{X_1\oplus X_2}$.
    \end{itemize}
\end{itemize}
A 2-category is additive if it has a zero object, direct sums, and if it is {\it locally additive}, meaning every Hom-category is an additive 1-category.\footnote{The exact details of this definition require a categorification of the data for an additive 1-category, but the heuristic picture is the same. For example, in the definition of direct sums for a 2-category, $\rho_i\circ\iota_i$ need only be isomorphic to ${\rm id}_{X_i}$ instead of equal, and $\rho_j\circ\iota_i$ must be isomorphic to a zero object of $\Hom(X_1,X_2)$ instead of identically zero.} Inductively, a general $n$-category is additive if it has a zero object, direct sums, and if its Hom-$(n-1)$-categories are additive. We have discussed these properties in the case of our symmetry operators in $\cT$ surrounding \eqref{eq:example_additivity}, namely, there is a zero operator at every level of ${\rm End}(\cT)$, as well as a notion of addition $\oplus$ at every level, thus making ${\rm End}(\cT)$ additive.

\paragraph{Composition of condensations:} A useful property of categorical condensations is that they can be composed. This is crucial for us to realize sequentially gauging in a physical theory as compositions of categorical condensations. More precisely, suppose there exist two $n$-condensations of objects in an $n$-category, i.e.
\begin{equation}
	\rho_{X,Y}: X \rightleftarrows Y :\iota_{X,Y} \, , \qquad \rho_{Y,Z}: Y \rightleftarrows Z :\iota_{Y,Z} \, ,
\end{equation}
such that the morphisms above satisfy
\begin{equation}
	e_{X,Y} \circ e_{X,Y} \cond e_{X,Y} \, , \quad \rho_{X,Y} \circ \iota_{X,Y} \cond 1_Y \, , \quad e_{Y,Z} \circ e_{Y,Z} \cond e_{Y,Z} \, , \quad \rho_{Y,Z} \circ \iota_{Y,Z} \cond 1_Z \, .
\end{equation}
If we define the compositions of these morphisms as (up to isomorphisms)
\begin{equation}
	\rho_{X,Z} \cong \rho_{Y,Z} \circ \rho_{X,Y} \, , \qquad \iota_{X,Z} \cong \iota_{X,Y} \circ \iota_{Y,Z} \, , \qquad e_{X,Z} \cong \iota_{X,Z} \circ \rho_{X,Z} \, ,
\end{equation}
then one obtains
\begin{align}
	e_{X,Z} \circ e_{X,Z} & \cong (\iota_{X,Y} \circ \iota_{Y,Z} \circ \rho_{Y,Z} \circ \rho_{X,Y}) \circ (\iota_{X,Y} \circ \iota_{Y,Z} \circ \rho_{Y,Z} \circ \rho_{X,Y})\nonumber\\
	& \cond (\iota_{X,Y} \circ \iota_{Y,Z} \circ \rho_{Y,Z}) \circ 1_Y \circ (\iota_{Y,Z} \circ \rho_{Y,Z} \circ \rho_{X,Y})\nonumber\\
	& \cond (\iota_{X,Y} \circ \iota_{Y,Z}) \circ 1_Z \circ (\rho_{Y,Z} \circ \rho_{X,Y})\nonumber\\
	& \cong \iota_{X,Z} \circ \rho_{X,Z}\nonumber\\
	& \cong e_{X,Z} \, ,
\end{align}
and also
\begin{equation}
	\rho_{X,Z} \circ \iota_{X,Z} \cong (\rho_{Y,Z} \circ \rho_{X,Y}) \circ (\iota_{X,Y} \circ \iota_{Y,Z}) \cond \rho_{Y,Z} \circ 1_Y \circ \iota_{Y,Z} \cond 1_Z \, ,
\end{equation}
which implies that
\begin{equation}
	\rho_{X,Z}: X \rightleftarrows Z :\iota_{X,Z}
\end{equation}
is an $n$-condensation. Therefore, the composition of two $n$-condensations is also an $n$-condensation.

\paragraph{Semisimplicity:} In an additive $n$-category $\fC$, an object $X$ is said to be {\it simple} if it is indecomposable under the direct sum, i.e.~there is no way it can be written as $X = \bigoplus_i X_i$ for $X_i$ simple. From here, we can define a semisimple $n$-category recursively.

An $n$-category $\fC$ is said to be {\it semisimple} if it is locally semisimple, and every object is either simple or can be written as a finite sum of simple objects. By locally semisimple, we mean that $\Hom(X,Y)$ is a semisimple $(n-1)$-category for any two objects $X$ and $Y$ in $\fC$. This compiles recursively for every categorical level in $\fC$.

\paragraph{Finiteness:} For an $n$-category $\fC$ to be finite it must have a finite set of connected components of simple objects and the $(n-1)$-category $\Hom(X,Y)$ must also be finite for any two objects $X$ and $Y$ in $\fC$.\footnote{As a remark, finite $n$-categories are equivalently referred to as being ``compact'' in \cite{2021arXiv211109080D}.} By connected components, we mean elements of
\begin{equation}
    \pi_0(\fC) \coloneqq \{\text{simple objects of }\fC\}/\sim\,,
\end{equation}
where two simple objects $X$ and $Y$ of $\fC$ are considered equivalent if there is a non-zero 1-morphism between them. In this case one says $X$ and $Y$ are ``Schur-connected'' \cite{2018arXiv181211933D,Johnson_Freyd_2023}. For higher categories this condition is equivalent to the statement that there is a {\it condensation} between them \cite{TheoSeminarTalk,ReutterConfTalk,Bartsch:2023wvv}. This must similarly hold for each Hom-$(n-1)$-category, and so on recursively at every categorical level in $\fC$. For $n$-morphisms, the condition is that the vector spaces are finite dimensional. In this way we see that finiteness is mostly a constraint on the original symmetry group $A^{(p)}$ we use to construct $\fC$ for a physical QFT. While the number of $m$-condensation defects is still finite, as discussed in the main text, this was not strictly necessary in order to claim $\fC$ is finite. It should also be noted that the fusion of defects does not add any complications to this claim. The defects of $\fC$ are closed under parallel fusion for all QFTs we discuss.

\paragraph{Rigidity:} A monoidal $n$-category $\fC$ is said to be {\it rigid} if every $m$-morphism has a dual in $\fC$ for $0\leq m\leq n$.\footnote{For $m \geq 1$, the duals are also called adjoints.} As introduced in the main text, an $m$-morphism $f:X\to Y$ has a dual $f^\vee:Y\to X$ if there exist $(m+1)$-morphisms
\begin{equation}\label{eq:birth_and_death}
    d_f: f^\vee\otimes f\to 1_X^{[m]}\,,\quad\text{ and }\quad b_f: 1_Y^{[m]}\to f\otimes f^\vee\,.
\end{equation}
These morphisms must additionally satisfy coherence relations known as the ``zig-zag'' relations. These relations state that the following diagrams commute, at least up to an invertible $(m+2)$-morphism.
\begin{equation}\label{eq:zig-zag_identity}
    \begin{gathered}
	\begin{tikzcd}
        (f\otimes f^\vee) \otimes f \arrow[r, "a_{f,f^\vee,f}"] &  f\otimes (f^\vee \otimes f) \arrow[d, "1_f^{[m+1]}\otimes b_f"] \\
		1^{[m]}\otimes f \arrow[u,"b_f\otimes 1_f^{[m+1]}"]& f\otimes 1^{[m]}\arrow[d,"R_f"]\\
		f \arrow[u, "L_f^{-1}"] \arrow[r, "1_f^{[m+1]}"] & f
	\end{tikzcd}\\
    \begin{tikzcd}
        f^\vee\otimes (f \otimes f^\vee) \arrow[r, "a^{-1}_{f^\vee,f,f^\vee}"] &  (f^\vee \otimes f) \otimes f^\vee \arrow[d, "d_f \otimes 1_{f^\vee}^{[m+1]}"] \\
		f^\vee \otimes 1^{[m]} \arrow[u,"1_{f^\vee}^{[m+1]}\otimes b_f"]& 1^{[m]}\otimes f^\vee\arrow[d,"L_{f^\vee}"]\\
		f^\vee \arrow[u, "R_{f^\vee}^{-1}"] \arrow[r, "1_{f^\vee}^{[m+1]}"] & f^\vee
	\end{tikzcd}
    \end{gathered}
\end{equation}
Additionally, all this data must be consistent with the data in the remaining categorical levels of the monoidal category $\fC$.

\paragraph{Full dualizability:} The final property for a monoidal $n$-category $\fC$ to be fusion is that it is a fully dualizable object in the $(n+2)$-category ${\rm Mor}_1(n{\rm KarCat}_\bC)$. Let us now dissect what this statement means. First, $n{\rm KarCat}_\bC$ is the $(n+1)$-category of $\bC$-linear, additive, and Karoubi-complete $n$-categories. This is a symmetric monoidal $(n+1)$-category, where the monoidal product comes from a Karoubi-completed Deligne tensor product of $n$-categories. Physically, this is coming from stacking two different symmetry structures and Karoubi-completing the result $\fC\boxtimes\fC'$. As a result of $n{\rm KarCat}_\bC$ being symmetric monoidal, one can further construct the monoidal $(n+2)$-category of monoid objects in $n{\rm KarCat}_\bC$. It is this category that is denoted ${\rm Mor}_1(n{\rm KarCat}_\bC)$, where the ${\rm Mor}_1$ denotes equivalences in this $(n+2)$-category are given by Morita equivalence.

In a generic monoidal $m$-category $\mathfrak{D}$, an object $X$ is 1-dualizable if it admits a dual $X^\vee$ when viewed as a 1-morphism in $B\mathfrak{D}$.\footnote{Here $B\mathfrak{D}$ is the {\it delooping} of $\mathfrak{D}$. This is an $(m+1)$-category with a single object $\bullet$, where ${\rm End}(\bullet) \cong \mathfrak{D}$.} If the 1-morphisms $d_X$ and $b_X$ in $\mathfrak{D}$, as defined in \eqref{eq:birth_and_death}, themselves admit duals, then $X$ is further said to be 2-dualizable. If this iterates up to $m$-morphisms, then $X$ is said to be {\it fully dualizable}. In this sense, the final requirement for a monoidal $n$-category $\fC$ to be fusion by the definition of \cite{Johnson-Freyd:2020usu} is that $\fC$ is a fully dualizable object of ${\rm Mor}_1(n{\rm KarCat}_\bC)$.

However, this requirement can also be understood as a combination of three other properties of $\fC$, namely, the requirement that $\fC$ is semisimple, finite, and rigid \cite{TheoPrivateEmail}. As such, it suffices to show that $\fC$ satisfies the necessary conditions for these three criteria as argued in the main body. When combined with the requirements of monoidality, $\bC$-linearity, additivity, and Karoubi completeness, we thus see that ${\rm End}(\cT)$ satisfies the necessary requirements to be fusion.

\subsection{Monoidal Product Coherence Relations}

In Section \ref{sec:monoidal_structure}, we briefly reviewed the definition of a monoidal category and quickly stated the necessary coherence relations that such a structure must satisfy. Here we expand on this description and offer some proof of additional claims. We do not claim any of the material in this subsection is new. Most of it can indeed be found in standard textbooks such as \cite{etingof2015tensor}. We simply include it for completeness and clarity.

A 1-category $\mathfrak{D}$ with functor $\otimes:\mathfrak{D}\times\mathfrak{D} \to \mathfrak{D}$ forms a monoidal 1-category $(\mathfrak{D},\otimes)$ if the following is true.
\begin{itemize}
    \item There is a distinguished object $1^{[1]}$ in $\mathfrak{D}$ known as the monoidal unit of $\mathfrak{D}$.
    \item There is a natural isomorphism defined for any $X,Y,$ and $Z$ in $\mathfrak{D}$,
    \begin{equation}
        a_{X,Y,Z}: (X\otimes Y)\otimes Z \xrightarrow{\raisebox{-0.65ex}{$\sim$}} X\otimes(Y\otimes Z)\,,
    \end{equation}
    called the associator.
    \item There are natural isomorphisms
    \begin{equation}
        L_X: 1^{[1]}\otimes X \xrightarrow{\raisebox{-0.65ex}{$\sim$}} X\,,\quad\text{ and } R_X: X\otimes 1^{[1]}\xrightarrow{\raisebox{-0.65ex}{$\sim$}} X\,,
    \end{equation}
    respectively called left and right unitors.
\end{itemize}
In order for $(\mathfrak{D},\otimes)$ to be consistently defined, so that the operation $\otimes$ behaves as an appropriate monoidal product, there must be additional coherence relations imposed on the above data.

The first of such data is known as the {\it pentagon axiom} or {\it pentagon identity}, and demands that the associator satisfies the following commutative diagram.
\begin{equation}\label{eq:pentagon_axiom}
	\begin{tikzcd}
		& (W\otimes X)\otimes(Y\otimes Z) \arrow[dr,"a_{W,X,Y\otimes Z}"]&\\
		((W\otimes X)\otimes Y)\otimes Z \arrow[ur, "a_{W\otimes X,Y,Z}"] \arrow[dd,"a_{W,X,Y}\otimes 1^{[2]}_Z"]& & W\otimes(X\otimes (Y\otimes Z))\\
		& & \\
		(W\otimes (X\otimes Y))\otimes Z \arrow[rr,"a_{W,X\otimes Y,Z}"] & & W\otimes((X\otimes Y)\otimes Z)\arrow[uu, "1^{[2]}_W\otimes a_{X,Y,Z}"]
	\end{tikzcd}
\end{equation}
The diagram must hold for any $W,X,Y,$ and $Z$ in $\mathfrak{D}$. This gives consistency between the associator and identity morphisms for each object.

The second demand is similarly known as the {\it triangle axiom} or {\it triangle identity}, and is such that the unitors satisfy the following diagram. 
\begin{equation}\label{eq:triangle_axiom}
    \begin{tikzcd}
    	& X\otimes(1^{[1]}\otimes Y) \arrow[dr,"1^{[2]}_X\otimes L_Y"]&\\
    	(X\otimes 1^{[1]})\otimes Y \arrow[ur, "a_{X,1^{[1]},Y}"] \arrow[rr,"R_X\otimes 1^{[2]}_Y"]& & X\otimes Y\\
    \end{tikzcd}
\end{equation}
This ensures consistency between the associator and the monoidal unit of $(\mathfrak{D},\otimes)$. As long as both \eqref{eq:pentagon_axiom} and \eqref{eq:triangle_axiom} are satisfied, then $(\mathfrak{D},\otimes)$ forms a consistent monoidal 1-category.

The monoidal operation $\otimes$ in a monoidal category $(\mathfrak{D},\otimes)$ may be equipped with additional structure. Such structure is a categorified analogue of commutativity, and is known as a {\it braiding}. A braiding structure is given by an isomorphism,
\begin{equation}
    c_{X,Y}: X\otimes Y \xrightarrow{\raisebox{-0.65ex}{$\sim$}} Y\otimes X\,,
\end{equation}
for any two objects $X$ and $Y$ in $\mathfrak{D}$. In order to be consistent with the rest of the data in $(\mathfrak{D},\otimes)$, the braiding structure must satisfy additional coherence data on top of that already demanded of a monoidal category. In particular, the braiding structure must be consistent with the associator morphisms, through what is known as the {\it hexagon axiom} or the {\it hexagon identity}. This is the statement that $c_{X,Y}$ must satisfy the following commutative diagram for any $X,Y,$ and $Z$.
\begin{equation}\label{eq:hexagon_axiom}
	\begin{tikzcd}
		& X\otimes (Y\otimes Z) \arrow[r, "c_{X,Y\otimes Z}"] & (Y\otimes Z) \otimes X \arrow[dr, "a_{Y,Z,X}"]&\\
		(X\otimes Y)\otimes Z \arrow[ur, "a_{X,Y,Z}"] \arrow[dr, "c_{X,Y}\otimes 1^{[2]}_Z"] & & & Y\otimes (Z\otimes X)\\
		& (Y\otimes X)\otimes Z \arrow[r, "a_{Y,X,Z}"] & Y\otimes (X\otimes Z) \arrow[ur, "1^{[2]}_Y \otimes c_{X,Z}"]&
	\end{tikzcd}
\end{equation}
Once this is satisfied, it follows without additional assumptions that the braiding is consistent with the unitor isomorphisms as well. If the operation $\otimes$ is coherent with the existence of $c_{X,Y}$ in this way, we say $(\mathfrak{D},\otimes)$ is a {\it braided} monoidal category.

\subsection{Homotopy Cardinality of Discrete Gauge Theory}\label{app:homotopy_cardinality}

Here we will further explain the equivalence of the two expressions \eqref{eq:homotopy_cardinality} and \eqref{eq:finite-group_partition_function} in the case of a $d$-dimensional QFT $\cT$ with a finite $p$-form symmetry $G^{(p)}$. After constructing $\fC$ as in the main body, let us focus on the operators in ${\rm End}(\cT)$, which will form a $(d-1)$-subcategory of $\fC$. For notational brevity, let us denote $\fT \coloneqq {\rm End}(\cT)$. In considering the ``size'' of $\fT$, we compute the homotopy cardinality as in \eqref{eq:homotopy_cardinality},
\begin{equation}\label{eq:app_homotopy_cardinality}
     \big|\fT\big| = \sum_{x\in\pi_0(\fT)}\prod_{k=1}^{d-1}\left|\pi_k (\fT,x)\right|^{(-1)^k} = \sum_{x\in\pi_0(\fT)} \frac{\big|\pi_2 (\fT,x)\big|\big|\pi_4 (\fT,x)\big|\cdots}{\big|\pi_1 (\fT,x)\big|\big|\pi_3 (\fT,x)\big|\cdots}\,.
\end{equation}
In order to proceed in evaluating this expression, we need to take a closer look at the category $\fT$. The objects (and higher morphisms) are given by extended operators in our field theory. Such operators can equivalently be viewed as maps from the spacetime $M^d$ to the classifying space of our symmetry group $B^{p+1}G = K(G,p+1)$. In other words, we can view our category $\fT$ as the mapping space ${\rm Map}(M^{d},B^{p+1}G)$.

Once we have reframed our understanding of $\fT$ in this way, the computation becomes almost automatic. We simply need to use one more result, originally from \cite{fbc52825-a403-38b8-87e5-64427df9db62},
\begin{equation}
    \pi_n{\rm Map}(M^d,B^{p+1}G) \cong H^{p+1-n}(M^d;G)\,.
\end{equation}
Applying this relation to \eqref{eq:app_homotopy_cardinality}, we thus find
\begin{equation}\label{eq:cardinality_of_T}
    \big|\fT\big| = \big|H^{p+1}(M^d;G)\big|\frac{\big|H^{p-1}(M^d;G)\big|\big|H^{p-3}(M^d;G)\big|\cdots}{\big|H^p(M^d;G)\big|\big|H^{p-2}(M^d;G)\big|\cdots}\,,
\end{equation}
where the first factor comes from the sum over connected components of ${\rm Map}(M^d,B^{p+1}G)$ in \eqref{eq:app_homotopy_cardinality}. Meanwhile, the partition function for a $(p+1)$-form gauge theory with gauge group $G$ can be computed as
\begin{equation}\label{eq:app_partition_function}
    Z[M^{d},G] = \frac{|H^{p+1}(M^{d};G)||H^{p-1}(M^{d};G)|\cdots}{|H^{p}(M^{d};G)||H^{p-2}(M^{d};G)|\cdots}\,,
\end{equation}
by counting the number of inequivalent gauge configurations, accounting for all (higher) gauge transformations. Hence, we see the desired equivalence between \eqref{eq:app_homotopy_cardinality} and \eqref{eq:app_partition_function}.




\bibliographystyle{./JHEP}
\bibliography{./refs}


\end{document}

%% file: auxx.tex
\newcommand{\Hom}{{\rm Hom}}

\definecolor{DiagramMagenta}{RGB}{189,16,224}
\definecolor{DiagramBlue}{RGB}{74,144,226}

\def\bC{\mathbb{C}}
\def\bZ{\mathbb{Z}}

\def\d{\delta}

\def\s{\sigma}

\newcommand{\ket}[1]{|{#1}\rangle}

\def\gcd{\mathop{\mathrm{gcd}}}

\def\fC{\mathfrak{C}}
\def\fT{\mathfrak{T}}

\def \d{\partial}

\newcommand{\beq}{\begin{equation}}
\newcommand{\eeq}{\end{equation}}

\newcommand\nn{\nonumber}

\newcommand{\cA}{\mathcal{A}}
\newcommand{\cB}{\mathcal{B}}
\newcommand{\cC}{\mathcal{C}}
\newcommand{\cD}{\mathcal{D}}

\newcommand{\cH}{\mathcal{H}}

\newcommand{\cO}{\mathcal{O}}

\newcommand{\cT}{\mathcal{T}}
\newcommand{\cU}{\mathcal{U}}

\newcommand{\cW}{\mathcal{W}}

\numberwithin{equation}{section}

\def\bea{\begin{eqnarray}}
\def\eea{\end{eqnarray}}

\DeclarePairedDelimiterX\MeijerM[3]{\lparen}{\rparen}%
{\begin{smallmatrix}#1 \\ #2\end{smallmatrix}\delimsize\vert\,#3}

\newcommand\MeijerG[8][]{%
  G^{\,#2,#3}_{#4,#5}\MeijerM[#1]{#6}{#7}{#8}}

\WithSuffix\newcommand\MeijerG*[7]{%
  G^{\,#1,#2}_{#3,#4}\MeijerM*{#5}{#6}{#7}}

\def\bC{\mathbb{C}}

\def\bZ{\mathbb{Z}}
\def\cH{\mathcal{H}}

\def\ket#1{|#1\rangle}

\def \beg#1{\begin{#1}} 

\def \bea{\beg{eqnarray}}
\def \eea{\end{eqnarray}}
\def \ee{\end{equation}}

\def \restr#1#2{{\left.\kern-\nulldelimiterspace#1\vphantom{\big|}\right|_{#2}}}

\def \bC{{\bf C}}

\def \nn{\nonumber}

\def \d{\partial}

\def \bC{\mathbb{C}}

\def \bZ{\mathbb{Z}}

